\documentclass[prd,showpacs,preprintnumbers,amsmath,amssymb, superscriptaddress,preprint, nofootinbib]{revtex4-1}
\usepackage{epsfig}
\usepackage{graphicx}
\usepackage{color}
\usepackage{float}
\usepackage{bm}
\usepackage{subcaption}
\usepackage{hyperref}
\usepackage[compat=1.1.0]{tikz-feynman}
\usepackage{tikz}

\definecolor{airforceblue}{rgb}{0.36, 0.54, 0.66}
\definecolor{blue(ncs)}{rgb}{0.0, 0.53, 0.74}
\definecolor{blue(blob)}{RGB}{40,94,140}
\definecolor{caribbeangreen}{rgb}{0.0, 0.8, 0.6}

\usepackage{stackengine}
\stackMath
\usepackage{graphicx}

\newcommand{\thickhline}{%
    \noalign {\ifnum 0=`}\fi \hrule height 1pt
    \futurelet \reserved@a \@xhline
}

\def\be{\begin{equation}}
\def\ee{\end{equation}}
\newcommand\bea{\begin{align}}
\newcommand\eea{\end{align}}
\def\nn{\nonumber}
\def\eqn#1{Equation~(\ref{#1})}

\def\sect#1{Section~{\ref{#1}}}

\begin{document}
\title{Extracting Einstein from the Loop-Level Double-Copy}
\author{John Joseph M. Carrasco}
\affiliation{Department of Physics \& Astronomy, Northwestern University, Evanston, Illinois 60208, USA}
\affiliation{Institut de Physique Theorique, Universite Paris Saclay, CEA, CNRS, F-91191 Gif-sur-Yvette, France}
\author{Ingrid A. Vazquez-Holm}
\affiliation{Institut de Physique Theorique, Universite Paris Saclay, CEA, CNRS, F-91191 Gif-sur-Yvette, France}
\date{\today}
\begin{abstract}
The naive double-copy of (multi) loop amplitudes involving massive matter coupled to gauge theories will generically produce  amplitudes in a gravitational theory that   contains additional contributions from propagating antisymmetric tensor and dilaton states even at tree-level.   We present a graph-based approach that combines the method of maximal cuts with double-copy construction to offer a systematic framework to isolate the pure Einstein-Hilbert gravitational contributions through loop level.  Indeed this allows for a bootstrap of pure-gravitational results from the double-copy of massive scalar-QCD.  We apply this to construct the novel result of the $D$-dimensional one-loop five-point QFT integrand relevant in the classical limit to generating observables associated with the radiative effects of massive black-hole scattering via pure Einstein-Hilbert gravity. 
\end{abstract}
\preprint{NUHEP-TH/21-03}

\maketitle
\newpage
\tableofcontents
\newpage

\section{Introduction}

The color-dual double copy construction~\cite{Bern:2008qj, BCJLoop} of quantum field theory predictions can be used to build scattering amplitudes for a wide variety of theories from phenomenological effective field theories to formal completions of Yang-Mills and Gravity like open and closed string theories\footnote{See, e.g., Section 5 of ref.~\cite{BCJreview} for a recent review of the color-dual web of theories.}.  The first concrete hint of this web of theories was the construction of closed string tree-level amplitudes from sums over permutations of Chan-Paton-stripped open-string amplitudes via the celebrated relations of Kawaii, Lewellen and Tye~\cite{KLT} (KLT). In the low energy limit these relations allow the expression of tree-level graviton amplitudes in terms of sums over permutations of ordered tree-level Yang-Mills amplitudes. 
 In the color-dual approach of Bern, Johansson, and one of the current authors (BCJ) the building blocks of gravitational predictions are gauge theory kinematic graph weights that satisfy the \textit{duality between color and kinematics}. This means that the gauge theory predictions can be expressed in a representation  where, graph-by-graph, kinematic weights obey the same algebraic relations as the color weights. The gravitational double copy prediction is then obtained by exchanging the gauge theory's color factors for a second set of kinematic weights.  This approach can be used to generate the field-theory limit of KLT relations at tree-level~\cite{Bern:2008qj}, by inverting the relationship between color-dual numerators and ordered-amplitudes, and generalizes to  integrand representations at multiloop levels~\cite{BCJLoop, GeneralizedDoubleCopy}. The theory of massive scalars minimally coupled to Yang-Mills, massive scalar-QCD, is compatible with this duality both at tree level \cite{Johansson:2015oia, Plefka:2019wyg}, as well as at the one-loop integrand level \cite{Carrasco:2020ywq}.

The advent of precision gravitational wave science has invited a renewed interest in applying quantum insights to classical predictions~(see, e.g.,~refs.\cite{Saotome2012vy, Monteiro2014cda, Luna2015paa,Ridgway2015fdl, Luna2016due, White2016jzc,   Goldberger2016iau, Cardoso2016amd, Luna2016hge, Goldberger2017frp, Adamo2017nia, DeSmet2017rve, BahjatAbbas2017htu, 
  CarrilloGonzalez2017iyj, Goldberger2017ogt, Li2018qap,Ilderton:2018lsf,  ShenWorldLine,  Lee:2018gxc, Plefka:2018dpa, CheungPM, Berman:2018hwd, Gurses:2018ckx,  Adamo:2018mpq, Bahjat-Abbas:2018vgo, Luna:2018dpt,Kosower:2018adc,Farrow:2018yni, Bern:2019nnu, 3PMLong,Antonelli:2019ytb,Damgaard:2019lfh,CarrilloGonzalez:2019gof,Maybee:2019jus, PV:2019uuv, Huang:2019cja, Alawadhi:2019urr, Emond:2020lwi, Berman:2020xvs, Bautista:2019evw, Herrmann:2021lqe, Mougiakakos:2020laz, Bern:2021dqo, Herrmann:2021tct, Bjerrum-Bohr:2021vuf, Bjerrum-Bohr:2021din}).  Indeed, the highest precision post-Minkowskian $\mathcal{O}(G^4)$ correction to the scattering of classical non-rotating black holes has involved a synthesis of effective field theory techniques, advanced multiloop integration innovation, and double-copy applied to tree-level scattering amplitudes~\cite{Bern:2021dqo}.  Scattering amplitudes in massive scalar-QCD, where massive scalars are coupled via Yang-Mills, can be used via  double-copy to construct the scattering of massive scalars coupled to gravitons, whose classical limit describes the evolution and interaction of non-rotating black holes.   At tree level, color-dual amplitudes in massive scalar QCD have been presented for up to six-points with three pairs of massive scalars \cite{Plefka:2019wyg}, and in ref.~\cite{Carrasco:2020ywq} it was conjectured that a simple bootstrap extends to all multiplicity and loop-level using only color-kinematics and factorization, verifying through 1-loop five-point.

The straightforward double copy of Yang-Mills scattering without any particular state management results in predictions of the so called  $\mathcal{N}=0$ supergravity, where -- in addition to gravitons -- a massless scalar (often called a dilaton) and an anti-symmetric two-form  (dual to an axion in four dimensions) can propagate.  These contributions can be avoided for external gravitons at tree-level for massless theories by coordinating the external states of gluons, but become relevant at loop level.  Such extra-state contributions affect even  tree-level when gravitons are coupled to massive matter. Such long-range mediators can survive the classical limit, and so must be projected out or otherwise removed if one is targeting classical predictions equivalent to that of pure Einstein-Hilbert gravity.  There can be additional differences in double-copy amplitudes, such as additional local contact terms only between massive states.  Such massive-state local contact terms are irrelevant to long range classical physics and so will not concern us here. 

There are many strategies for removing the extra states arising from naive double-copy.  The first examples of  gravitational multi-loop cut-construction involved tree-level cuts built from ordered gauge-theory cuts by applying  field theory KLT relations (see, e.g.~\cite{BDDPR}).  Cut construction for pure-gravity theories can progress by explicitly subtracting out contributions from unwanted states (c.f. refs.~\cite{Bern:1993wt, Dunbar:1994bn}).   At loop level, the classical limit of amplitude integrands for binary black holes  has been calculated up to three-loops using generalized unitarity cuts and the double copy of tree-level amplitudes to construct pure-graviton  integrands \cite{Bern:2021dqo} via the method of maximal cuts~\cite{Bern:2007ct} .

The method of maximal cuts~\cite{Bern:2007ct} offers a hierarchical approach to perturbative quantization. Here we combine it with double-copy construction at the off-shell integrand level.  Thus we construct pure Einstein-Hilbert gravity integrands by starting with compact expressions for the $\mathcal{N}=0$ supergravity integrands and systematically project out any non-gravitational propagating modes.   This projective double-copy offers a systematic isolation of local gravitational contributions.

Of course, if one is only interested in classical gravitation, it is not necessary to build consistent loop-level integrands at all, not to mention the generation of off-shell $\mathcal{N}=0$ supergravity integrands. One can employ~\cite{Bern:2021dqo} a variant of the method of maximal cuts to directly target only the relevant classical contributions.  If we aspire, however, to apprehend the connections between theories in the web of theories at the integrand level, it is natural to develop techniques to map out and relate graph by graph the relevant integrands of both $\mathcal{N}=0$ supergravity as well as Einstein-Hilbert as we present here.

It is a good question as to whether it is possible to  double-copy directly to pure-gravity integrands.  In the massless case Johansson and Ochirov introduced~{\cite{Johansson2014zca} a prescription involving ghostly matter to remove unwanted states at loop-level.   Inspired by this approach, ref.~\cite{Luna:2017dtq} constructed Einstein-Hilbert gravitation amplitudes  involving massive matter through five-points at tree level by introducing a massless, ghostly scalar in the gauge theory.  A side benefit of our approach is the generation of data which may clarify the path forward at loop-level involving massive matter. Starting from loop-level integrands of $\mathcal{N}=0$ supergravity,  we  generate {\em difference integrands} corresponding to the propagation of unwanted states.  This data may be helpful as a future loop-level double-copy target for state-ghost targeting of double-copy amplitudes with massive states. 

We will review the color-kinematics duality and the method of maximal cuts in \sect{review}. We introduce our approach to projective double-copy construction via the method of maximal cuts in \sect{projectiveDC}. In \sect{gravitycuts} we will   demonstrate our approach first at tree-level in \sect{treelevel},  then the one-loop correction to two distinct-mass scalar particles scattering gravitationally in \sect{loopfourpoint}, as well as including gravitational radiation in the related five-point one-loop amplitude integrand in \sect{loopfivepoint}. We conclude and discuss potential next steps in \sect{conclusion}.

\section{Review}
\label{review}

\subsection{Gravity amplitudes from the double copy}

The general form of a $m$-point $L$-loop gauge theory amplitude $\mathcal{A}^{(L)}_m$ in $D$ space-time dimensions with gluons and massive particles can be written as
\begin{equation}\label{Eq: General amplitude fraction}
\mathcal{A}^{(L)}_m = i^L g^{m-2+2L} \sum_{i \in \Gamma} \int \prod_{l=1}^L \frac{d^Dp_l}{(2 \pi)^D} \frac{1}{S_i} \frac{n_iC_i}{\prod_{\alpha_i}(p_{\alpha_i}^2 - m_{\alpha_i}^2)},
\end{equation}
where $g$ is the gauge coupling constant, and $m_{\alpha_i}^2$ is the on-shell mass of the particle with momentum $p_{\alpha_i}$. The sum runs over the complete set $\Gamma$ of $m$-point $L$-loop graphs with only cubic vertices, including all possible permutations of external legs.  The integrals are over the independent loop momenta $p_l$, and each graph is dressed with a kinematic numerator $n_i$ dependent upon the graph topology, a color factor $C_i$, and the propagator structure of the graph.  

Contact terms are encoded by allowing the cubic-graphs to be dressed with inverse-propagators. Due to the ambiguity from assigning contact terms to different cubic graphs, there is some freedom in choosing the kinematic numerators $n_i$. A fortuitous representation is one where the kinematic numerator weights $n_i$ are color-dual~\cite{Bern:2008qj, BCJLoop} -- where they obey the same algebraic relations as the color (or flavor) weights $C_i$.  At four points this is responsible for so-called radiation zeros~\cite{EarlyDualJacobi, Goebel:1980es} and happens for all assignations to only cubic graphs at this multiplicity.  It requires specific generalized gauge choice at higher multiplicity and loop order, and can be made manifest in a wide web of theories~\cite{BCJreview}, including non-supersymmetric matter in the fundamental~\cite{Johansson:2015oia, Luna:2017dtq, Plefka:2019wyg,Carrasco:2020ywq}.  This means that for Jacobi-triple color weights, $C_i$, the kinematic numerators, $n_i$, satisfy the same anti-symmetry properties and Jacobi-like relations as the color factors. If the kinematic graph weights of a given amplitude satisfy the duality, we can replace the color factors in \eqn{Eq: General amplitude fraction} with another set of kinematic numerators
\begin{equation}\label{Eq: Gravitational amplitude L-loop}
\mathcal{M}^{(L)}_m = i^{L+1} \Big( \frac{\kappa}{2} \Big)^{m-2+2L} \sum_{i \in \Gamma} \int \prod^L_{l=1} \frac{d^D p_l}{(2 \pi)^D} \frac{1}{S_i} \frac{n_i \tilde{n}_i}{\prod_{\alpha_i} (p_{\alpha_i}^2 - m_{\alpha_i}^2)},
\end{equation}
where the tilde on $\tilde{n}$ signifies that there can be two distinct gauge theories.  Any gauge-invariance that was predicated on the algebraic properties of the color-weights are automatically inherited by the double-copy.  Indeed, it is sufficient for only one of the numerator weights to make  manifest  the color kinematics duality. The expression obtained in Equation~\ref{Eq: Gravitational amplitude L-loop}  is a gravitational amplitude. This  generation of gravitational amplitudes is called \textit{double copy construction}.

There are many possible combinations of kinematic numerators from different gauge theories, with varying degrees of supersymmetry, which lead to amplitudes in different gravitational theories---for a recent review of the relevant literature see, e.g.~\cite{BCJreview}. In this paper we will start with the double copy of massive scalar-QCD theory predictions given in ref.~\cite{Carrasco:2020ywq}.  This generates predictions in a gravitational theory with propagating dilatons as well as antisymmetric tensor modes. In order to obtain amplitudes in Einstein-Hilbert gravity we will modify the method of maximal cuts to remove any non-gravitational propagating massless states.  Additionally in the gravitational double-copy there can be novel local contact terms between massive matter.  Such massive matter contact contributions are irrelevant to long-range predictions in any classical limit.

\subsection{Method of maximal cuts}

The method of maximal cuts iteratively constructs tree-level and the integrands of multi-loop scattering amplitudes by inverting the sufficient verification condition of a loop-integrand satisfying all unitarity cuts.  It is important to note that cut construction alone is only sufficient when there are no contact terms invisible to all factorization channels that one has not accounted for. Since unitarity cuts involving trees with uncut propagators span any unitarity cuts with additionally consistent cut propagators there is a natural hierarchy for construction.  Namely one can build weights  associated with graphs that first satisfy all possible cut conditions (all propagators put on shell), then consider what is required when when  cut conditions are relaxed to satisfy more inclusive cuts.  Any cut-constructible missing information (encodable potentially in local Feynman rules) must be proportional to newly uncut propagators in order to have vanished when such newly-relevant propagators were on-shell.

Consider a general unitarity cut with $k$ uncut propagators:
\begin{equation}\label{Eq:GeneralUnitarityCut}
\mathcal{C}^{N^kMC} = \sum_{\text{states}}\mathcal{M}^{\text{tree}}_{m(1)}...\mathcal{M}^{\text{tree}}_{m(p)}\hspace{2cm}k \equiv\sum_{i=1}^p m(i)-3,
\end{equation}
where $\mathcal{M}^{\text{tree}}_{m(i)}$ are tree-level $m(i)$ multiplicity gravity amplitudes, $k$ is the number of propagators that remain off-shell, and the sum runs over the possible states of the cut. 

The first step in the method of maximal cuts is therefore to perform the maximal cuts of all relevant topologies of the amplitude, and compare with the sum over states of the product of trees described above. Thus for each maximal cut, which can be represented by a cubic graph, all propagators are put on-shell.  The tree amplitudes in a maximal cut \eqn{Eq:GeneralUnitarityCut} are all three-point amplitudes, and so only one cubic graph contributes to each cut.  The contributions of maximal cuts can be unambiguously assigned to those/ cubic topologies.  For local representations there is no choice other than how to represent conservation of momentum zeros under cut conditions.  The only potential missing contributions must be proportional to inverse-propagators (momentum zeros for all maximal cuts).  To identify them we subsequently relax cut conditions. This subsequent step involves performing the next-to-maximal cuts (NMC) and assigning their contributions to any of the cubic graphs that contribute to that cut. For each next to maximal cut, only one of the cut conditions in a maximal cut is released, effectively returning propagators off-shell, so for full (orderless) cuts --- as relevant for gravitational or color-dressed theories --- at most\footnote{Flavor conservation can further restrict the number of potential graphs contributing to any cut.} three graphs could contribute to each cut.  For each such cut the only potentially new information will be proportional to the inverse of the newly uncut propagator, and this can be assigned or distributed amongst the three cubic topologies.   One proceeds similarly, for the N$^2$MC two propagators are held off-shell, and so on, until all potentially missing information is accounted for.


\section{Projective double-copy}
\label{projectiveDC}
If one were to attempt to bootstrap pure Einstein-Hilbert gravitational amplitudes for four external gravitons from three-point amplitudes, using the method of maximal cuts alone, one runs into trouble immediately due to the four-point contact term required for linearized diffeomorphism invariance.  Famously, gravitation requires additional contact terms at all multiplicity.  At tree-level one can exploit analyticity properties to guarantee gauge-invariance via on-shell recursion~\cite{BCFW,Benincasa:2007qj}.  Naive double-copy, which generalizes straightforwardly to loop-level, alleviates such worry, at the cost of potentially extra modes relative to pure-Einstein-Hilbert gravitation, and fixing a choice of theory dependent contact terms between massive-scalars.  Any weight given to massive-scalar contacts are of course irrelevant in the classical limit.  All supergravity theories, including the so-called $\mathcal{N}=0$ supergravity double-copy of pure Yang-Mills, share tree-level all-graviton amplitudes with pure Einstein-Hilbert gravitation. This of necessity changes at loop-level, where every state in the theory can contribute to loops.  Famously, even at tree-level, this universality is also no longer the case when coupling to massive massive matter~\cite{Luna:2017dtq}. Critically, identifying these extra-mode contributions on any given amplitude is accessible through the method of maximal cuts.

We bootstrap pure-gravitational scattering between massive scalars via double-copy and the method of maximal cuts by only allowing gravitons to flow through the cuts involving massless propagators. To this end we use the $D_s$-dimensional physical state projector, described in \cite{3PMLong} and references therein, to sew together the tree amplitudes,
\begin{equation}\label{Eq: Big gravity projector}
P^{\mu \nu \rho \sigma} (p,q) = \sum_{\text{pols.}}\varepsilon^{\mu \nu}(-p) \varepsilon^{\rho \sigma}(p) = \frac{1}{2} (P^{\mu \rho}P^{\nu \sigma} + P^{\mu \sigma} P^{\nu \rho}) - \frac{1}{D_s - 2} P^{\mu \nu}P^{\rho \sigma},
\end{equation}
where
\begin{equation}
P^{\mu \nu} = \eta^{\mu \nu} - \frac{q^{\mu} p^{\nu} + p^{\mu}q^{\nu}}{p \cdot q},
\end{equation}
where $q$ is an arbitrary null reference momenta that should factor out of any physical expression once the cut conditions and conservation of momenta have been imposed.   In certain cases the state projector is equivalent to the simpler gauge-dependent projector,
\begin{equation}\label{Eq: Nice gravity projector}
\sum_{\text{pols.}} \varepsilon_{\mu \nu}(-p) \varepsilon_{\rho \sigma}(p) = \frac{1}{2} \eta_{\mu \rho} \eta_{\nu \sigma} + \frac{1}{2} \eta_{\mu \sigma} \eta_{\nu \rho} - \frac{1}{D_s - 2} \eta_{\mu \nu} \eta_{\rho \sigma}.
\end{equation}

The state projector projects out non-gravitational states so that only gravitons are allowed to cross the cuts. We can then attribute any discrepancies between products of trees and the cut amplitude, or amplitude integrand, to excess states. We correct for these contributions by subtracting them from new higher-level contact terms as necessary.  One can assign these contact contributions to cubic parent graphs in a manner that obeys the symmetries of those graphs (via ansatz and explicit symmetrization).

The initial contact contribution, involving non gravitationally mediated massless states, to each graph is given by that graph's maximal cut -- where all propagators are cut: 
\begin{equation}
\Delta^{MC}_{g} = \mathcal{M}_g\big|^{MC} - \sum_{\text{projected states}}\mathcal{M}^{\text{tree}}_{3} \cdots \mathcal{M}^{\text{tree}}_{3},
\end{equation}
where $\mathcal{M}_g\big|^{MC}$ is the maximal cut of graph $g$.  The Einstein-Hilbert numerators, $n^\text{G}$, that satisfy these cut conditions are given by subtracting off a topology-symmetrized $\tilde{\Delta}^{MC}$ from the $\mathcal{N}=0$ numerator of the graph
\begin{equation}
n^{\text{G}}_g = n_g^{\mathcal{N}} - \tilde{\Delta}_g^{MC}.
\end{equation}
Any errors in $n^{\text{G}}$ must be proportional to contact terms, but critically can not involve contact terms due to all-graviton interaction, as those are correctly accounted for in the $\mathcal{N}=0$ supergravity numerators. 

Higher cuts proceed as above, but the initial contribution is a $\Delta^{N^kMC}_\text{contact}|$ which must be assigned to cubic parent graph $n^{\text{G}}_g$ via inverse propagators.   For these higher point contact corrections we use $\tilde{\Delta}_g$ to represent the symmetrized corrections including the inverse propagators that target a particular graph $g$. The procedure terminates when all physical cut conditions are released.

\section{Extracting Einstein-Hilbert at Tree and Loop Level}
\label{gravitycuts}

Here we carry out the program of bootstrapping pure gravity amplitudes from double copies of massive scalars coupled to gluons, by projecting out excess states at the level of the numerator dressings.   In \cite{Carrasco:2020ywq} a set of color-dual kinematic numerators was determined for amplitudes with gluons and massive scalar particles in scalar QCD. The `naive double copy' numerators are simply squares of these
\begin{equation} 
n^{\mathcal{N}= 0}_g = \left(n^{QCD}_g \right)^2,
\end{equation}
where $g$ is a cubic graph topology contributing to the amplitude, and $n_g^{QCD}$ is the kinematic numerator dressing from the gauge theory amplitude. The pure gravity three-point graph numerators are simply given by the double copy of the corresponding kinematic dressings found in \cite{Carrasco:2020ywq}
\begin{align}
\label{oneMassDC}
n^G \left(
\begin{gathered}
\begin{tikzpicture}
\begin{feynman}
\vertex (mid) at (0,0);
\vertex (a1) at (-0.8, -0.6) {\(k_2\)};
\vertex (a2) at (0, 1) {\(k_3\)};
\vertex (a3) at (0.8, -0.6) {\(k_1\)};
\diagram{(mid) --[gluon] (a2), (a1) --[thick, color=red](mid) --[thick, color=red](a3)};
\end{feynman}
\end{tikzpicture}
\end{gathered} \right) &= ~\left(k_1 \cdot \varepsilon_3\right)^2, \\
\label{pureGravDC}
n^G \left(
\begin{gathered}
\begin{tikzpicture}
\begin{feynman}
\vertex (mid) at (0,0);
\vertex (a1) at (-0.8, -0.6) {\(k_1\)};
\vertex (a2) at (0, 1) {\(k_2\)};
\vertex (a3) at (0.8, -0.6) {\(k_3\)};
\diagram{(mid) --[gluon] {(a1), (a2), (a3)}};
\end{feynman}
\end{tikzpicture}
\end{gathered} \right) &=
 \Big((k_3 \cdot \varepsilon_1)(\varepsilon_2 \cdot \varepsilon_3) - (k_3 \cdot \varepsilon_2)(\varepsilon_1 \cdot \varepsilon_3) + (k_2 \cdot \varepsilon_3)(\varepsilon_1 \cdot \varepsilon_2)\Big)^2 ,
\end{align}

We can now bootstrap from the above results to find the scattering amplitudes for massive scalars coupled gravitationally, including all appropriate gravitational self-interaction terms. 

\subsection{Tree-level}
\label{treelevel}

In addition to the graviton, the naive double-copy of the four-point tree-level amplitude with two massive scalars of distinct mass admits non-gravitational mediators, which we  project out on cuts using the physical graviton state projector as described in \sect{projectiveDC}. The tree-level case is relatively simple even at higher multiplicity, as only one channel is ever cut at a time.  The entire procedure can be understood by simply considering the four-point example in detail.  We tabulate all results with external massive scalars through six-point, including full amplitudes in auxiliary Mathematica files.  

The graph representation used here consists solely of cubic graphs,  so the four-point amplitude involves a single graph topology,
\begin{equation}
n \left(\begin{gathered}
\begin{tikzpicture}[scale=1.5]
\begin{feynman}
\vertex (a1) at (-1, -0.5) {\(a\)};
\vertex (a2) at (-1, 0.5) {\(b\)};
\vertex (mid1) at (-0.3,0);
\vertex (mid2) at (0.3,0);
\vertex (a3) at (1, 0.5) {\(c\)};
\vertex (a4) at (1, -0.5) {\(d\)};
\diagram{
(a1)--[thick, color=red] (mid1),
(a2) -- [thick, color=red] (mid1) ,
(mid1)--[gluon](mid2), 
(mid2)--[thick, color=blue(ncs)](a3), 
(a4) -- [thick, color=blue(ncs)](mid2)
};
\end{feynman}
\end{tikzpicture}
\end{gathered} \right) = n_{4,2}(a,b,c,d),
\end{equation}
which is dressed with a \textit{functional kinematic numerator} $n_{4,2}(a,b,c,d)$. The numerator function takes the ordered external legs $a,b,c,d$ as arguments, by which it is completely determined. The functional expression consists of Lorentz products of kinematic variables -- such as momenta and polarization vectors -- and must obey the same isomorphisms, or symmetries, as the corresponding graph topology. In \cite{Carrasco:2020ywq} this kinematic numerator was determined for massive scalars coupled to gluons
\begin{equation}
n^{QCD}_{4,2}(a, b, c, d) = -\frac{1}{2} \left( k_a \cdot k_b+ 2k_b \cdot k_c + m_1^2\right), 
\end{equation}
where $m_1^2$ is the squared mass of the scalar $a,b$, and $p_i \cdot p_j$ is the Lorentz product of momenta. The $\mathcal{N}=0$ supergravity numerator is the square of the QCD numerator
\begin{equation}
n^{\mathcal{N}=0}_{4,2}(a, b, c, d) = \left(n^{QCD}_{4,2}(a, b, c, d) \right)^2 = \frac{1}{4} \left( k_a \cdot k_b+ 2k_b \cdot k_c + m_1^2\right)^2.
\end{equation}

For this amplitude the projective double-copy requires fixing the only factorization channel of the graph. We find the product of trees as described above using the state projector in \eqn{Eq: Big gravity projector},
\begin{equation}
\begin{aligned}
\begin{gathered}
\begin{tikzpicture}[scale=1.5]
\begin{feynman}
\vertex (a1) at (-1, -0.5) {\(a\)};
\vertex (a2) at (-1, 0.5) {\(b\)};
\vertex (mid1) at (-0.35,0);
\vertex (mid11) at (-0.07,0);
\vertex (mid2) at (0.35,0);
\vertex (mid22) at (0.07,0);
\vertex (cut1) at (0, 0.55);
\vertex (cut2) at (0, -0.55);
\vertex (a3) at (1, 0.5) {\(c\)};
\vertex (a4) at (1, -0.5) {\(d\)};
\diagram{
(a1)--[thick, color=red] (mid1),
(a2) -- [thick, color=red] (mid1) ,
(mid1)--[gluon](mid11),
(mid22)--[gluon](mid2), 
(cut1) -- [dashed, red] (cut2),
(mid2)--[thick, color=blue(ncs)](a3), 
(a4) -- [thick, color=blue(ncs)](mid2)
};
\end{feynman}
\end{tikzpicture}
\end{gathered} 
=&~
\sum_s\mathcal{M}^{\text{trees}}_3(a, b, q^{s})\mathcal{M}^{\text{trees}}_3(-q^{\overline{s}}, c,d)\\
=&~(k_b \cdot k_d)^2 - \frac{m_1^2m_2^2}{(D_s-2)}.
\end{aligned}
\end{equation}
Recall,  $D_s$ is the number of space-time dimensions arising from the trace of the metric, $\delta^{\mu}_{ \mu} = D_s$, in the physical state projector.
The cut amplitude is given by
\begin{equation}
\mathcal{M}^{\text{tree}}_{4,2}(a, b, c, d)\big|_{q^2=0}^{\text{cut}} =
n^{\mathcal{N}=0}_{4,2}(a, b, c, d)\big|_{q^2=0}^{\text{cut}} = (k_b \cdot k_d)^2.
\end{equation}

The extra-state difference between the factorized numerators is identified,
\begin{equation}
\Delta_{4,2} =\mathcal{M}^{\text{tree}}_{4,2}(a, b, c, d)\big|_{q^2=0}^{\text{cut}} -  \sum_s\mathcal{M}^{\text{trees}}_3(a^{m_1}, b^{m_1}, q^{s})\mathcal{M}^{\text{trees}}_3(-q^{\overline{s}}, c^{m_2},d^{m_2}) = \frac{m_1^2m_2^2}{(D_s-2)},
\end{equation}
where massive particles are denoted in the tree-labels by a superscript $m_i$, and the two $\mathcal{M}^{\text{tree}}_3$ are the $3$-point gravity tree amplitudes given in \eqn{oneMassDC}. We then proceed to update the numerator function for pure-gravity exchange by subtracting the extra-state dimension
\begin{equation}
n^G_{4,2}(a, b, c, d) = n^{\mathcal{N}=0}_{4,2}(a, b, c, d) - \Delta_{4,2} = \frac{1}{4} \left( k_a \cdot k_b+ 2k_b \cdot k_c + m_1^2\right)^2 - \frac{m_1^2m_2^2}{(D_s-2)}.
\end{equation}
The full amplitude for two-to-two scattering at tree level in Einstein-Hilbert gravity (c.f. ref.~\cite{Luna:2017dtq})  is then found to be,
\begin{equation}
\mathcal{M}^{\text{tree}}_4 (a^{m_1}, b^{m_1}, c^{m_2}, d^{m_2}) = \frac{\frac{1}{4} \left( k_a \cdot k_b+ 2k_b \cdot k_c + m_1^2\right)^2 - \frac{m_1^2m_2^2}{(D_s-2)}}{(a+b)^2}.
\end{equation}

\begin{table}[t]
\begin{align*}
\begin{array}{c|c|c}
\noalign{\hrule height 1.2pt}
\text{Graph topology} & \mathcal{N}=0 \text{ supergravity numerator} & ~\Delta~ \\
\noalign{\hrule height 1.2pt}
\begin{gathered}
\begin{tikzpicture}[scale=1.5]
\begin{feynman}
\vertex (a1) at (-1, -0.5) {\(a\)};
\vertex (a2) at (-1, 0.5) {\(b\)};
\vertex (mid1) at (-0.3,0);
\vertex (mid2) at (0.3,0);
\vertex (a3) at (1, 0.5) {\(c\)};
\vertex (a4) at (1, -0.5) {\(d\)};
\diagram{
(a1)--[gluon] (mid1),
(a2) -- [gluon] (mid1),
(mid1)--[gluon](mid2), 
(mid2)--[gluon](a3), 
(a4) -- [gluon](mid2)
};
\end{feynman}
\end{tikzpicture}
\end{gathered} &
\begin{gathered}
\begin{aligned}
\frac{1}{4} &\Big[ 2 (\varepsilon_a\cdot \varepsilon_b) \big((k_b\cdot \varepsilon_c)( k_c\cdot \varepsilon_d)- (k_b\cdot \varepsilon_d)(k_d\cdot \varepsilon_c) \big)\\
&- (\varepsilon_a\cdot \varepsilon_d) \big(2 (k_d\cdot \varepsilon_c)(k_a\cdot \varepsilon_b)
+(k_a \cdot k_b) (\varepsilon_b\cdot \varepsilon_c)\big)\\
&-2 (k_c\cdot \varepsilon_d) ((\varepsilon_b\cdot \varepsilon_c) (k_b\cdot \varepsilon_a)-(\varepsilon_a\cdot \varepsilon_c)( k_a\cdot \varepsilon_b))\\
&+ (\varepsilon_b\cdot \varepsilon_d)\big((k_a \cdot k_b) (\varepsilon_a\cdot \varepsilon_c) +2 (k_d\cdot \varepsilon_c)( k_b\cdot \varepsilon_a) \big)\\
&-(\varepsilon_c\cdot \varepsilon_d) (2 (k_d\cdot \varepsilon_a)(k_c\cdot \varepsilon_b)-2 (k_c\cdot \varepsilon_a)(k_d\cdot \varepsilon_b)\\
&+((k_a \cdot k_b)+2 (k_b \cdot k_c)) (\varepsilon_a\cdot \varepsilon_b))\Big]^2
\end{aligned}
\end{gathered}
& 0\\
\begin{gathered}
\begin{tikzpicture}[scale=1.5]
\begin{feynman}
\vertex (a1) at (-1, -0.5) {\(b\)};
\vertex (a2) at (-1, 0.5) {\(c\)};
\vertex (mid1) at (-0.3,0);
\vertex (mid2) at (0.3,0);
\vertex (a3) at (1, 0.5) {\(d\)};
\vertex (a4) at (1, -0.5) {\(a\)};
\diagram{
(a1)--[gluon] (mid1),
(a2) -- [thick, color=red] (mid1) ,
(mid1)--[thick, color=red](mid2), 
(mid2)--[thick, color=red](a3), 
(a4) -- [gluon](mid2)
};
\end{feynman}
\end{tikzpicture}
\end{gathered} & 
\Big[(k_c\cdot \varepsilon_b) \left(k_d\cdot \varepsilon_a\right)+\frac{1}{2} (k_b \cdot k_c) (\varepsilon_a\cdot \varepsilon_b)\Big]^2
 & 0\\
\begin{gathered}
\begin{tikzpicture}[scale=1.5]
\begin{feynman}
\vertex (a1) at (-1, -0.5) {\(c\)};
\vertex (a2) at (-1, 0.5) {\(d\)};
\vertex (mid1) at (-0.3,0);
\vertex (mid2) at (0.3,0);
\vertex (a3) at (1, 0.5) {\(a\)};
\vertex (a4) at (1, -0.5) {\(b\)};
\diagram{
{(a1), (a2)} --[thick, color=red] (mid1), 
(mid1)--[gluon](mid2), 
{(a3), (a4)}--[gluon](mid2)
};
\end{feynman}
\end{tikzpicture}
\end{gathered} &
\begin{gathered}
\begin{aligned}
\Big[&(k_a\cdot \varepsilon_b)(k_c\cdot \varepsilon_a)- (k_b\cdot \varepsilon_a)(k_c\cdot \varepsilon_b)
-\frac{1}{2}\left(k_{ab}^-\cdot k_c\right)(\varepsilon_a\cdot \varepsilon_b)\Big]^2
\end{aligned}
\end{gathered}
 & 0\\
\begin{gathered}
\begin{tikzpicture}[scale=1.5]
\begin{feynman}
\vertex (a1) at (-1, -0.5) {\(d\)};
\vertex (a2) at (-1, 0.5) {\(e\)};
\vertex (mid1) at (-0.3,0);
\vertex (mid2) at (0.3,0);
\vertex (mid3) at (0.9, 0);
\vertex (a5) at (0.3, 0.5) {\(a\)};
\vertex (a3) at (1.6, 0.5) {\(b\)};
\vertex (a4) at (1.6, -0.5) {\(c\)};
\diagram{
(mid2) --[gluon] (a5),
(mid2) -- [gluon](mid3),
(a1)--[thick, color=red] (mid1),
(a2) -- [thick, color=red] (mid1) ,
(mid1)--[gluon](mid2), 
(mid3)--[gluon](a3), 
(a4) -- [gluon](mid3)
};
\end{feynman}
\end{tikzpicture}
\end{gathered} &
 (k_b\cdot \varepsilon_a)^2 (k_b\cdot \varepsilon_c)^2 (k_d\cdot \varepsilon_b)^2+...+\frac{1}{4}(k_c\cdot k_e)^2 (k_e\cdot \varepsilon_a)^2 (\varepsilon_b\cdot \varepsilon_c)^2
 & 0 \\
\begin{gathered}
\begin{tikzpicture}[scale=1.5]
\begin{feynman}
\vertex (a1) at (-1, -0.5) {\(e\)};
\vertex (a2) at (-1, 0.5) {\(a\)};
\vertex (mid1) at (-0.3,0);
\vertex (mid2) at (0.3,0);
\vertex (mid3) at (0.9, 0);
\vertex (a5) at (0.3, -0.5) {\(d\)};
\vertex (a3) at (1.6, 0.5) {\(b\)};
\vertex (a4) at (1.6, -0.5) {\(c\)};
\diagram{
(mid2) --[thick, color=red] (a5),
(mid2) -- [gluon](mid3),
(a1)--[thick, color=red]  (mid1),
(a2) --[gluon] (mid1) ,
(mid1)--[thick, color=red](mid2), 
(mid3)--[gluon](a3), 
(a4) -- [gluon](mid3)
};
\end{feynman}
\end{tikzpicture}
\end{gathered}  & 
(k_b\cdot \varepsilon_a)^2(k_b\cdot \varepsilon_c)^2 (k_d\cdot  \varepsilon_b)^2+...+\frac{1}{4}(k_c\cdot k_d)^2 (k_d\cdot \varepsilon_a)^2 (\varepsilon_b\cdot \varepsilon_c)^2
& 0 \\
\begin{gathered}
\begin{tikzpicture}[scale=1.5]
\begin{feynman}
\vertex (a1) at (-1, -0.5) {\(e\)};
\vertex (a2) at (-1, 0.5) {\(a\)};
\vertex (mid1) at (-0.3,0);
\vertex (mid2) at (0.3,0);
\vertex (mid3) at (0.9, 0);
\vertex (a5) at (0.3, 0.5) {\(b\)};
\vertex (a3) at (1.6, 0.5) {\(c\)};
\vertex (a4) at (1.6, -0.5) {\(d\)};
\diagram{
(mid2) --[gluon] (a5),
(mid2) -- [thick, color=red](mid3),
(a1)--[thick, color=red] (mid1),
(a2) -- [gluon] (mid1) ,
(mid1)--[thick, color=red](mid2), 
(mid3)--[gluon](a3), 
(a4) -- [thick, color=red](mid3)
};
\end{feynman}
\end{tikzpicture}
\end{gathered}
& 
(k_b\cdot \varepsilon_a)^2 (k_c\cdot \varepsilon_b)^2 (k_d\cdot \varepsilon_c)^2+...+\frac{1}{4} (k_c\cdot k_d)^2 (k_d\cdot \varepsilon_a)^2 (\varepsilon_b\cdot \varepsilon_c)^2
& 0 
\end{array}
\end{align*}
\caption{Tree amplitude graphs for one massive scalar at four and five-point, where  $k_{ij}=k_i+k_j$ and $k_{ij}^- = k_i - k_j$. The correction $\Delta$ should be subtracted from the $\mathcal{N}=0$ supergravity dressing to obtain the pure graviton numerator.}
\label{Tab: 1 massive scalar trees dilatons}
\end{table}

Higher multiplicity proceeds as above where first one considers any corrections from maximal cuts and proceeds to release all cut conditions. The $\mathcal{N}=0$ supergravity dressings and excess state contributions $\Delta$ are shown in Table~\ref{Tab: 1 massive scalar trees dilatons} for amplitudes with one massive scalar. As seen from the table there are no excess state corrections for the tree amplitudes with a single scalar line. This is not unexpected, as the states are expected to propagate between scalar lines. 

\begin{table}[t]
\begin{align*}
\begin{array}{c|c|c}
\noalign{\hrule height 1.2pt}
\text{Graph topology}  & \mathcal{N}=0\text{ supergravity numerator} & \Delta\\
\noalign{\hrule height 1.2pt}
\begin{gathered}
\begin{tikzpicture}[scale=1.5]
\begin{feynman}
\vertex (a1) at (-1, -0.5) {\(a\)};
\vertex (a2) at (-1, 0.5) {\(b\)};
\vertex (mid1) at (-0.3,0);
\vertex (mid2) at (0.3,0);
\vertex (a3) at (1, 0.5) {\(c\)};
\vertex (a4) at (1, -0.5) {\(d\)};
\diagram{
(a1)--[thick, color=red] (mid1),
(a2) -- [thick, color=red] (mid1) ,
(mid1)--[gluon](mid2), 
(mid2)--[thick, color=blue(ncs)](a3), 
(a4) -- [thick, color=blue(ncs)](mid2)
};
\end{feynman}
\end{tikzpicture}
\end{gathered} & 
\begin{gathered}
\frac{1}{4}  \big(k_a \cdot k_b+2~ k_b \cdot k_c+m_1^2\big)^2
\end{gathered} & 
\begin{gathered}
\frac{m_1^2 m_2^2}{D_s-2}
\end{gathered}
\\
\begin{gathered}
\begin{tikzpicture}[scale=1.5]
\begin{feynman}
\vertex (a1) at (-1, -0.5) {\(d\)};
\vertex (a2) at (-1, 0.5) {\(e\)};
\vertex (mid1) at (-0.3,0);
\vertex (mid2) at (0.3,0);
\vertex (mid3) at (0.9, 0);
\vertex (a5) at (0.3, 0.5) {\(a\)};
\vertex (a3) at (1.6, 0.5) {\(b\)};
\vertex (a4) at (1.6, -0.5) {\(c\)};
\diagram{
(mid2) --[gluon] (a5),
(mid2) -- [gluon](mid3),
(a1)--[thick, color=red] (mid1),
(a2) -- [thick, color=red] (mid1) ,
(mid1)--[gluon](mid2), 
(mid3)--[thick, color=blue(ncs)](a3), 
(a4) -- [thick, color=blue(ncs)](mid3)
};
\end{feynman}
\end{tikzpicture}
\end{gathered} & 
\begin{gathered}
\begin{aligned}
\frac{1}{16}\Big[(k_{bc}^- \cdot k_1)(k_{de}^-\cdot \varepsilon_1)
&+ 2(k_c \cdot k_{de}^-)(k_b \cdot \varepsilon_a)\\
&- 2(k_b \cdot k_{de}^-)(k_c \cdot \varepsilon_a)\Big]{}^2
\end{aligned}
\end{gathered}
 & 
 \begin{gathered}
\frac{ m_1^2m_2^2}{D_s-2} \left(k_{de} \cdot \varepsilon_a\right)^2
\end{gathered}
\\
\begin{gathered}
\begin{tikzpicture}[scale=1.5]
\begin{feynman}
\vertex (a1) at (-1, -0.5) {\(a\)};
\vertex (a2) at (-1, 0.5) {\(b\)};
\vertex (mid1) at (-0.3,0);
\vertex (mid2) at (0.3,0);
\vertex (mid3) at (0.9, 0);
\vertex (a5) at (0.3, 0.5) {\(c\)};
\vertex (a3) at (1.6, 0.5) {\(d\)};
\vertex (a4) at (1.6, -0.5) {\(e\)};
\diagram{
(mid2) --[thick, color=blue(ncs)] (a5),
(mid2) -- [gluon](mid3),
(a1)-- [gluon] (mid1),
(a2) --[thick, color=blue(ncs)] (mid1) ,
(mid1)--[thick, color=blue(ncs)](mid2), 
(mid3)--[thick, color=red](a3), 
(a4) -- [thick, color=red](mid3)
};
\end{feynman}
\end{tikzpicture}
\end{gathered} & 
\begin{gathered}
\frac{1}{16}\Big[\left(k_{ce} \cdot k_{de}\right) (k_{de}^- \cdot \varepsilon_a)+2~ (k_c \cdot
k_{de}^-) (k_b \cdot \varepsilon_a)\Big]^2
\end{gathered}
 & 
 \begin{gathered}
\frac{m_1^2m_2^2 }{D_s-2}(k_b \cdot \varepsilon_a)^2
\end{gathered}
\end{array}
\end{align*}
\caption{Tree amplitude graphs for two massive scalars, at four and five-point, where  $k_{ij}=k_i+k_j$ and $k_{ij}^- = k_i - k_j$. The correction $\Delta$ should be subtracted from the $\mathcal{N}=0$ supergravity dressing to obtain the pure graviton numerator.}
\label{Tab: 2 massive scalars trees dilatons}
\end{table}

The tree-level dressings for four-point and five-point amplitudes with two scalar lines are shown in Table~\ref{Tab: 2 massive scalars trees dilatons}, and the corrections for the six-point amplitude with two massive scalars are shown in Table~\ref{Tab: 2 massive scalars 6p trees dilatons}. These graphs \textit{do} require corrections from $\mathcal{N}=0$ SG. We can now easily calculate the first correction to massive scalar scattering with radiation. The amplitude gets contributions from the second topology in Table~\ref{Tab: 2 massive scalars trees dilatons} and four differently-labeled versions of the third topology in Table~\ref{Tab: 2 massive scalars trees dilatons}, using $k_{ij}^-\equiv k_i-k_j$,
\begin{multline}
\mathcal{M}^{\text{tree}}_5(k_1, k_2^{m_1}, k_3^{m_1}, k_4^{m_2}, k_5^{m_2}) = \\~\frac{\left(2 \left(k_3\cdot k_{45}^-\right) \left(k_2 \cdot\varepsilon_1\right)+\left(k_1\cdot k_2\right) \left(k_{45}^-\cdot \varepsilon_1\right)\right){}^2-\frac{1}{D_s-2}16 m_1^2 m_2^2 \left(k_2\cdot \varepsilon_1\right){}^2}{16 \left(k_4+k_5\right){}^2 \left(\left(k_1+k_2\right){}^2-m_1^2\right)}
\\
+\frac{\left(2 \left(k_2\cdot k_{45}^-\right) (k_3\cdot \varepsilon_1)+\left(k_1\cdot k_3\right) \left(k_{45}^-\cdot \varepsilon_1\right)\right){}^2-\frac{1}{D_s-2}16 m_1^2 m_2^2 \left(k_3\cdot \varepsilon_1\right){}^2}{16 \left(k_4+k_5\right){}^2 \left(\left(k_1+k_3\right){}^2-m_1^2\right)}
\\
+\frac{\left(2 \left( k_4 \cdot k_{23}^-\right) (k_5\cdot \varepsilon_1)+\left(k_1 \cdot k_5\right) \left(k_{23}^-\cdot \varepsilon_1\right)\right){}^2-\frac{1}{D_s-2}16 m_1^2 m_2^2 \left(k_5\cdot \varepsilon_1\right){}^2}{16 \left(k_2+k_3\right){}^2 \left(\left(k_1+k_5\right){}^2-m_2^2\right)}
\\
+\frac{\left(2 \left( k_5 \cdot k_{23}^-\right) (k_4\cdot \varepsilon_1)+(k_1 \cdot k_4) \left(k_{23}^-\cdot \varepsilon_1\right)\right){}^2-\frac{1}{D_s-2}16 m_1^2 m_2^2 \left(k_4\cdot \varepsilon_1\right){}^2}{16 \left(k_2+k_3\right){}^2 \left(\left(k_1+k_4\right){}^2-m_2^2\right)}
\\
+\frac{\left((k_{23}^- \cdot k_1)(k_{45}^-\cdot \varepsilon_1)+ 2(k_3 \cdot k_{45}^-)(k_2 \cdot \varepsilon_1)
- 2(k_2 \cdot k_{45}^-)(k_3 \cdot \varepsilon_1)\right){}^2- \frac{1}{D_s-2}16m_1^2 m_2^2 \left(k_{45}\cdot \varepsilon_1\right){}^2}{16 \left(k_2+k_3\right){}^2 \left(k_4+k_5\right){}^2}.
\end{multline}

\begin{table}[H]
\begin{align*}
\begin{array}{c|c}
\noalign{\hrule height 1.2pt}
\text{Graph topology}   &  \Delta\\
\noalign{\hrule height 1.2pt}
\begin{gathered}
\begin{tikzpicture}[scale=1.5]
\begin{feynman}
\vertex (a0) at (-1, 0.1) {\(b\)};
\vertex (a1) at (-1, -0.5) {\(a\)};
\vertex (a2) at (-1, 0.5) {\(f\)};
\vertex (mid1) at (-0.3,0);
\vertex (mid1m) at (-0.5,-0.2);
\vertex (mid2) at (0.3,0);
\vertex (mid3) at (0.9, 0);
\vertex (a5) at (0.3, 0.5) {\(e\)};
\vertex (a3) at (1.6, 0.5) {\(c\)};
\vertex (a4) at (1.6, -0.5) {\(d\)};
\diagram{
(mid2) --[thick, color=blue(ncs)] (a5),
(mid2) -- [gluon](mid3),
(a1)-- [gluon] (mid1m)--[gluon] (mid1),
(mid1m) -- [gluon](a0),
(a2) --[thick, color=blue(ncs)] (mid1) ,
(mid1)--[thick, color=blue(ncs)](mid2), 
(mid3)--[thick, color=red](a3), 
(a4) -- [thick, color=red](mid3)
};
\end{feynman}
\end{tikzpicture}
\end{gathered} 
 & 
\begin{gathered} 
\begin{aligned}
 \frac{1}{4} \frac{m_1^2 m_2^2}{D_s-2} 
 \Big[ &\left(k_b \cdot k_f-k_a\cdot k_f\right)(\varepsilon_a \cdot \varepsilon_b) \\
 &-2\left((k_{cde} \cdot \varepsilon_b)~ k_f \cdot \varepsilon_a -(k_{cde}\cdot \varepsilon_a)~k_f\cdot \varepsilon_b \right) \Big]^2
\end{aligned}
\end{gathered}
 \\
 \begin{gathered}
\begin{tikzpicture}[scale=1.5]
\begin{feynman}
\vertex (a0) at (-1, 0.1) {\(a\)};
\vertex (a1) at (-1, -0.5) {\(f\)};
\vertex (a2) at (-1, 0.5) {\(b\)};
\vertex (mid1) at (-0.3,0);
\vertex (mid1m) at (-0.5,-0.2);
\vertex (mid2) at (0.3,0);
\vertex (mid3) at (0.9, 0);
\vertex (a5) at (0.3, 0.5) {\(e\)};
\vertex (a3) at (1.6, 0.5) {\(c\)};
\vertex (a4) at (1.6, -0.5) {\(d\)};
\diagram{
(mid2) --[thick, color=blue(ncs)] (a5),
(mid2) -- [gluon](mid3),
(a1)-- [thick, color=blue(ncs)] (mid1m)--[thick, color=blue(ncs)] (mid1),
(mid1m) -- [gluon](a0),
(a2) --[gluon] (mid1) ,
(mid1)--[thick, color=blue(ncs)](mid2), 
(mid3)--[thick, color=red](a3), 
(a4) -- [thick, color=red](mid3)
};
\end{feynman}
\end{tikzpicture}
\end{gathered} 
&
\begin{gathered} 
\begin{aligned}
\frac{1}{64} \frac{m_1^2 m_2^2}{D_s-2} 
\Big[\left(k_{cd} \cdot k_{de}-4 k_a\cdot k_f\right)(\varepsilon_a\cdot \varepsilon_b)
-8 (k_f\cdot \varepsilon_a) (k_{cde} \cdot \varepsilon_b)
\Big]^2
\end{aligned}
\end{gathered}
\\
\begin{gathered}
\begin{tikzpicture}[scale=1.5]
\begin{feynman}
\vertex (a1) at (-1, -0.5) {\(c\)};
\vertex (a2) at (-1, 0.5) {\(d\)};
\vertex (mid1) at (-0.3,0);
\vertex (mid2) at (0.3,0);
\vertex (mid3) at (0.9, 0);
\vertex (a5) at (0.3, 0.2) ;
\vertex (t1) at (0,0.7) {\(a\)};
\vertex (t2) at (0.6,0.7) {\(b\)};
\vertex (a3) at (1.6, 0.5) {\(e\)};
\vertex (a4) at (1.6, -0.5) {\(f\)};
\diagram{
(mid2) --[gluon] (a5) --[gluon](t1),
(a5) --[gluon](t2),
(mid2) -- [gluon](mid3),
(a1)--[thick, color=blue(ncs)] (mid1),
(a2) -- [thick, color=blue(ncs)] (mid1) ,
(mid1)--[gluon](mid2), 
(mid3)--[thick, color=red](a3), 
(a4) -- [thick, color=red](mid3)
};
\end{feynman}
\end{tikzpicture}
\end{gathered} 
 & 
 \begin{gathered} 
 \begin{aligned}
\frac{1}{4} \frac{m_1^2 m_2^2}{D_s-2} 
\Big[&\left(2~k_{be} \cdot k_{ef}+k_{cd} \cdot k_{ef}\right)(\varepsilon_a\cdot \varepsilon_b) \\
&-2 (k_{ef}\cdot \varepsilon_a) (k_{cd}\cdot \varepsilon_b)+2 (k_{cd}\cdot \varepsilon_a) (k_{ef} \cdot \varepsilon_b)\Big]^2
\end{aligned}
\end{gathered}
 \\
  \begin{gathered}
\begin{tikzpicture}[scale=1.5]
\begin{feynman}
\vertex (a0) at (-1, 0.1) {\(a\)};
\vertex (a1) at (-1, -0.5) {\(f\)};
\vertex (a2) at (-1, 0.5) {\(e\)};
\vertex (mid1) at (-0.3,0);
\vertex (mid1m) at (-0.5,-0.2);
\vertex (mid2) at (0.3,0);
\vertex (mid3) at (0.9, 0);
\vertex (a5) at (0.3, 0.5) {\(b\)};
\vertex (a3) at (1.6, 0.5) {\(c\)};
\vertex (a4) at (1.6, -0.5) {\(d\)};
\diagram{
(mid2) --[gluon] (a5),
(mid2) -- [gluon](mid3),
(a1)-- [thick, color=red] (mid1m)--[thick, color=red] (mid1),
(mid1m) -- [gluon](a0),
(a2) --[thick, color=red] (mid1) ,
(mid1)--[gluon](mid2), 
(mid3)--[thick, color=blue(ncs)](a3), 
(a4) -- [thick, color=blue(ncs)](mid3)
};
\end{feynman}
\end{tikzpicture}
\end{gathered} 
&
\begin{gathered} 
\begin{aligned}
\frac{1}{64} \frac{m_1^2 m_2^2}{D_s-2}
\Big[\left(k_{bcd}\cdot k_e+k_{cd}\cdot k_d-3k_a \cdot k_f \right)(\varepsilon_a\cdot \varepsilon_b )
-8 (k_f\cdot \varepsilon_a) (k_{cd}\cdot \varepsilon_b)\Big]^2
\end{aligned}
\end{gathered}
\\
\begin{gathered}
\begin{tikzpicture}[scale=1.5]
\begin{feynman}
\vertex (a1) at (-1, -0.5) {\(f\)};
\vertex (a2) at (-1, 0.5) {\(a\)};
\vertex (mid1) at (-0.3,0);
\vertex (mid2) at (0.3,0);
\vertex (mid3) at (0.9, 0);
\vertex (a5) at (0.3, 0.2) ;
\vertex (t1) at (0,0.7) {\(c\)};
\vertex (t2) at (0.6,0.7) {\(d\)};
\vertex (a3) at (1.6, 0.5) {\(b\)};
\vertex (a4) at (1.6, -0.5) {\(e\)};
\diagram{
(mid2) --[gluon] (a5) --[thick, color=blue(ncs)](t1),
(a5) --[thick, color=blue(ncs)](t2),
(mid2) -- [thick, color=red](mid3),
(a1)--[thick, color=red] (mid1),
(a2) -- [gluon] (mid1) ,
(mid1)--[thick , color=red](mid2), 
(mid3)--[gluon](a3), 
(a4) -- [thick, color=red](mid3)
};
\end{feynman}
\end{tikzpicture}
\end{gathered} 
 & 
 \begin{gathered} 
 \begin{aligned}
\frac{1}{64} \frac{m_1^2 m_2^2}{D_s-2}
\Big[\left(k_b\cdot k_e + k_a \cdot k_f\right)(\varepsilon_a \cdot \varepsilon_b)+8 (k_f\cdot \varepsilon_a) (k_e\cdot \varepsilon_b)\Big]^2
\end{aligned}
\end{gathered}
\\
\begin{gathered}
\begin{tikzpicture}[scale=1.5]
\begin{feynman}
\vertex (a1) at (-1, -0.5) {\(c\)};
\vertex (a2) at (-1, 0.5) {\(d\)};
\vertex (mid1) at (-0.3,0);
\vertex (mid2a) at (0.1,0);
\vertex (mid2b) at (0.5,0);
\vertex (mid3) at (0.9, 0);
\vertex (a5) at (0.3, 0.2) ;
\vertex (t1) at (-0.1,0.5) {\(b\)};
\vertex (t2) at (0.6,0.5) {\(a\)};
\vertex (a3) at (1.6, 0.5) {\(e\)};
\vertex (a4) at (1.6, -0.5) {\(f\)};
\diagram{
(a1)--[thick, color=blue(ncs)] (mid1),
(a2) -- [thick, color=blue(ncs)] (mid1) ,
(mid1) -- [gluon](mid3),
(mid2a) -- [gluon](t1),
(mid2b) -- [gluon](t2),
(mid3)--[thick, color=red](a3), 
(a4) -- [thick, color=red](mid3)
};
\end{feynman}
\end{tikzpicture}
\end{gathered} 
 & 
\begin{gathered}
\begin{aligned}
\frac{1}{16} \frac{m_1^2 m_2^2}{D_s-2} \Big[
 \left(2 k_{bcd}\cdot k_{ef}+k_c\cdot k_{cd}+3k_e\cdot k_{ef}\right)(\varepsilon_a\cdot \varepsilon_b)-4 (k_{ef}\cdot \varepsilon_a) (k_{cd}\cdot\varepsilon_b)
\Big]^2
\end{aligned}
\end{gathered}
 \\
 \begin{gathered}
\begin{tikzpicture}[scale=1.5]
\begin{feynman}
\vertex (a0) at (-1, 0.1) {\(b\)};
\vertex (a5) at (1.6, 0.1) {\(a\)};
\vertex (a1) at (-1, -0.5) {\(c\)};
\vertex (a2) at (-1, 0.5) {\(d\)};
\vertex (mid1) at (-0.2,0);
\vertex (mid1m) at (-0.5,-0.2);
\vertex (mid3) at (0.8, 0);
\vertex (mid3m) at (1.1,-0.2);
\vertex (a3) at (1.6, 0.5) {\(e\)};
\vertex (a4) at (1.6, -0.5) {\(f\)};
\diagram{
(mid1) -- [gluon](mid3),
(a1)-- [thick, color=blue(ncs)] (mid1m)--[thick, color=blue(ncs)] (mid1),
(mid1m) -- [gluon](a0),
(mid3m) -- [gluon](a5),
(a2) --[thick, color=blue(ncs)] (mid1) ,
(mid3)--[thick, color=red](a3), 
(mid3m) -- [thick, color=red](mid3),
(mid3m) -- [thick, color=red](a4)
};
\end{feynman}
\end{tikzpicture}
\end{gathered} 
&
\begin{gathered}
\begin{aligned}
\frac{1}{64} \frac{m_1^2 m_2^2}{D_s-2} \Big[
 \left(k_{bcd} \cdot k_{de}-2 k_a\cdot k_f\right)(\varepsilon_a\cdot \varepsilon_b)
 -8 (k_f\cdot \varepsilon_a)( k_c\cdot \varepsilon_b)\Big]^2
\end{aligned}
\end{gathered}
\end{array}
\end{align*}
\caption{The graph topologies for the six-point tree amplitude for two distinct-mass scalars, and the corresponding contributions from non-gravitational states $\Delta$, where  $k_{ij}=k_i+k_j$.}
\label{Tab: 2 massive scalars 6p trees dilatons}
\end{table}

\subsection{Loop Level}

The physical cuts relevant to one-loop are relatively straightforward. For both four-point one-loop and five-point one-loop  reduction to standard integral basis means satisfying cuts to tree-level under the following restrictions:
\begin{itemize}
\item  Every bubble cut (two-particle cut) involves physical tree amplitudes (multiplicity four or higher). These are N$^2$MC1-N$^2$MC4 in Figure~\ref{Fig: 4p1lN2MC} for the four-point amplitude, and N$^3$MC1-N$^3$MC5 in Figure~\ref{Fig: 5p n3max cuts} for the five-point amplitude.
\item  Every triangle cut (three-particle cut) must involve the loop momenta flowing through each tree. These are cuts NMC1, NMC2 and NMC3 in Figure~\ref{Fig: 4p1lNMC} for the four-point amplitude, and N$^2$MC1-N$^2$MC11 in Figure~\ref{Fig: 5p n2max triangle}.
\item Every box cut (four-particle cut) for the five-point amplitude, must involve the loop-momenta flowing through each tree. These are cut MC1 in Figure~\ref{Fig: 4p1lMC} for the four-point amplitude and cuts NMC1-NMC7 in Figure~\ref{Fig: 5pNMax cuts box}.
\end{itemize}
This set of restrictions precludes one-loop snail\footnote{Apparently one-particle reducible topologies can contribute in an all cubic diagram basis, but only when dressed  with appropriate inverse-propagators to actually be encoding one-particle irreducible contact-diagrams~\cite{SimplifyingBCJ}.} diagrams as well as tadpole diagrams from contributing to the cuts. Any such contributions are not cut constructible with standard (non-regulated) approaches --- even in arbitrary dimensions --- representing a distinct set of considerations, but one absolutely complementary and compatible with unitary methods as per ref.~\cite{BernMorgan}.  As such for each multiplicity at one loop, we only concern ourselves with the series of maximal cuts which can contribute to these physical cuts.  This entirely spans any information relevant to the classical $\hbar\to0$ limit.

\subsubsection{Four-point one-loop}
\label{loopfourpoint}

\begin{figure}
\begin{subfigure}[b]{0.25\textwidth}
\includegraphics[scale=0.4]{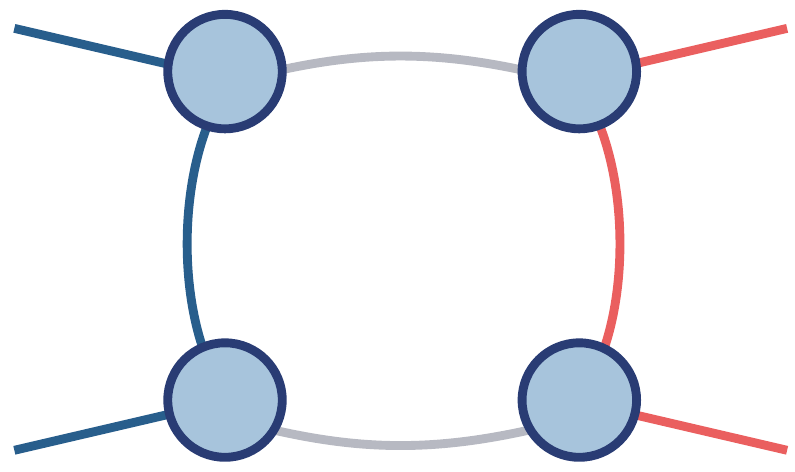}
\caption*{MC1}
\end{subfigure}
~
\begin{subfigure}[b]{0.25\textwidth}
\includegraphics[scale=0.4]{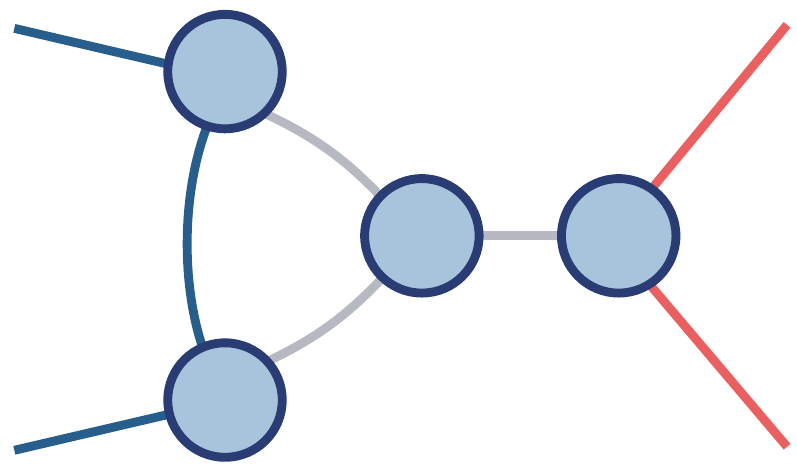}
\caption*{MC2}
\end{subfigure}
~
\begin{subfigure}[b]{0.25\textwidth}
\includegraphics[scale=0.4]{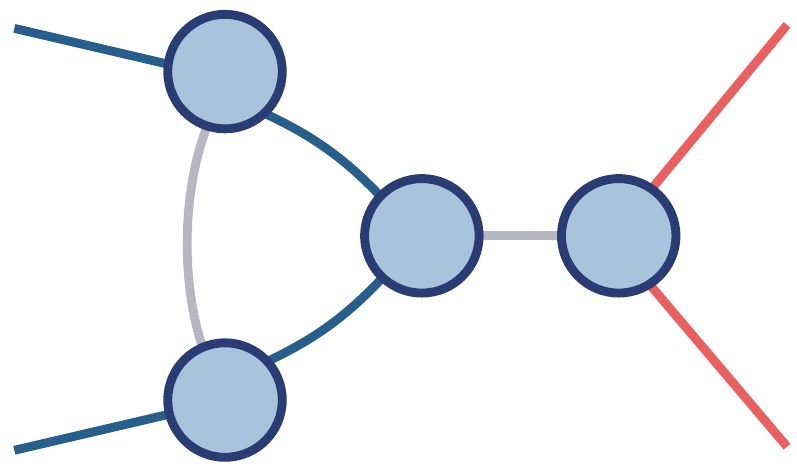}
\caption*{MC3}
\end{subfigure}
 
\begin{subfigure}[b]{0.25\textwidth}
\includegraphics[scale=0.4]{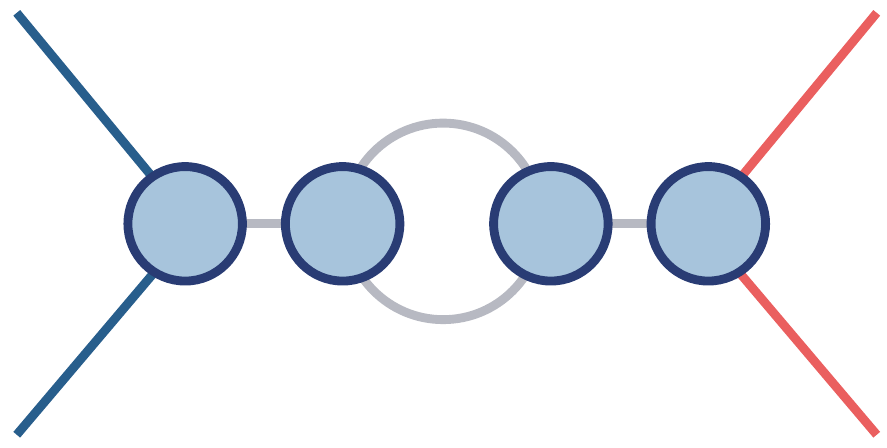}
\caption*{MC4}
\end{subfigure}
~
\begin{subfigure}[b]{0.25\textwidth}
\includegraphics[scale=0.4]{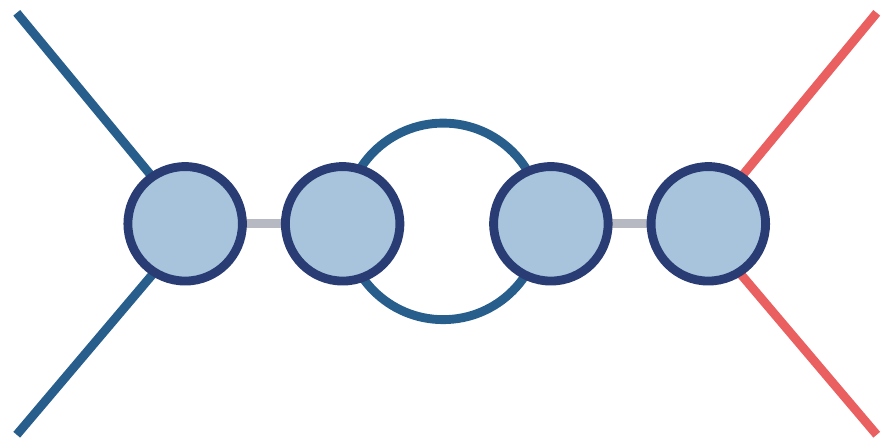}
\caption*{MC5}
\end{subfigure}
\caption{All distinct maximal cuts of the one-loop four-point amplitude, not including tadpole graphs. Blue blobs represent three point amplitudes, and exposed legs represent on-shell propagators. }
\label{Fig: 4p1lMC}
\end{figure}

Projective double copy allows us to construct the Einstein-Hilbert gravity numerator dressings for the graph topologies of the one-loop four-point amplitude that could possibly relevant in the classical limit. The maximal cuts expose the information that is inherent to each graph topology, \textit{i.e.} terms that cannot be moved to other cubic graphs. The N-maximal cuts in turn provide contact terms that can be distributed amongst the uncut propagators of the given cut. At this point a representation choice can thus be made as to where to assign these contributions. Some N-maximal cuts are delicate, in that they have infinite poles that blow up as cut conditions are released. The information that is missed by omitting these cuts can, however, be retrieved at the level of N$^2$-maximal cuts. This type of information shows up as a propagator with some kinematic coefficient, and must therefore be attributed to the corresponding graph topology. We find that the Einstein-Hilbert numerators are corrected by up to N$^2$-maximal cuts relative to the $\mathcal{N}=0$ supergravity kinematic weights.

The first step is to calculate all maximal cuts.  At one-loop the four-point amplitude has five cubic graph topologies which can ever contribute to the physical cuts\footnote{We do not include here contacts associated with cubic graphs like bubbles on external legs nor massive tadpole graphs.}, whose topologies can label potential maximal cuts.  We present these in Figure~\ref{Fig: 4p1lMC}.  
 
For the first consideration of physically relevant information, we now provide details for calculation of the cut MC1 (cf. Figure~\ref{Fig: 4p1lMC}) as an example. All propagators are on-shell and so the left-hand side of Equation~\ref{Eq:GeneralUnitarityCut} is simply the numerator of the box 
\begin{equation}\label{Eq: 4p1l MC1 graphs}
\begin{aligned}
\mathcal{M}_4^{1-\text{loop}}\big|^{\text{cut}}_{\text{MC1}} &= n_{\mathcal{N}=0}\left(
\begin{gathered}
\begin{tikzpicture}[scale=0.5]
\begin{feynman}
\vertex (a1m) at (-0.5, -0.5); 
\vertex (a1) at (-1.5, -0.5) {\(1\)};
\vertex (a2m) at (-0.5, 0.5); 
\vertex (a2) at (-1.5, 0.5) {\(2\)};
\vertex (a3m) at (0.5, 0.5); 
\vertex (a3) at (1.5, 0.5) {\(3\)};
\vertex (a4m) at (0.5, -0.5); 
\vertex (a4) at (1.5, -0.5) {\(4\)};
\diagram{
(a1m) -- [thick, color=blue(ncs)]  (a2m),
(a1) - - [thick, color=blue(ncs)] (a1m) -- [thick, color=blue(ncs)] (a2m) --[thick, color=blue(ncs)]  (a2),
(a1m) --  (a4m),
(a2m) -- (a3m),
(a3) --[thick, color=red]  (a3m) --[thick, color=red]  (a4m) --[thick, color=red]  (a4)
};
\end{feynman}
\end{tikzpicture}
\end{gathered}\right)\Bigg|_{\text{MC1}}^{\text{cut}} = \left( k_1 \cdot k_4 \right)^4 ,
\end{aligned}
\end{equation}
where $\ell$ is the loop momenta and we have imposed the cut conditions $\ell^2 =0$, $(\ell -k_1)^2 = m_1^2$, $(\ell-k_1-k_2)^2 = 0$ and $(\ell+k_4)^2 = m_2^2$. The right-hand side of Equation~\ref{Eq:GeneralUnitarityCut} is the state sum over the product of three-point tree amplitudes, 
\begin{equation}
\begin{aligned}
\mathcal{M}^{\text{trees}}\big|_{MC1} &=
\sum_{\text{states}}
\mathcal{M}^{\text{tree}}_3(k_1^{m_1}, l_1^{m_1}, l_3^{s_3})
\mathcal{M}^{\text{tree}}_3(-l_1^{m_1}, k_2^{m_1}, l_2^{s_2})\\
&\hspace{1cm} \times
\mathcal{M}^{\text{tree}}_3 (-l_2^{\overline{s}_2}, k_3^{m_2}, -l_4^{m_2})
\mathcal{M}^{\text{tree}}_3(l_4^{m_2}, k_4^{m_2}, -l_3^{\overline{s}_3})\\
&= \sum_{\text{states}}(l_1 \cdot \varepsilon(l_3))^2~(l_1 \cdot \varepsilon(l_2))^2~(l_4 \cdot \varepsilon(-l_2))^2~(l_4 \cdot \varepsilon(-l_3))^2.
\end{aligned}
\end{equation}
 Note that $\varepsilon (l_i) \neq \varepsilon (-l_i)$. Using the state projector to sum over polarizations, we find
\begin{equation}\label{Eq: 4p1l MC1 trees}
\mathcal{M}^{\text{trees}}\big|_{MC1} = \left(\frac{m_1^2m_2^2}{D_s-2}-\left(k_1 \cdot k_4\right)^2\right)^2.
\end{equation}

Similarly to tree-level, the contribution from non-gravitational states is given by the difference between \eqn{Eq: 4p1l MC1 graphs} and \eqn{Eq: 4p1l MC1 trees}
\begin{equation}
\begin{aligned}
\Delta^{\text{MC1}} &= \mathcal{M}_4^{1-\text{loop}}\big|^{\text{cut}}_{
MC1} - \mathcal{M}^{\text{trees}}\big|_{\text{MC1}}\\
&= 2~m_1^2m_2^2 ~ \frac{\left(k_1 \cdot k_4 \right)^2}{D_s-2}-\frac{ (m_1^2m_2^2)^2}{(D_s -2)^2}.
\end{aligned}
\end{equation}
As mentioned, the functional numerators must obey all the isomorphisms of the graph topologies they describe -- such as $k_1 \leftrightarrow k_2, k_3 \leftrightarrow k_4$ for the box graph -- and $\Delta^{MC1}$ is symmetrized accordingly. In this particular case the expression already obeys the relevant isomorphisms, and we have
\begin{equation}
\hat{\Delta}^{MC1} = -\frac{(m_1^2 m_2^2)^2}{(D_s-2)^2}+
\frac{2 m_1^2 m_2^2}{ D_s-2}
\left(k_1\cdot k_4\right)^2.
\end{equation}
The constructed Einstein-Hilbert numerator is then given by
\begin{equation}
n_{G}^{\text{box}} = n^{\text{box}}_{\mathcal{N}=0} - \hat{\Delta}^{MC1}.
\end{equation}
The $\Delta$ contributions before symmetrization from the maximal cuts of the one-loop amplitude can be found in Table~\ref{Tab: 4p1lMC}.

\begin{table}[H]
\begin{align*}
\begin{array}{c|c}
\noalign{\hrule height 1.2pt}
\text{Graph topology}   & \text{Maximal cut }\Delta\\
\noalign{\hrule height 1.2pt}
\begin{gathered}
\includegraphics[scale=0.4]{figs/maxCut4p1.pdf}
\end{gathered}
& 
\begin{gathered} 
\frac{2}{D_s-2}  m_1^2 m_2^2 \left(k_1\cdot k_4\right)^2-\frac{m_1^4 m_2^4}{(D_s-2)^2}
\end{gathered}
\\
\begin{gathered}
\includegraphics[scale=0.4]{figs/maxCut4p2.pdf}
\end{gathered} 
& \frac{1}{D_s-2}m_1^4 (\ell \cdot k_4)^2 \\
\begin{gathered}
\includegraphics[scale=0.4]{figs/maxCut4p3.pdf}
\end{gathered} 
& 
\begin{gathered}
\frac{m_1^4}{D_s-2}  \left[\left(\ell\cdot k_4+k_2\cdot k_4\right)^2+ m_1^2 m_2^2\right]-\frac{m_1^6 m_2^2}{(D_s-2)^2} 
\end{gathered} \\
\begin{gathered}
\includegraphics[scale=0.4]{figs/maxCut4p4.pdf}
\end{gathered}
& 
\begin{gathered}
\frac{1}{2} \left((D_s-2)^2-(D_s-2)+2\right) (\ell\cdot k_2)^2 (\ell\cdot k_4)^2
\end{gathered}
 \\
\begin{gathered}
\includegraphics[scale=0.4]{figs/maxCut4p5.pdf}
\end{gathered}
&
\begin{gathered} 
\frac{m_1^4 \left(\ell \cdot k_4\right)^2+m_1^2 m_2^2 \left(\ell \cdot k_2\right){}^2}{D_s-2}-\frac{m_1^6 m_2^2}{\left(D_s-2\right){}^2}
\end{gathered}
\end{array}
\end{align*}
\caption{The $\Delta$ contributions from the maximal cuts of the one-loop four-point amplitude, not including tadpole graphs. Blue blobs represent three point amplitudes, and exposed legs represent on-shell propagators. }
\label{Tab: 4p1lMC}
\end{table}

Updating numerators on the N$^k$-maximal cuts is a similar procedure, and the next-to-maximal cuts (NMC) can be found in Figure~\ref{Fig: 4p1lNMC}. In the Figure, the light blue blobs represent three-point tree amplitudes, and darker blobs represent higher-point trees where cut conditions have been released.  The maximal cuts fix all the information inherent to each graph, and excess state contributions from higher-level cuts are therefore contact terms. As we have chosen to represent the amplitudes exclusively in terms of cubic graphs, these terms are associated with uncut propagators of the cut graph topologies. For example,  consider the next-to-maximal cut NMC1, which yields the following additional information
\begin{equation}
\begin{aligned}
\Delta^{\text{NMC1}}
=&
\frac{
n_G\left(
\begin{gathered}
\begin{tikzpicture}[scale=0.5]
\begin{feynman}
\vertex (a1m) at (-0.5, -0.5); 
\vertex (a1) at (-1.5, -0.5) {\(1\)};
\vertex (a2m) at (-0.5, 0.5); 
\vertex (a2) at (-1.5, 0.5) {\(2\)};
\vertex (a3m) at (0.5, 0.5); 
\vertex (a3) at (1.5, 0.5) {\(3\)};
\vertex (a4m) at (0.5, -0.5); 
\vertex (a4) at (1.5, -0.5) {\(4\)};
\diagram{
(a1m) -- [thick, color=blue(ncs)]  (a2m),
(a1) - - [thick, color=blue(ncs)] (a1m) -- [thick, color=blue(ncs)] (a2m) --[thick, color=blue(ncs)]  (a2),
(a1m) --  (a4m),
(a2m) -- (a3m),
(a3) --[thick, color=red]  (a3m) --[thick, color=red]  (a4m) --[thick, color=red]  (a4)
};
\end{feynman}
\end{tikzpicture}
\end{gathered}\right)\Bigg|_{\text{NMC1}}}{(\ell - k_3)^2}
+
\frac{
n_G\left(
\begin{gathered}
\begin{tikzpicture}[scale=0.5]
\begin{feynman}
\vertex (a1m) at (-0.5, -0.5); 
\vertex (a1) at (-1.5, -0.5) {\(1\)};
\vertex (a2m) at (-0.5, 0.5); 
\vertex (a2) at (-1.5, 0.5) {\(2\)};
\vertex (a3m) at (0.5, 0.5); 
\vertex (a3) at (1.5, 0.5) {\(4\)};
\vertex (a4m) at (0.5, -0.5); 
\vertex (a4) at (1.5, -0.5) {\(3\)};
\diagram{
(a1m) -- [thick, color=blue(ncs)]  (a2m),
(a1) - - [thick, color=blue(ncs)] (a1m) -- [thick, color=blue(ncs)] (a2m) --[thick, color=blue(ncs)]  (a2),
(a1m) --  (a4m),
(a2m) -- (a3m),
(a3) --[thick, color=red]  (a3m) --[thick, color=red]  (a4m) --[thick, color=red]  (a4)
};
\end{feynman}
\end{tikzpicture}
\end{gathered}\right)\Bigg|_{\text{NMC1}}}{(\ell - k_4)^2}
+
\frac{
n_G \left(
\begin{gathered}
\begin{tikzpicture}[scale=.5]
\begin{feynman}
\vertex (a1) at (1.5, -0.5){\(4\)};
\vertex (a2) at (1.5, 0.5) {\(3\)};
\vertex (a3m) at (-0.5, 0.5); 
\vertex (a3) at (-1.5, 0.5) {\(2\)};
\vertex (a4m) at (-0.5, -0.5); 
\vertex (a4) at (-1.5, -0.5) {\(1\)};
\vertex (s1) at (0.5, 0);
\vertex (s2) at (0,0);
\diagram{
{(a1), (a2)} --[thick, color=red] (s1), 
(s1) --  (s2),
(s2) --  (a3m),
(s2) --  (a4m),
(a3) - -[thick, color=blue(ncs)] (a3m) --[thick, color=blue(ncs)] (a4m) --[thick, color=blue(ncs)] (a4)
};
\end{feynman}
\end{tikzpicture}
\end{gathered} \right)\Bigg|_{\text{NMC1}}}{(k_3+k_4)^2}\\
&\hspace{0.5cm}-
\sum_{\text{states}}
\mathcal{M}^{\text{tree}}_3(k_1^{m_1}, l_1^{m_1}, l_3^{s_3})
\mathcal{M}^{\text{tree}}_3(-l_1^{m_1}, k_2^{m_1}, l_2^{s_2})
\mathcal{M}^{\text{tree}}_4 (-l_2^{\overline{s}_2}, -l_3^{\overline{s}_3}, k_3^{m_2}, k_4^{m_2})\\
=& -\frac{m_1^2  \left(m_1^2-4 m_2^2\right)}{8 \left(D_s-2\right)}\left(k_3+ k_4\right)^2-\frac{m_1^4 m_2^2}{\left(D_s-2\right){}^2},
\end{aligned}
\end{equation}
where $k_{ij} = k_i + k_j$. The resulting expression is a contact term which we are free to attribute to an uncut propagator in the box or triangle graph. We choose the triangle graph, whose uncut propagator is $(k_3 + k_4)^2$. The correction to the Einstein-Hilbert numerator of the MC2 topology is then given by
\begin{equation}
\hat{\Delta}_{\text{NMC}} = - \textcolor{blue(blob)}{\bm{(k_3 + k_4)^2}}~ \left[ \frac{ m_1^2\left(m_1^2-4 m_2^2\right)\left(k_3 +k_4\right)^2}{8 (D_s-2)}+\frac{m_2^2 m_1^4}{(D_s-2)^2} \right],
\end{equation}
where we have highlighted the propagator prefactor of the expression, which is kept explicit to ensure the vanishing of this term in the maximal cut MC2. 

Here one can choose representations that satisfy either\footnote{Or both.} adjoint-type or fundamental-type scalar single-copy origins. For expedience in this paper and associated ancillary files we collapse to adjoint-type scalars as the expressions are relatively compact. The remaining next-to-maximal cuts in Figure~\ref{Fig: 4p1lNMC} give similar corrections to the gravity numerators, and are shown in Table~\ref{Tab: 4p1lNMC} with the corresponding $\hat{\Delta}_{\text{NMC}}$ terms and the topology to which the new term is attributed.   

\begin{figure}
\begin{subfigure}[b]{0.22\textwidth}
\includegraphics[scale=0.4]{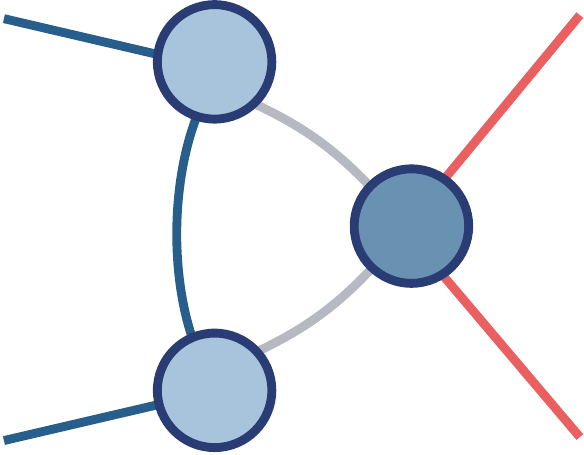}
\caption*{NMC1}
\end{subfigure}
~
\begin{subfigure}[b]{0.22\textwidth}
\includegraphics[scale=0.4]{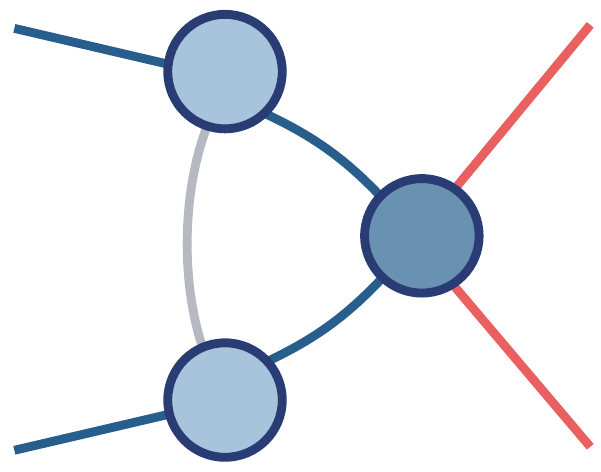}
\caption*{NMC2}
\end{subfigure}
~
\begin{subfigure}[b]{0.22\textwidth}
\includegraphics[scale=0.4]{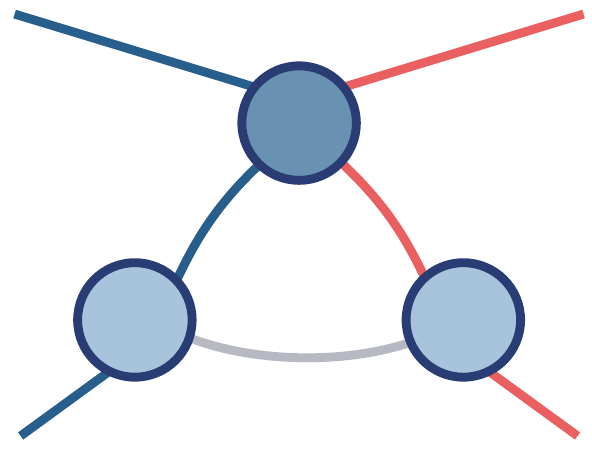}
\caption*{NMC3}
\end{subfigure}
\caption{Next-to-maximal cuts of the one-loop four-point amplitude. Light blue blobs represent three-point amplitudes, dark blobs represent four-point amplitudes, and exposed legs represent on-shell propagators. }
\label{Fig: 4p1lNMC}
\end{figure}

\begin{table}[H]
\begin{align*}
\begin{array}{c|c|c}
\noalign{\hrule height 1.2pt}
\text{Graph topology}   
& \text{N-Maximal cut }\hat{\Delta} 
& \text{Attributed to} \\
\noalign{\hrule height 1.2pt}
\begin{gathered}
\includegraphics[scale=0.4]{figs/nmaxCut4p1.pdf}
\end{gathered}
&
\begin{gathered} 
\begin{aligned}
- \textcolor{blue(blob)}{\bm{\left(k_3+k_4\right)^2}} \times \Bigg[ \frac{ m_1^2\left(m_1^2-4 m_2^2\right) \left( k_3+ k_4\right)^2}{8 (D_s-2)}+\frac{ m_1^4 m_2^2 }{(D_s-2)^2}\Bigg]
\end{aligned}
\end{gathered}
& 
\begin{gathered}
\includegraphics[scale=0.3]{figs/maxCut4p2.pdf}
\end{gathered}
\\
&&
\\
\begin{gathered}
\includegraphics[scale=0.4]{figs/nmaxCut4p4.pdf}
\end{gathered}
&
\begin{gathered} 
\begin{aligned}
\textcolor{blue(blob)}{\bm{\left(k_1+k_2+\ell\right)^2}} \times& \Bigg[m_1^2 m_2^2\frac{ (\ell\cdot k_1- \ell\cdot k_4)+2 \left(3k_1\cdot k_4-k_2\cdot k_4\right) }{16 (D_s-2)}\Bigg]\\
+\textcolor{blue(blob)}{\bm{\ell^2}} &\times \Bigg[m_1^2 m_2^2 \frac{  (\ell\cdot k_3- \ell \cdot k_2)+8 k_1 \cdot k_4}{16 (D_s-2)}\Bigg]
\end{aligned}
\end{gathered} 
&
\begin{gathered}
\includegraphics[scale=0.3]{figs/maxCut4p1.pdf}
\end{gathered}
\\
&&
\\
\begin{gathered}
\includegraphics[scale=0.4]{figs/nmaxCut4p3.pdf}
\end{gathered}
& 
\begin{gathered} 
\begin{aligned}
\textcolor{blue(blob)}{\bm{\left(k_3+k_4\right)^2}}\times \Bigg[m_1^2 \frac{  4 m_2^2 \left(k_3+ k_4\right)^2-m_1^2 \left((k_3+ k_4)^2+16 m_2^2\right)}{16 (D_s-2)} \Bigg]
\end{aligned}
\end{gathered}
& 
\begin{gathered}
\includegraphics[scale=0.3]{figs/maxCut4p3.pdf}
\end{gathered}
\end{array}
\end{align*}
\caption{The symmetrized $\hat{\Delta}$ contributions from the well-defined next-to-maximal cuts of the one-loop four-point amplitude attributed to inverse propagators. Blue blobs represent three point amplitudes, and exposed legs represent on-shell propagators. }
\label{Tab: 4p1lNMC}
\end{table}

Some next-to-maximal cuts are delicate. Cuts involving one collapsed propagator of bubble graphs such as NMC4-6 in Figure~\ref{Fig: 4p1lNMC ill-defined}, have infinite poles that blow up as cut conditions are imposed on the fully dressed graphs. By construction the numerator dressings are made up of a minimal kinematic basis where conservation of momentum has been imposed. This means that inverse propagators in the numerators -- which should cancel against propagators in the denominator -- are rewritten in a form where these cancellations do not occur. The cuts, as written, therefore naively go to infinity as propagators in the denominator are sent to zero. Fortunately any potentially physically relevant information in these types of cuts can instead be extracted using the N$^2$-maximal cuts (cf. Figure~\ref{Fig: 4p1lN2MC}). 

\begin{figure}
\begin{subfigure}[b]{0.22\textwidth}
\includegraphics[scale=0.4]{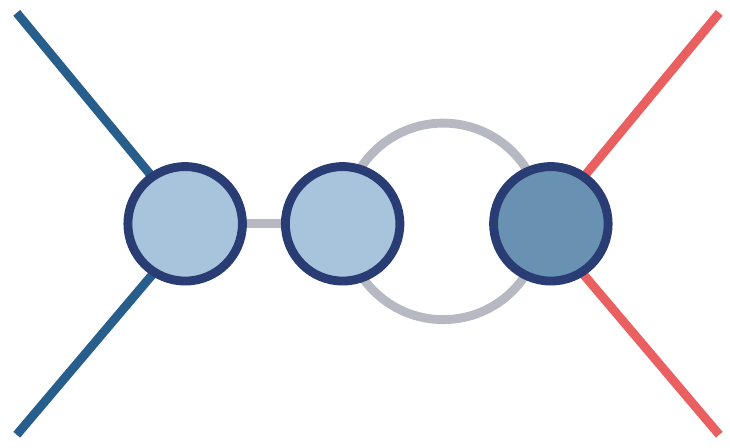}
\caption*{NMC4}
\end{subfigure}
~
\begin{subfigure}[b]{0.22\textwidth}
\includegraphics[scale=0.4]{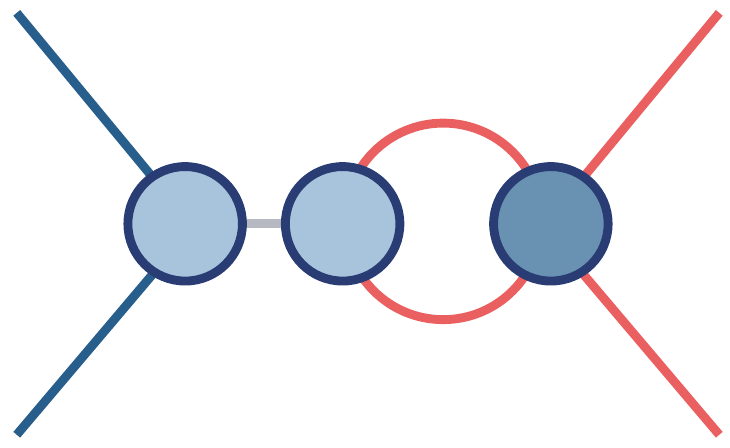}
\caption*{NMC5}
\end{subfigure}
~
\begin{subfigure}[b]{0.22\textwidth}
\includegraphics[scale=0.4]{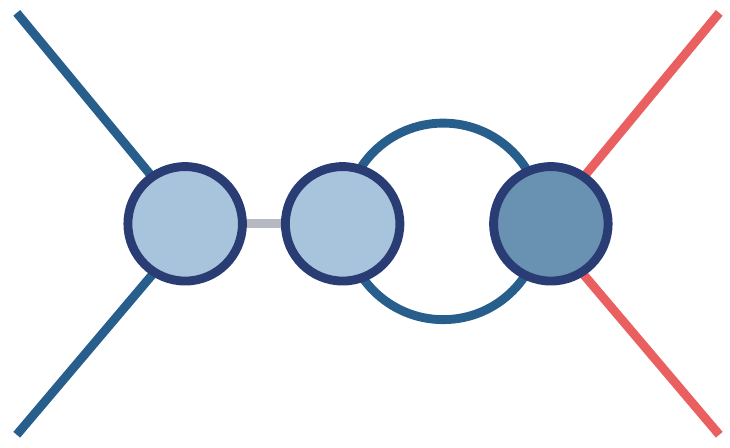}
\caption*{NMC6}
\end{subfigure}

\caption{Delicate next-to-maximal cuts of the one-loop four-point amplitude excluding snail and tadpole cuts. Light blue blobs represent three point amplitudes, and darker blobs represent four-point amplitudes. Exposed legs represent on-shell propagators. }
\label{Fig: 4p1lNMC ill-defined}
\end{figure}

\begin{figure}
\begin{subfigure}[b]{0.2\textwidth}
\includegraphics[scale=0.4]{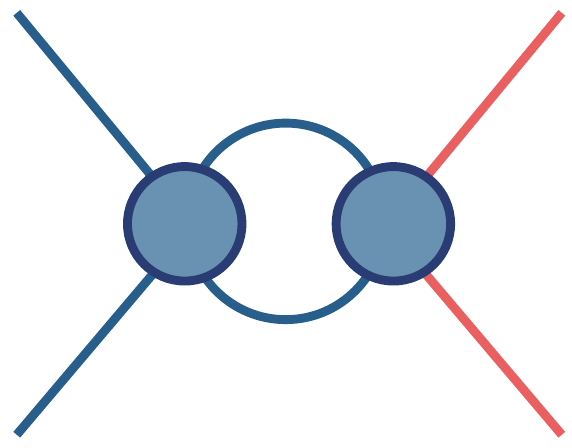}
\caption*{N$^2$MC1}
\end{subfigure}
~
\begin{subfigure}[b]{0.2\textwidth}
\includegraphics[scale=0.4]{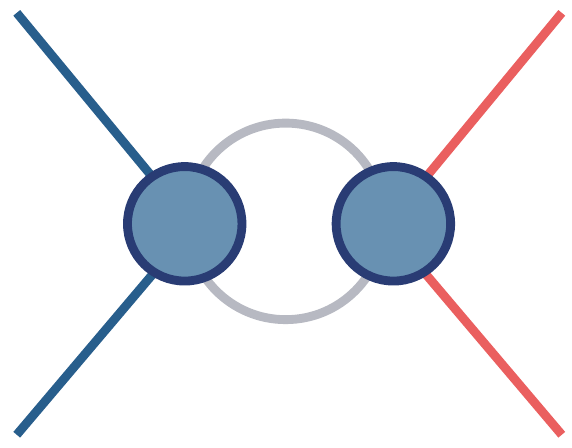}
\caption*{N$^2$MC2}
\end{subfigure}
~
\begin{subfigure}[b]{0.2\textwidth}
\includegraphics[scale=0.4]{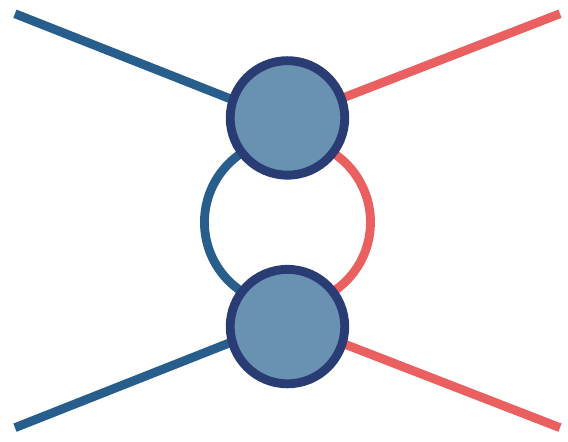}
\caption*{N$^2$MC3}
\end{subfigure}
\caption{All next-to-next-to-maximal cuts of the one-loop four-point amplitude, not including tadpole graphs. Light blue blobs represent three point amplitudes, and darker blobs represent four and five-point amplitudes. Exposed legs represent on-shell propagators.}
\label{Fig: 4p1lN2MC}
\end{figure}

Consider for example the N$^2$-maximal cut N$^2$MC1 in Figure~\ref{Fig: 4p1lN2MC}. This cut is equivalent to NMC6 with an additional propagator, $(k_1+k_2)^2$, put off-shell. The excess state term $\Delta$ for this cut is
\begin{equation}
\begin{aligned}
\Delta^{\text{N}^2\text{MC1}} &=- \frac{m_1^2}{16 (D_s-2)} \left[ \frac{12 m_1^2m_2^2}{(k_3 + k_4)^2}-6m_2^2 +m_1^2\right],
\end{aligned}
\end{equation}
where we see that this is not a pure contact term.  We see the contribution of a manifest propagator, so in this $N^2MC$ we have found the excess state contribution to one of the propagators in a previously skipped NMC cut. We attribute this information to the topology MC5 (massive bubble), by multiplying by appropriate combinations of the two uncut propagators $(k_1+k_2)^2=(k_3+k_4)^2$ which yields two distinct contact terms:
\begin{align}
\hat{\Delta}^{\text{N}^2\text{MC1}} =& \textcolor{blue(blob)}{\bm{\left(k_3+k_4\right)^2}}\times \frac{-3 m_1^4 m_2^2 }{4(D_s-2)}\\
&+\textcolor{blue(blob)}{\bm{{\left(k_1+k_2\right)^2}{\left(k_3+k_4\right)^2}}} \times \frac{ \left(6 m_1^2 m_2^2-m_1^4\right)}{16(D_s-2)} \nn. 
\end{align}
As the first  is proportional to a single uncut propagator, we see that it contains information relevant to a cut like NMC6.  Of course because of conservation of momentum it vanishes when $(k_1+k_2)^2$ is on-shell, which is why it was sensible to wait until this N$^2$-maximal cut to figure out where it belongs. The second term is proportional to the two propagators whose cut conditions are released at N$^2$-maximal level, and so is inherently a N$^2$MC correction.

Of the N$^2$-maximal cuts shown in Figure~\ref{Fig: 4p1lN2MC} the cuts N$^2$MC1 and N$^2$MC2 yield numerator corrections, and the remaining cut can be used for verification.  The symmetrized corrections for the N$^2$MC are shown in Table~\ref{Tab: 4p1lN2MC}, with the propagator prefactors that correspond to attributing the cut information to the topologies shown in the right column.

\begin{table}[H]
\begin{align*}
\begin{array}{c|c|c}
\noalign{\hrule height 1.2pt}
\text{Graph topology}   & \text{N$^2$-Maximal cut }\hat{\Delta} & \text{Attributed to}\\
\noalign{\hrule height 1.2pt}
\begin{gathered}
\includegraphics[scale=0.4]{figs/n2maxCut4p2.pdf}
\end{gathered}
&
\begin{gathered} 
\begin{aligned}
&-\textcolor{blue(blob)}{\bm{\left(  k_{34}\right)^2}} \times \frac{3 m_1^4 m_2^2 }{4(D_s-2)} 
-\textcolor{blue(blob)}{\bm{\left[\left(k_{34}\right)^2\right]^2}}\times \frac{m_1^2 \left(m_1^2-6 m_2^2\right)}{16(D_s-2)}
\end{aligned}
\end{gathered}
& 
\begin{gathered}
\includegraphics[scale=0.3]{figs/maxCut4p5.pdf}
\end{gathered}
\\
&&
\\
&&
\\
\begin{gathered}
\includegraphics[scale=0.4]{figs/n2maxCut4p4.pdf}
\end{gathered}
& 
\begin{gathered} 
\begin{aligned}
&\textcolor{blue(blob)}{\bm{\left(k_{34}\right)^2}} \times
 \Bigg[-\frac{s P_{14}-4 \left(m_1^2 P_{34}+m_2^2 P_{12}\right)}{8 \left(D_s-2\right)}\\
&\hspace{1cm}+\frac{D_s-2}{16}  \Big(s \left(4 (k_4 \cdot k_{12}^-) (\ell\cdot k_{13}^-)+8 \ell_{13}
+P_{14}\right)\\
&\hspace{1cm} +4 \left(8 (k_2\cdot k_4 )\ell_{13}-(m_1^2 P_{34}+m_2^2 P_{12})\right)\\
&\hspace{5cm}-2 (k_2\cdot k_4) s^2\Big)\\
 &\hspace{1cm}
 -\frac{1}{16} \Big(s \left(8 (k_4 \cdot k_{12}^-)( \ell\cdot k_{13}^-)+16 \ell_{13}
 +3 P_{14}\right)\\
 &\hspace{1cm}+4 \left(16 (k_2\cdot k_4)\ell_{13}-3( m_1^2 P_{34}+ m_2^2 P_{12})\right)\\
 &\hspace{5cm}-4 (k_2\cdot k_4) s^2\Big)
 \Bigg]
\\
&+\textcolor{blue(blob)}{\bm{\left[\left(k_{34}\right)^2\right]^2}}\times
 \Bigg[\frac{\left(D_s-2\right){}^2}{128}  \left(s^2-4 P_{14}\right)\\
 &\hspace{1cm}+\frac{D_s-2}{128} \big(8 s u-9 s^2+32 \left(k_2\cdot k_4\right){}^2+12 P_{14}\big)\\
 &\hspace{1cm}-\frac{1}{64} \big(32 \left(k_2\cdot k_4\right){}^2-16 m_1^2 m_2^2+8P_{14}\\
 &\hspace{1.5cm}+8 s u+4 s m_{12}^2-9 s^2\big)\\
 &\hspace{1cm}-\frac{s m_{12}^2-16 m_1^2m_2^2-P_{14}}{8 \left(D_s-2\right)}
 -\frac{m_1^2 m_2^2}{\left(D_s-2\right)^2}
\Bigg]
\end{aligned}
\end{gathered}
& 
\begin{gathered}
\includegraphics[scale=0.3]{figs/maxCut4p4.pdf}
\end{gathered}
\end{array}
\end{align*}
\caption{The symmetrized $\hat{\Delta}$ contributions from the well-defined N$^2$-maximal cuts of the one-loop four-point amplitude attributed to inverse propagators. Blue blobs represent three point amplitudes, and exposed legs represent on-shell propagators. Here $\ell_{13}=(\ell \cdot k_1)(\ell \cdot k_3)$, $m_{12}^2 \equiv m_1^2+m_2^2$,  $s \equiv (k_3+k_4)^2 $, $u \equiv (k_2+k_4)^2$, $k_{ij}=k_i+k_j$, $k_{ij}^-=k_i - k_j$, $P_{12} = (\ell \cdot k_1)^2+(\ell \cdot k_2)^2$, $P_{34} = (\ell \cdot k_3)^2+(\ell \cdot k_4)^2$, and  $P_{14} = (\ell \cdot k_1)^2+(\ell \cdot k_2)^2+(\ell \cdot k_3)^2+(\ell \cdot k_4)^2$.  }
\label{Tab: 4p1lN2MC}
\end{table}

 All five maximal cuts of the four-point one-loop amplitude, not including snail or tadpole graphs, are shown in Figure~\ref{Fig: 4p1lMC}. All the cuts in the figure give corrections which are inherent to the gravity numerators of these topologies. The three next-to-maximal cuts shown in Figure~\ref{Fig: 4p1lNMC} are well-defined, and removing cut conditions do not result in infinite poles. The first of these cuts, NMC1, exposes a correction that can be attributed to one of two graph topologies (corresponding to MC1 and MC2 in Figure~\ref{Fig: 4p1lMC}). The remaining two well-defined cuts -- NMC2, NMC3 -- also give corrections, but they involve four-point tree amplitudes with two massive scalars, which means only one topology contributes to each of these cuts. In Figure~\ref{Fig: 4p1lNMC ill-defined} we show three delicate next to maximal cuts, whose relevant information can be more easily extracted by further relaxing cut conditions. Finally, the N$^2$-maximal cuts are shown in Figure~\ref{Fig: 4p1lN2MC}. The two first cuts -- N$^2$MC1 and N$^2$MC2 -- give corrections to the numerators, including information from the skipped N-maximal cuts. The third cut, N$^2$MC3, receives no correction.

\subsubsection{Five-point one-loop}
\label{loopfivepoint}

We proceed with the projective double copy to find the one-loop gravitational radiative correction to the two-to-two scattering of massive scalars. The procedure at five-points is the same as demonstrated at four-points, again starting with the maximal cuts. The 24 maximal cuts relevant to the physical cuts for this five-point amplitude are shown in Figures~\ref{Fig: 5p Max Cuts}-\ref{Fig: 5p Max Cuts 2},  excluding both tadpole and snail topologies. All the maximal cuts give excess state corrections to the corresponding graph numerators. 

The 30 next-to-maximal cuts can be found in Figures~\ref{Fig: 5pNMax cuts box}-\ref{Fig: 5p NMax non-box}, and all of these cuts give corrections to the numerators. There are 19 N$^2$-maximal cuts which are shown in Figures~\ref{Fig: 5p n2max triangle}-\ref{Fig: 5p n2max non-triangle}. In contrast to the next-to-maximal cuts, not all of these yield corrections. The cuts N$^2$MC1, N$^2$MC3, N$^2$MC11 and N$^2$MC19 give no contributions, while the remaining 15 cuts do correct for excess states. At five-point one-loop the highest order of corrections come from N$^2$-maximal cuts. However, some delicate cuts must be ignored at NMC and N$^2$MC and the information extracted using N$^3$MC, as described in the previous section for the four-point amplitude.  All N$^3$-max cuts are bubble cuts, and can be found in Figure~\ref{Fig: 5p n3max cuts}. The corrected gravitational numerators can be found in auxiliary files.

\begin{figure}
\begin{subfigure}[b]{0.25\textwidth}
\includegraphics[scale=0.4]{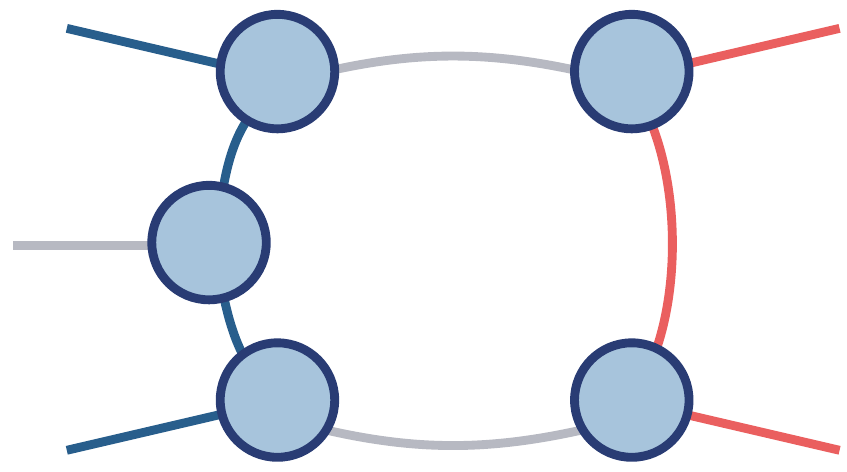}
\caption*{MC1}
\end{subfigure}
~
\begin{subfigure}[b]{0.25\textwidth}
\includegraphics[scale=0.4]{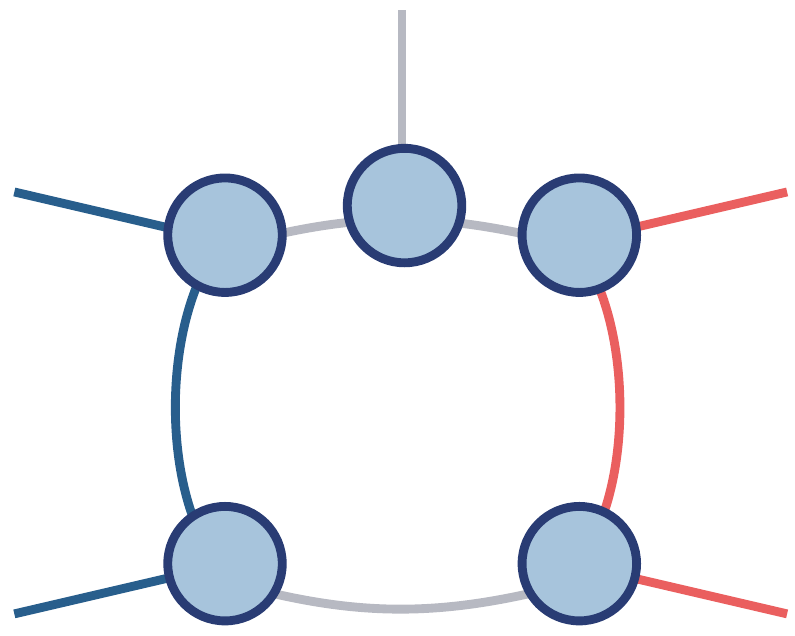}
\caption*{MC2}
\end{subfigure}
~
\begin{subfigure}[b]{0.25\textwidth}
\includegraphics[scale=0.4]{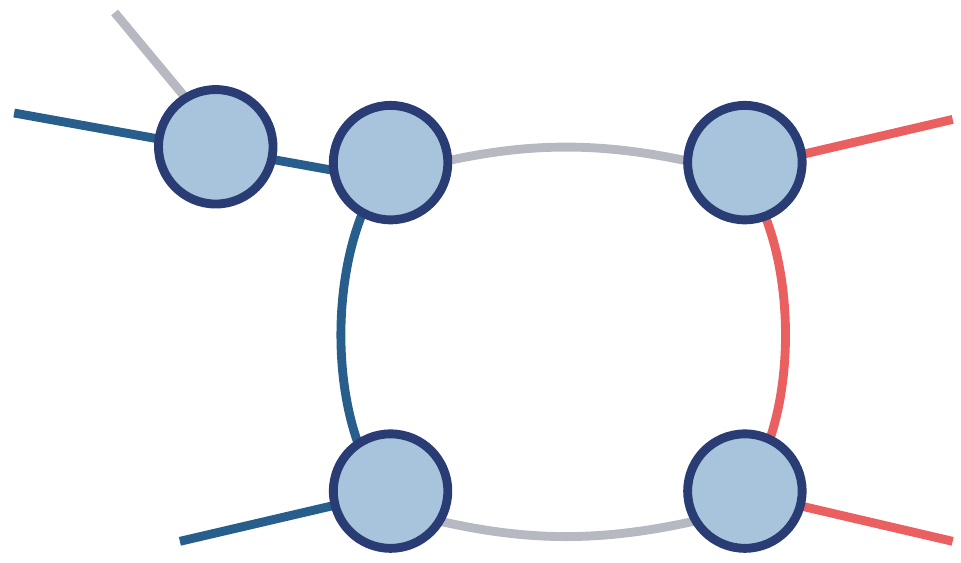}
\caption*{MC3}
\end{subfigure}

\begin{subfigure}[b]{0.3\textwidth}
\includegraphics[scale=0.4]{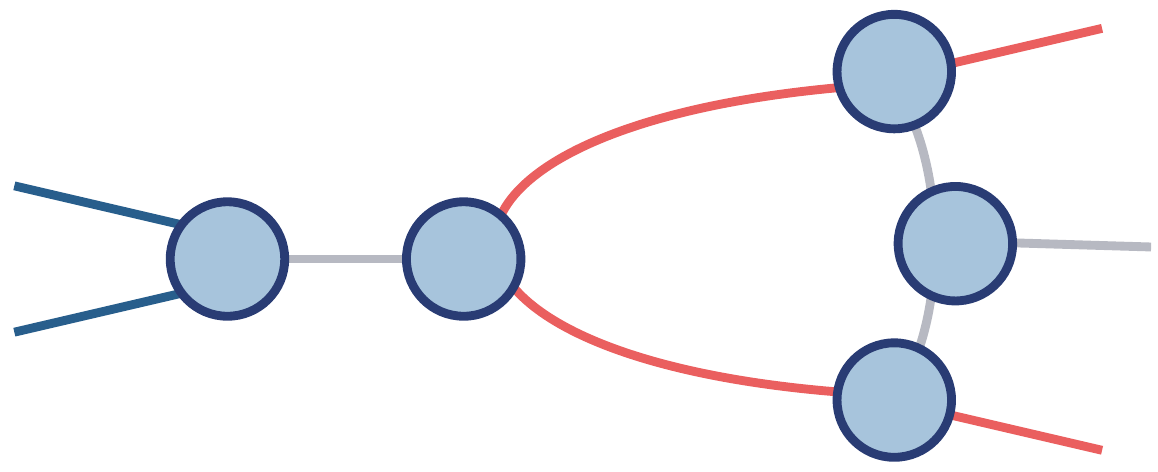}
\caption*{MC4}
\end{subfigure}
~
\begin{subfigure}[b]{0.3\textwidth}
\includegraphics[scale=0.4]{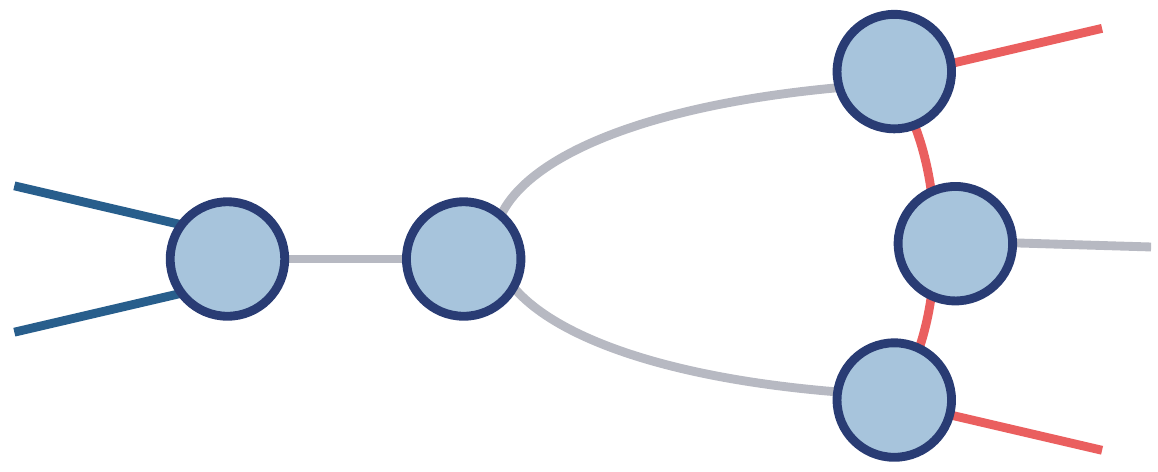}
\caption*{MC5}
\end{subfigure}
~
\begin{subfigure}[b]{0.3\textwidth}
\includegraphics[scale=0.4]{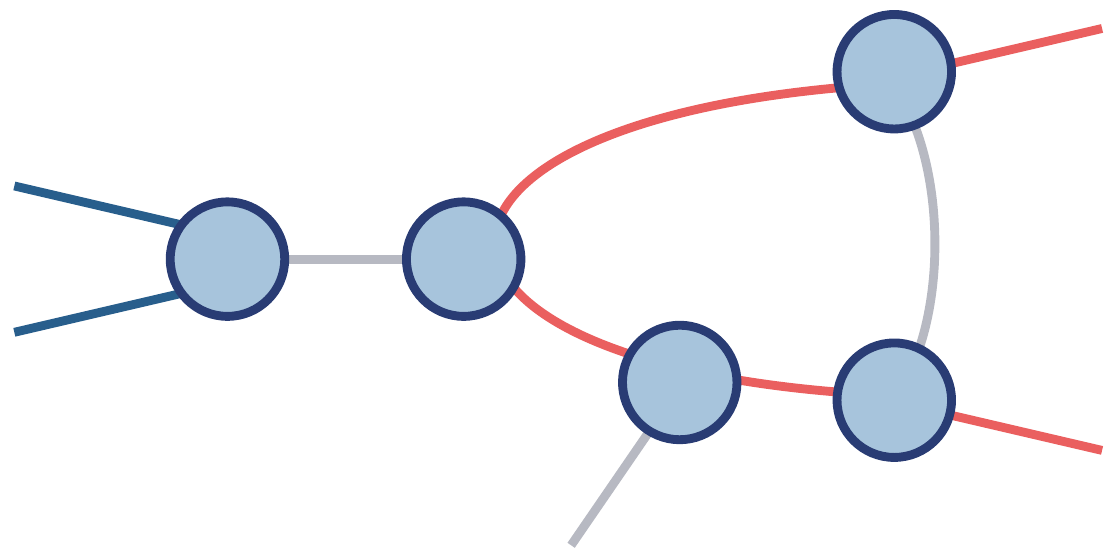}
\caption*{MC6}
\end{subfigure}

\begin{subfigure}[b]{0.3\textwidth}
\includegraphics[scale=0.4]{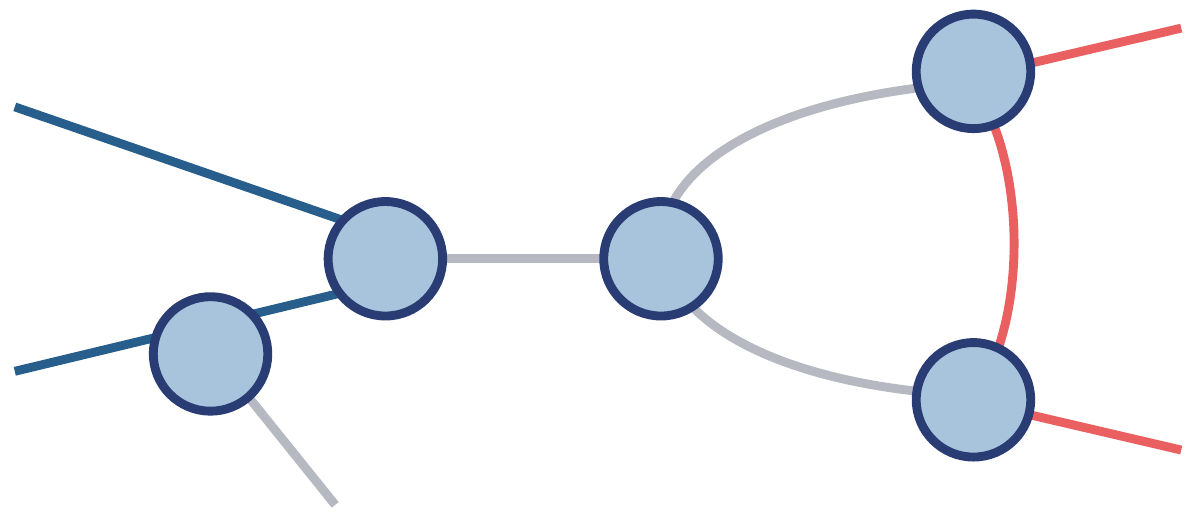}
\caption*{MC7}
\end{subfigure}
~
\begin{subfigure}[b]{0.3\textwidth}
\includegraphics[scale=0.4]{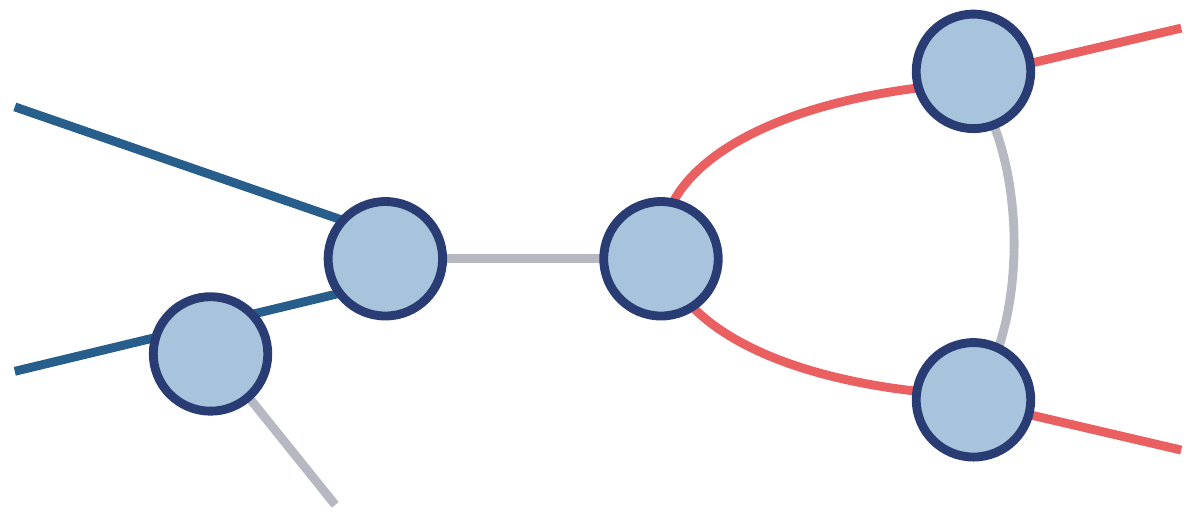}
\caption*{MC8}
\end{subfigure}
~
\begin{subfigure}[b]{0.3\textwidth}
\includegraphics[scale=0.4]{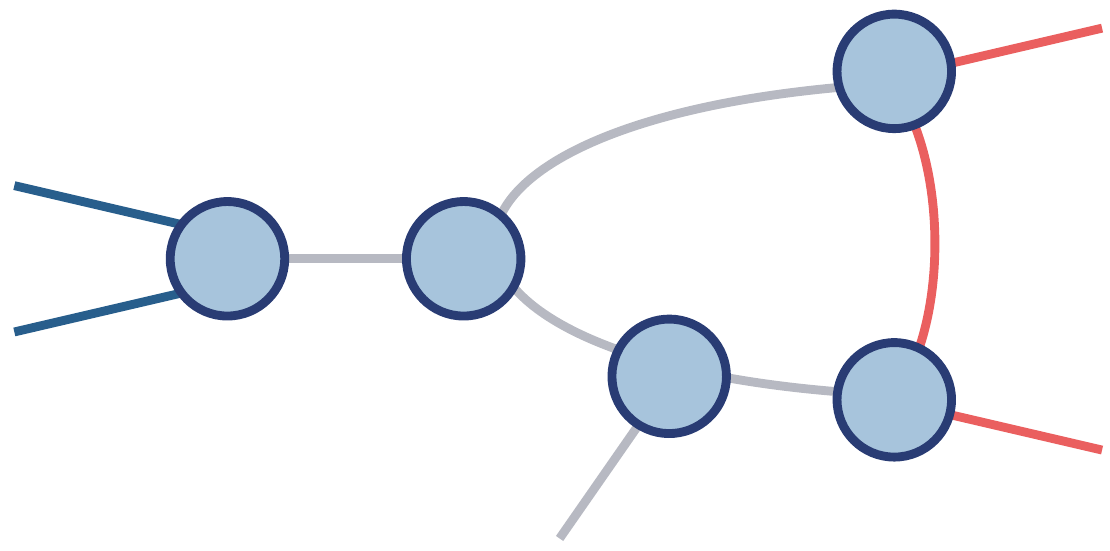}
\caption*{MC9}
\end{subfigure}

\begin{subfigure}[b]{0.3\textwidth}
\includegraphics[scale=0.4]{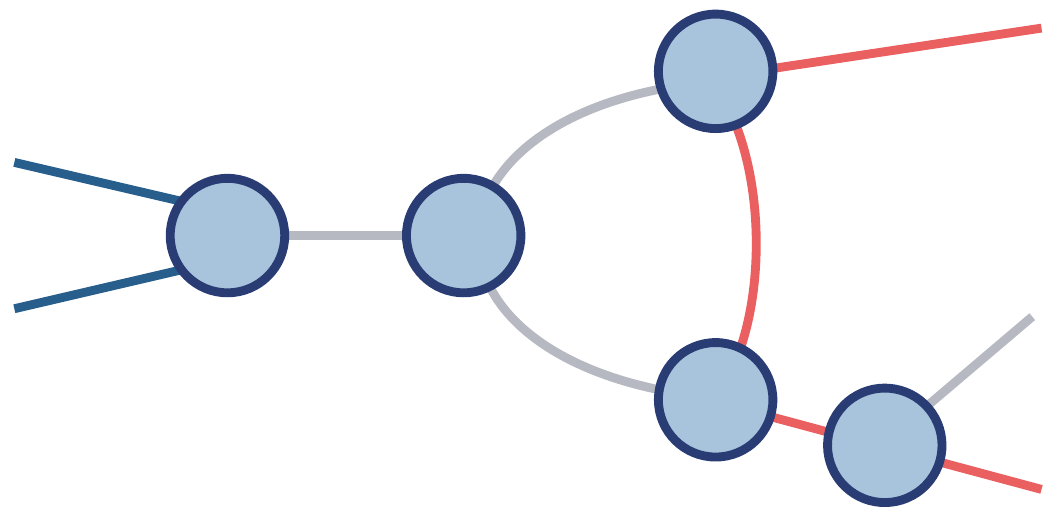}
\caption*{MC10}
\end{subfigure}
~
\begin{subfigure}[b]{0.3\textwidth}
\includegraphics[scale=0.4]{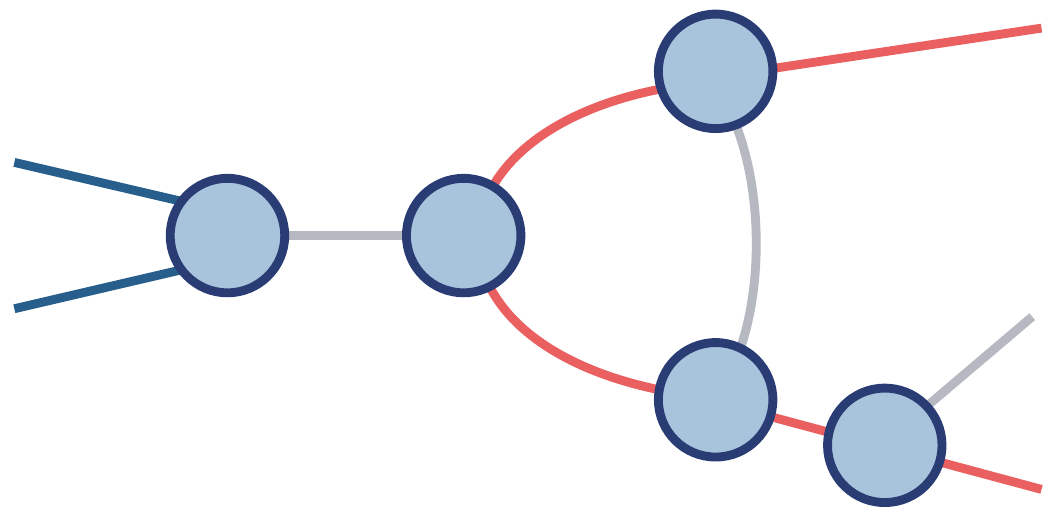}
\caption*{MC11}
\end{subfigure}
~
\begin{subfigure}[b]{0.3\textwidth}
\includegraphics[scale=0.4]{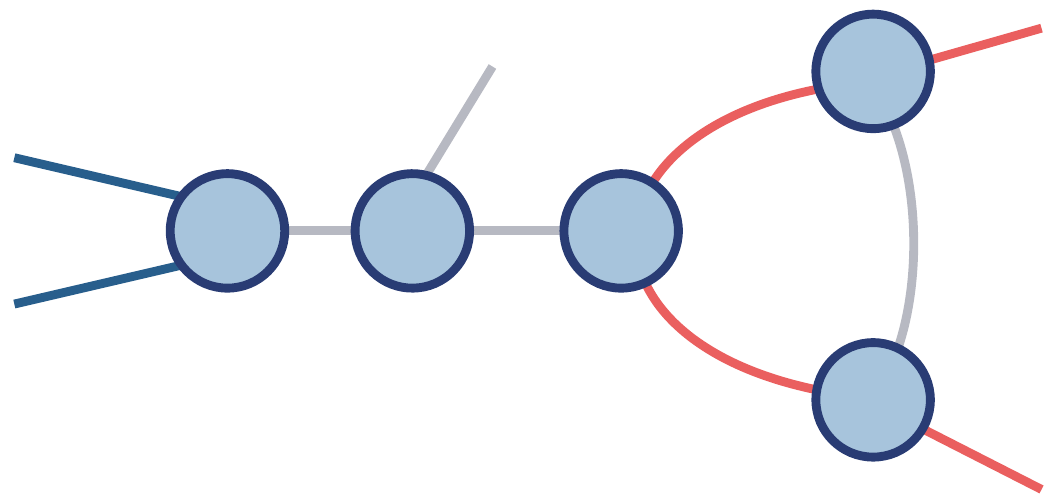}
\caption*{MC12}
\end{subfigure}

\caption{Maximal cuts MC1-MC12 of the one-loop five-point amplitude. Light blue blobs represent three point amplitudes, and darker blobs represent four and five-point amplitudes. Exposed legs represent on-shell propagators.}
\label{Fig: 5p Max Cuts}
\end{figure}

\begin{figure}
\begin{subfigure}[b]{0.3\textwidth}
\includegraphics[scale=0.4]{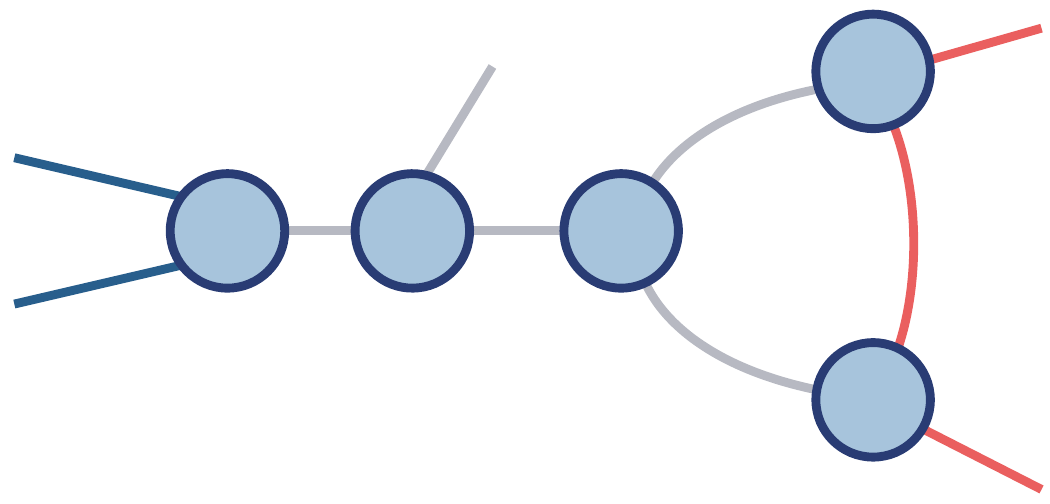}
\caption*{MC13}
\end{subfigure}
~
\begin{subfigure}[b]{0.3\textwidth}
\includegraphics[scale=0.4]{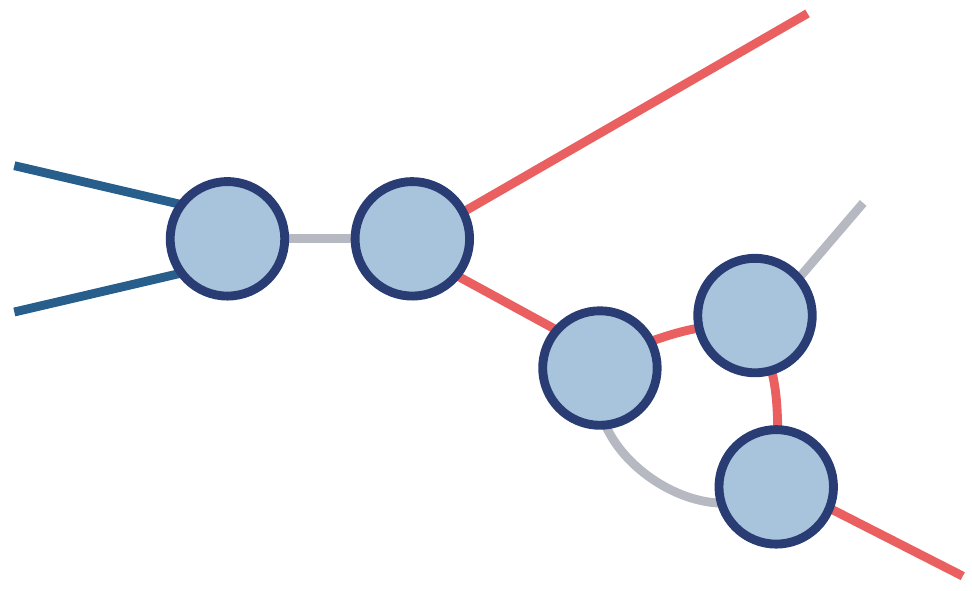}
\caption*{MC14}
\end{subfigure}
~
\begin{subfigure}[b]{0.3\textwidth}
\includegraphics[scale=0.4]{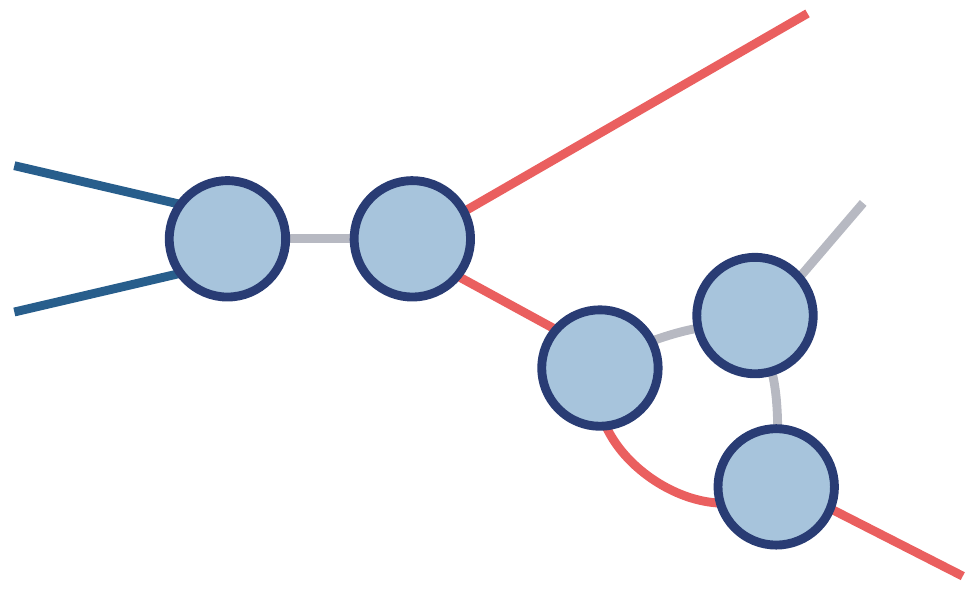}
\caption*{MC15}
\end{subfigure}

\begin{subfigure}[b]{0.3\textwidth}
\includegraphics[scale=0.4]{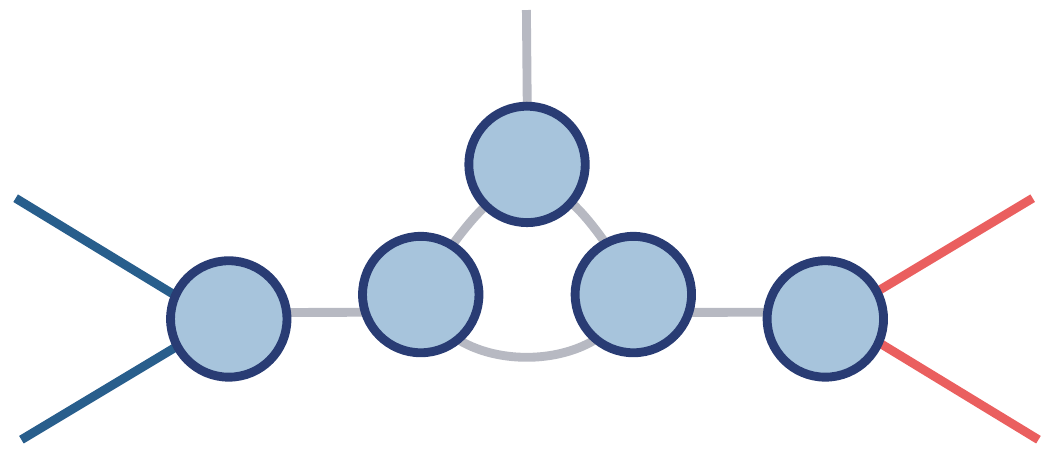}
\caption*{MC16}
\end{subfigure}
~
\begin{subfigure}[b]{0.3\textwidth}
\includegraphics[scale=0.4]{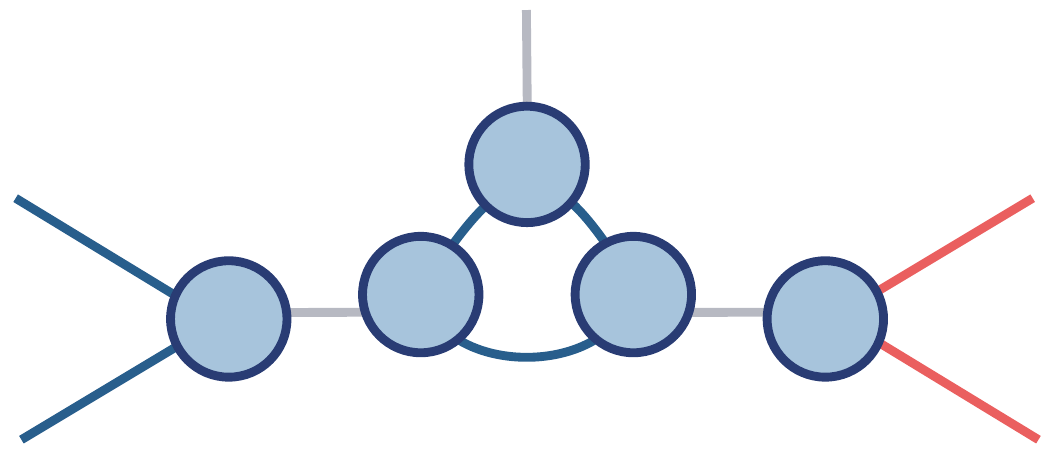}
\caption*{MC17}
\end{subfigure}
~
\begin{subfigure}[b]{0.3\textwidth}
\includegraphics[scale=0.4]{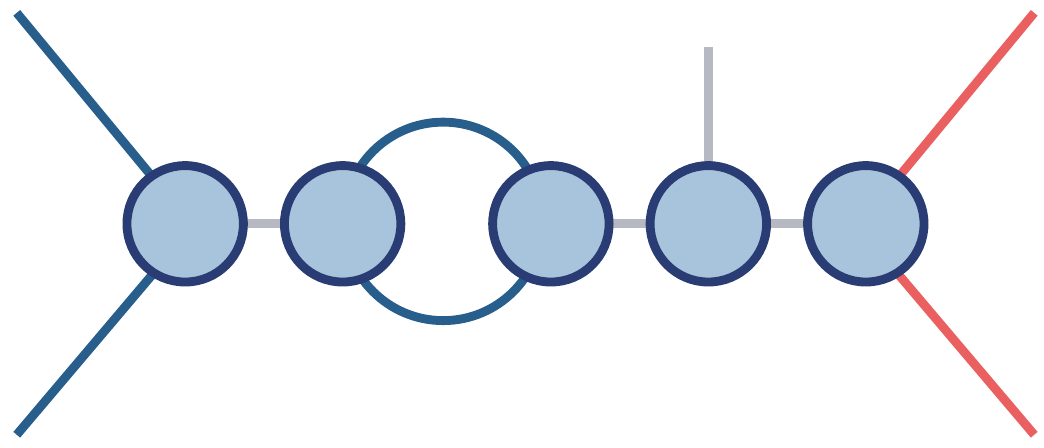}
\caption*{MC18}
\end{subfigure}

\begin{subfigure}[b]{0.3\textwidth}
\includegraphics[scale=0.4]{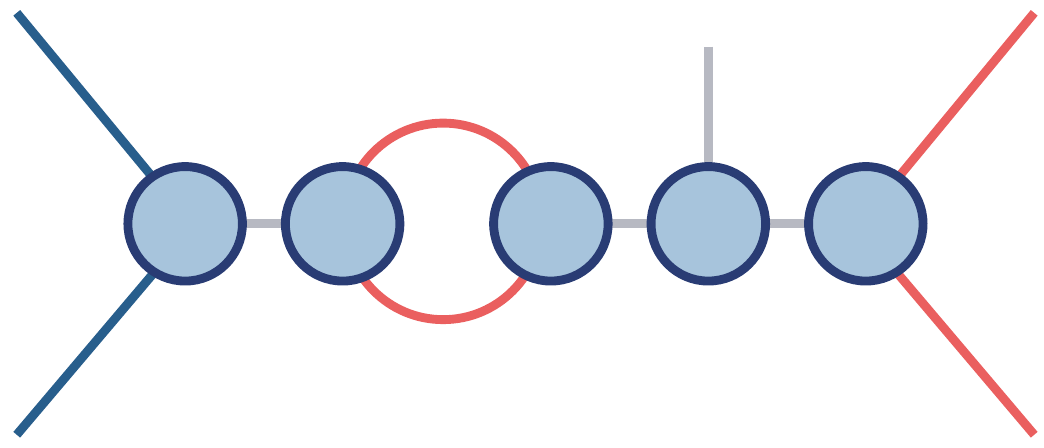}
\caption*{MC19}
\end{subfigure}
~
\begin{subfigure}[b]{0.3\textwidth}
\includegraphics[scale=0.4]{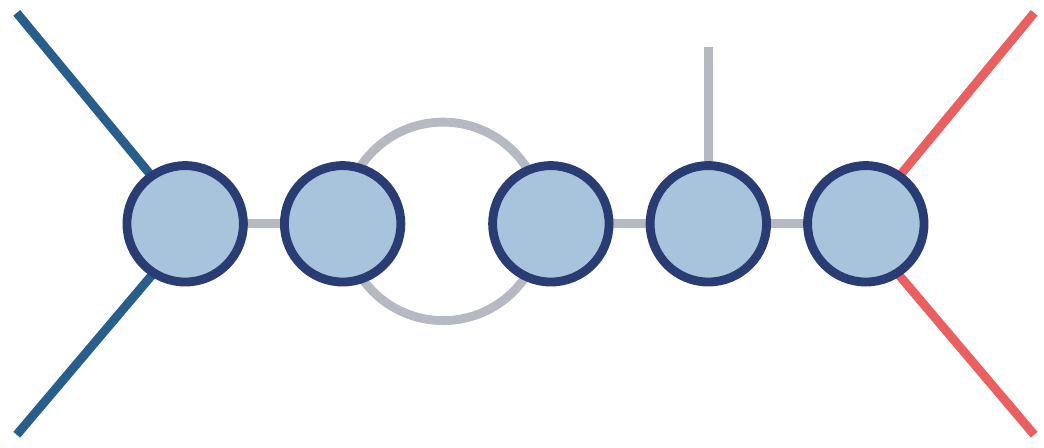}
\caption*{MC20}
\end{subfigure}
~
\begin{subfigure}[b]{0.3\textwidth}
\includegraphics[scale=0.4]{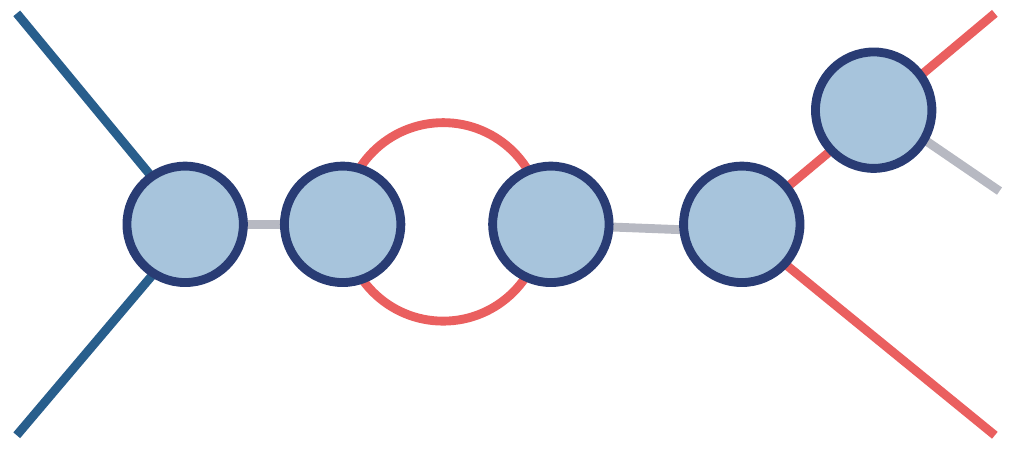}
\caption*{MC21}
\end{subfigure}

\begin{subfigure}[b]{0.3\textwidth}
\includegraphics[scale=0.4]{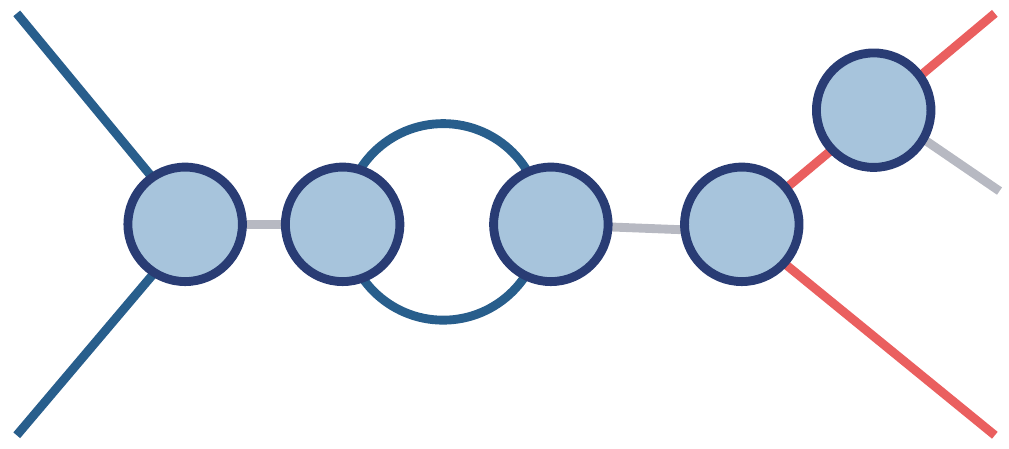}
\caption*{MC22}
\end{subfigure}
~
\begin{subfigure}[b]{0.3\textwidth}
\includegraphics[scale=0.4]{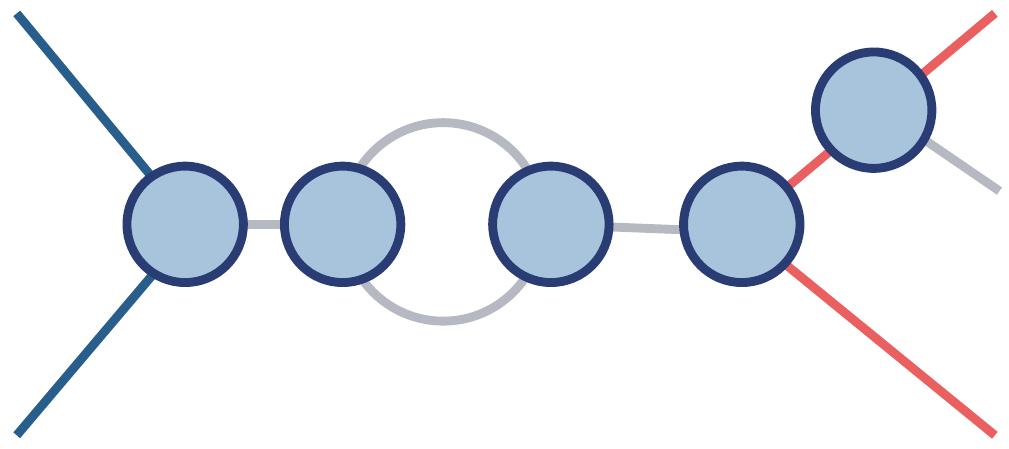}
\caption*{MC23}
\end{subfigure}
~
\begin{subfigure}[b]{0.3\textwidth}
\includegraphics[scale=0.4]{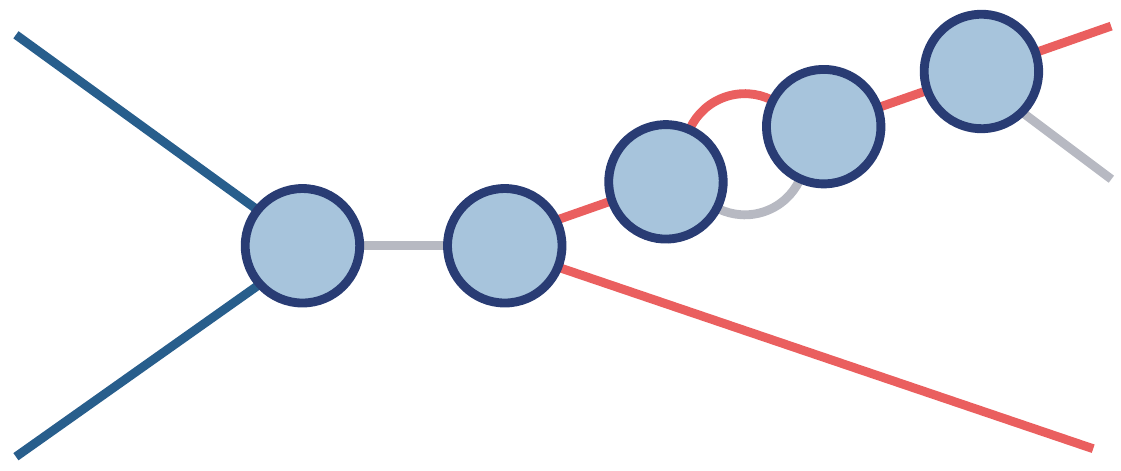}
\caption*{MC24}
\end{subfigure}
\caption{Maximal cuts MC13-MC24 of the one-loop five-point amplitude. Light blue blobs represent three point amplitudes, and darker blobs represent four and five-point amplitudes. Exposed legs represent on-shell propagators.}
\label{Fig: 5p Max Cuts 2}
\end{figure}

\begin{figure}
\begin{subfigure}[b]{0.22\textwidth}
\includegraphics[scale=0.4]{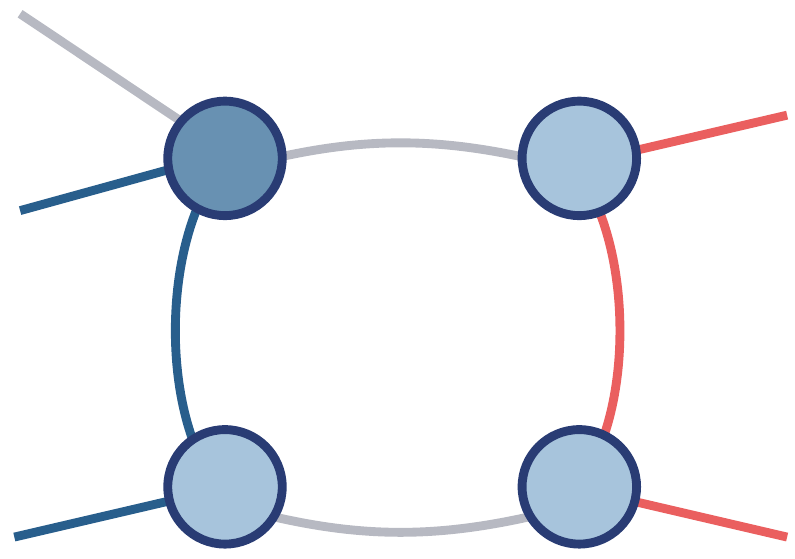}
\caption*{NMC1}
\end{subfigure}
~
\begin{subfigure}[b]{0.22\textwidth}
\includegraphics[scale=0.4]{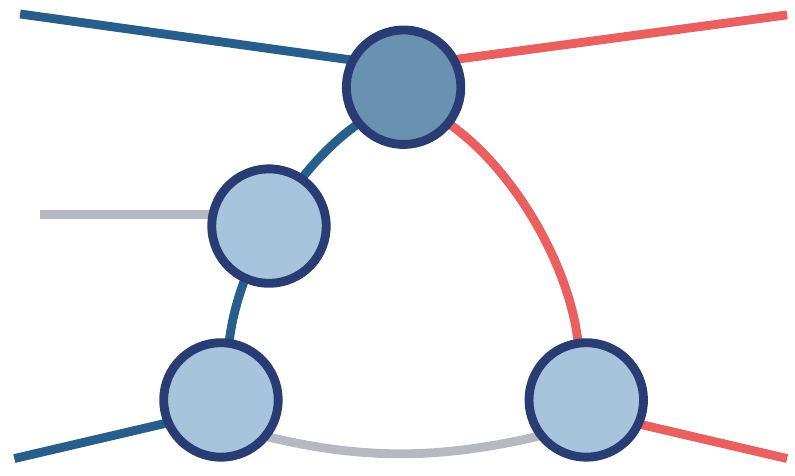}
\caption*{NMC2}
\end{subfigure}
~
\begin{subfigure}[b]{0.22\textwidth}
\includegraphics[scale=0.4]{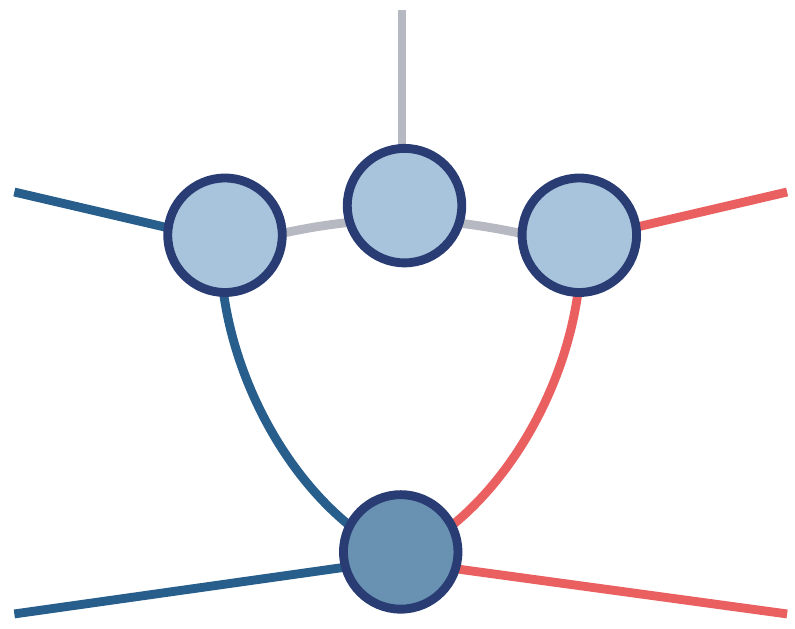}
\caption*{NMC3}
\end{subfigure}

\begin{subfigure}[b]{0.22\textwidth}
\includegraphics[scale=0.4]{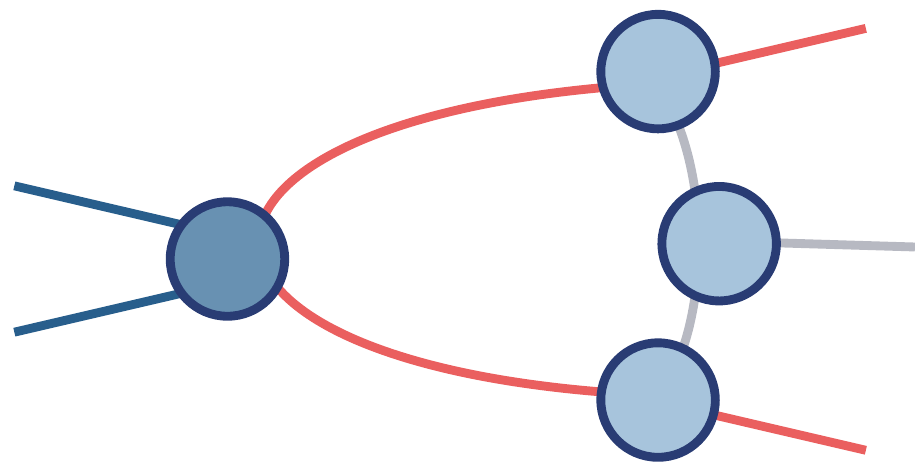}
\caption*{NMC4}
\end{subfigure}
~
\begin{subfigure}[b]{0.22\textwidth}
\includegraphics[scale=0.4]{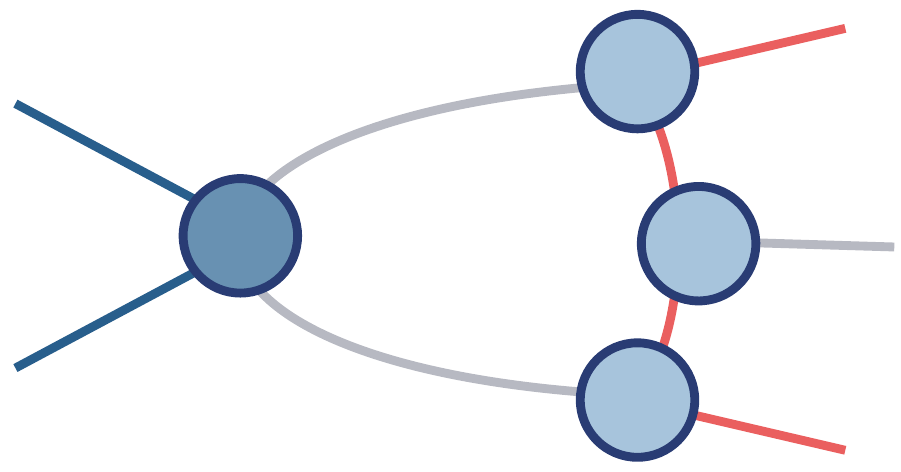}
\caption*{NMC5}
\end{subfigure}
~
\begin{subfigure}[b]{0.22\textwidth}
\includegraphics[scale=0.4]{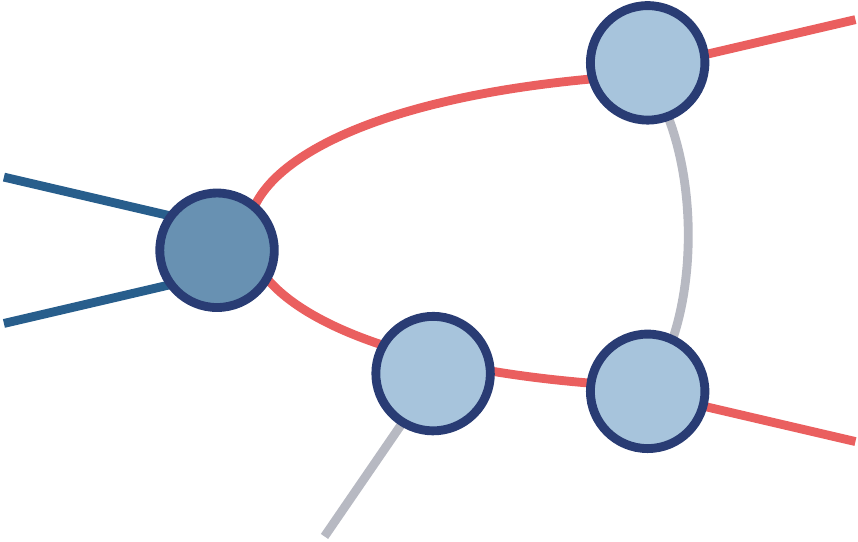}
\caption*{NMC6}
\end{subfigure}
~
\begin{subfigure}[b]{0.22\textwidth}
\includegraphics[scale=0.4]{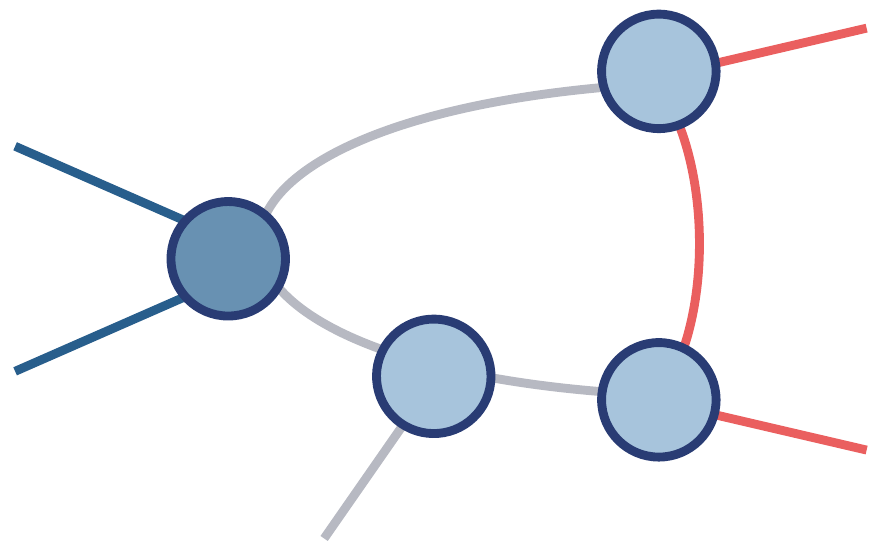}
\caption*{NMC7}
\end{subfigure}

\caption{Next-to-maximal box cuts of the one-loop five-point amplitude. Light blue blobs represent three-point amplitudes, and darker blobs represent four and five-point amplitudes. Exposed legs represent on-shell propagators.}
\label{Fig: 5pNMax cuts box}
\end{figure}

\begin{figure}

\begin{subfigure}[b]{0.22\textwidth}
\includegraphics[scale=0.4]{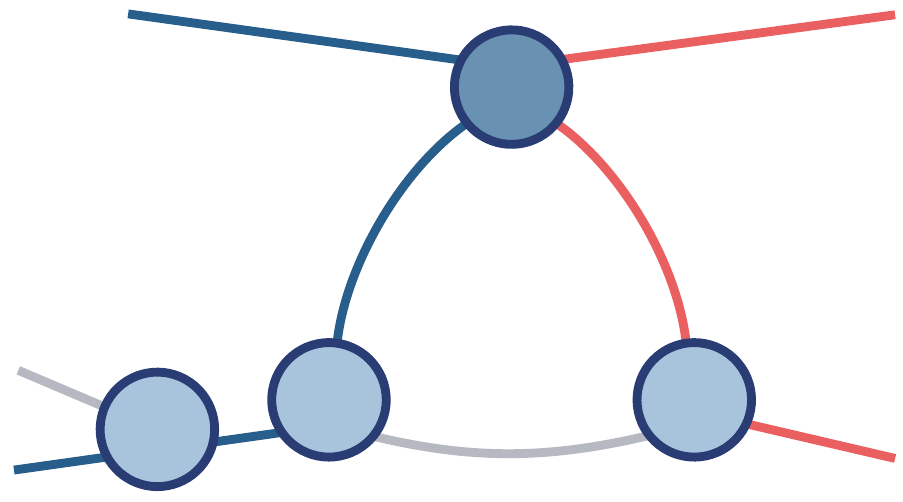}
\caption*{NMC8}
\end{subfigure}
~
\begin{subfigure}[b]{0.22\textwidth}
\includegraphics[scale=0.4]{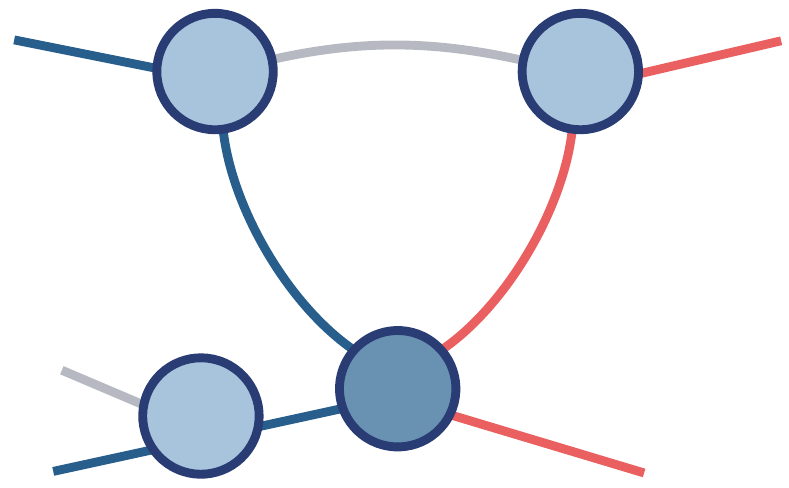}
\caption*{NMC9}
\end{subfigure}
~
\begin{subfigure}[b]{0.22\textwidth}
\includegraphics[scale=0.4]{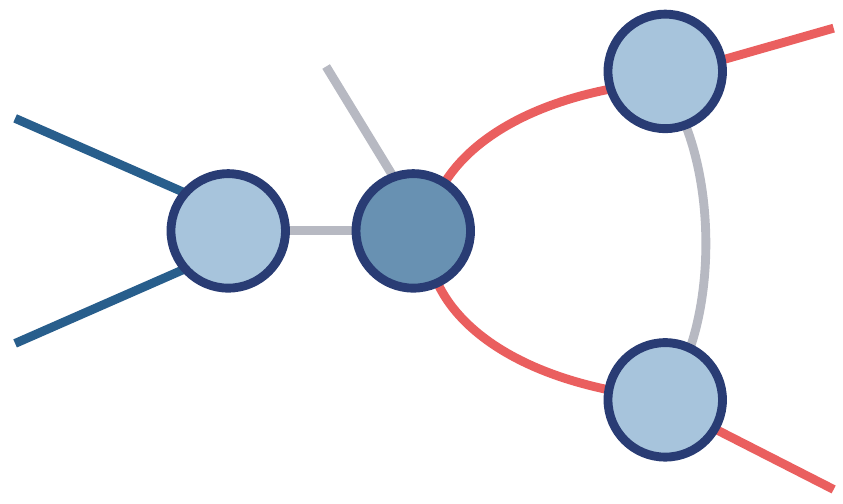}
\caption*{NMC10}
\end{subfigure}
~
\begin{subfigure}[b]{0.22\textwidth}
\includegraphics[scale=0.4]{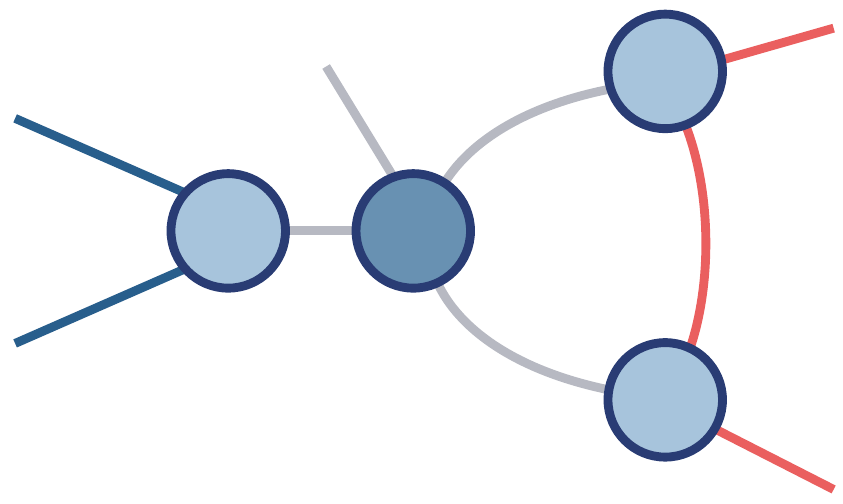}
\caption*{NMC11}
\end{subfigure}

\begin{subfigure}[b]{0.22\textwidth}
\includegraphics[scale=0.4]{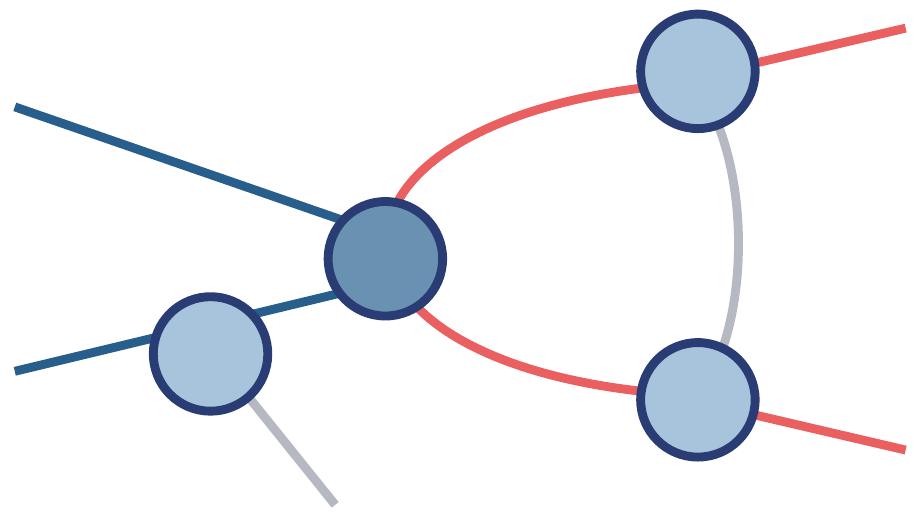}
\caption*{NMC12}
\end{subfigure}
~
\begin{subfigure}[b]{0.22\textwidth}
\includegraphics[scale=0.4]{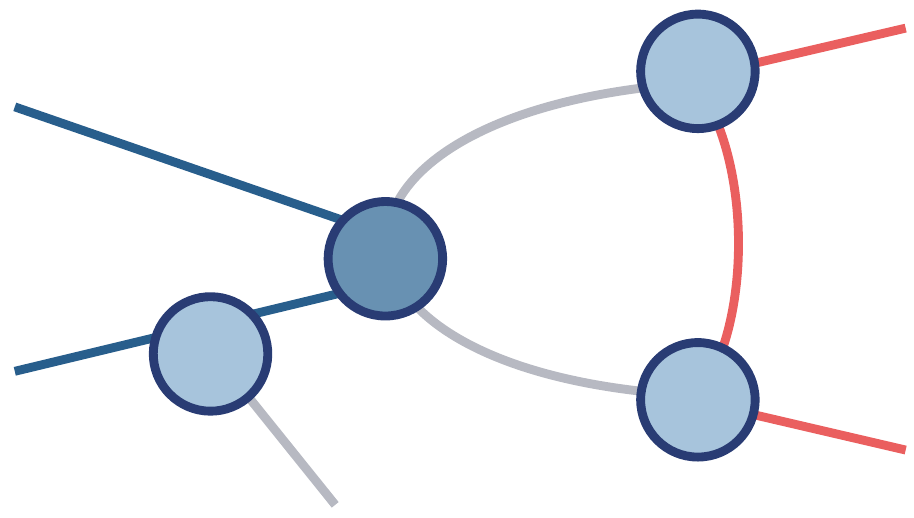}
\caption*{NMC13}
\end{subfigure}
~
\begin{subfigure}[b]{0.22\textwidth}
\includegraphics[scale=0.4]{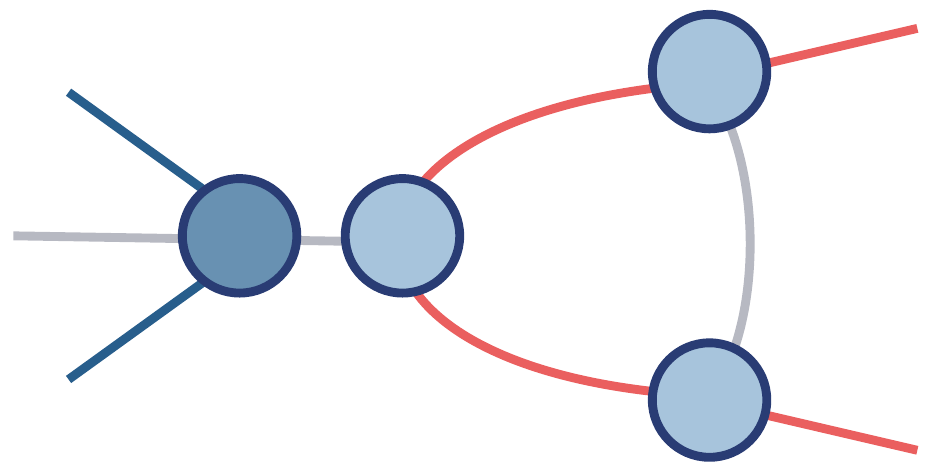}
\caption*{NMC14}
\end{subfigure}
~
\begin{subfigure}[b]{0.22\textwidth}
\includegraphics[scale=0.4]{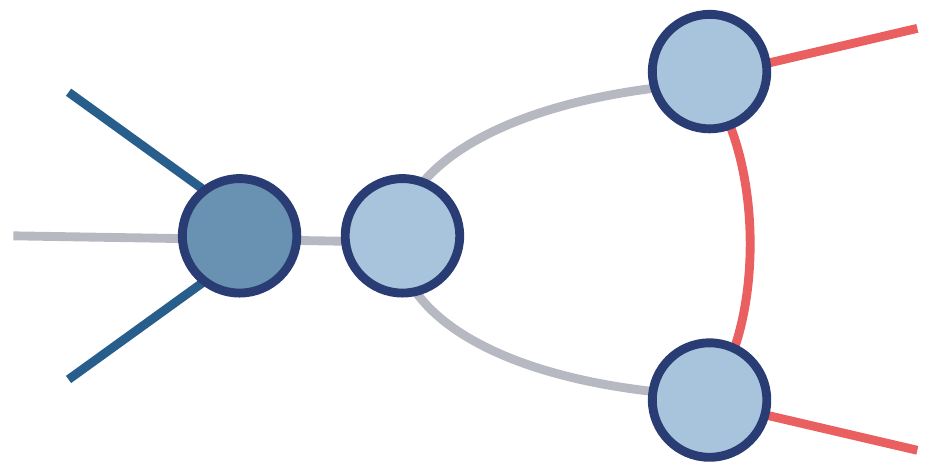}
\caption*{NMC15}
\end{subfigure}

\begin{subfigure}[b]{0.22\textwidth}
\includegraphics[scale=0.4]{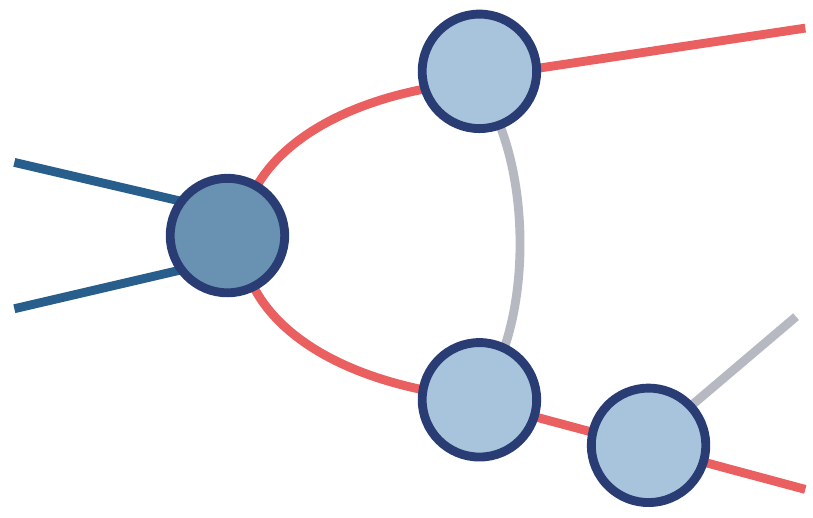}
\caption*{NMC16}
\end{subfigure}
~
\begin{subfigure}[b]{0.22\textwidth}
\includegraphics[scale=0.4]{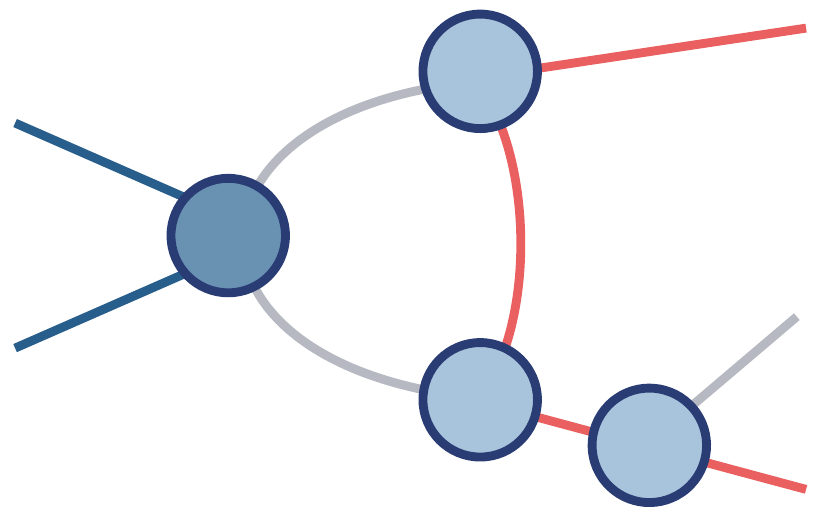}
\caption*{NMC17}
\end{subfigure}
~
\begin{subfigure}[b]{0.22\textwidth}
\includegraphics[scale=0.4]{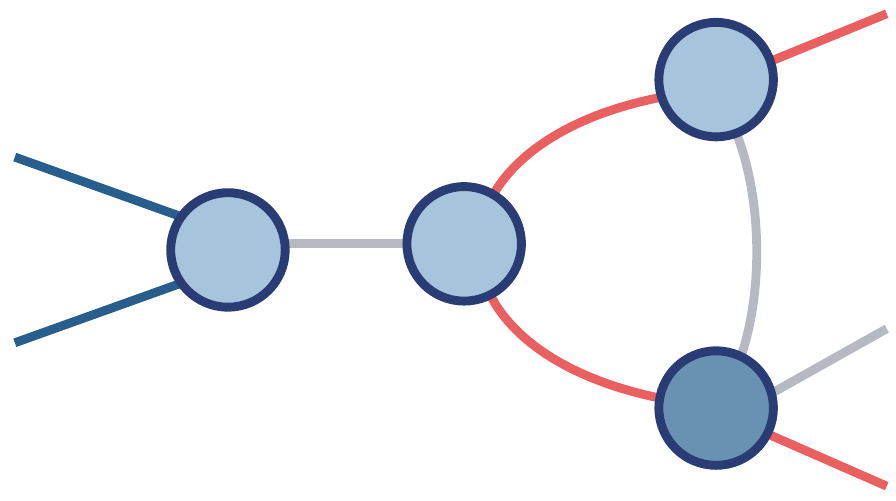}
\caption*{NMC18}
\end{subfigure}
~
\begin{subfigure}[b]{0.22\textwidth}
\includegraphics[scale=0.4]{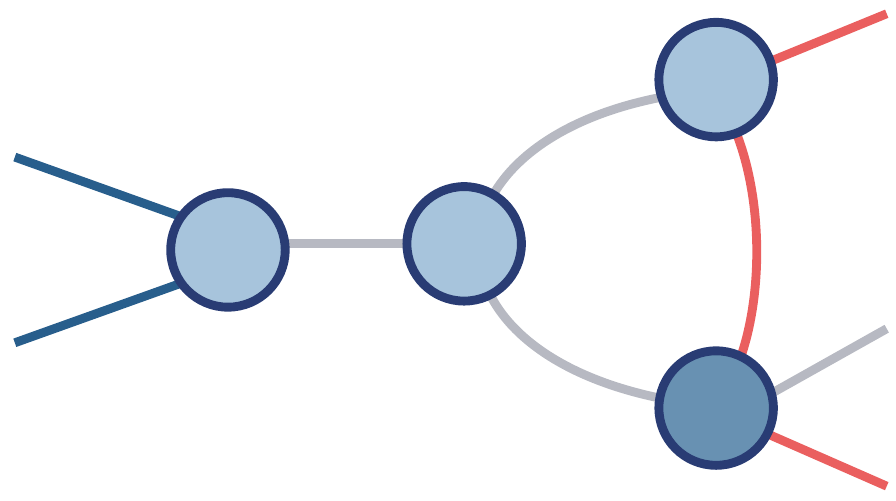}
\caption*{NMC19}
\end{subfigure}

\begin{subfigure}[b]{0.22\textwidth}
\includegraphics[scale=0.4]{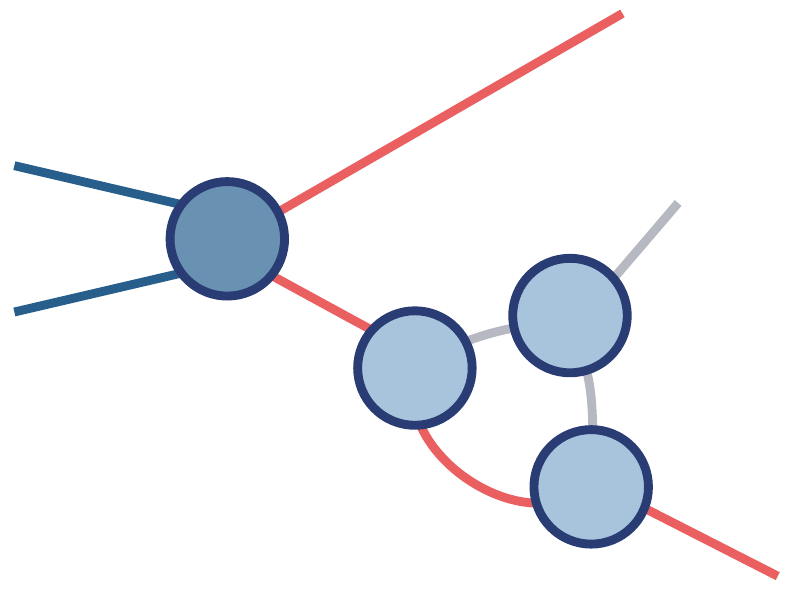}
\caption*{NMC20}
\end{subfigure}
~
\begin{subfigure}[b]{0.22\textwidth}
\includegraphics[scale=0.4]{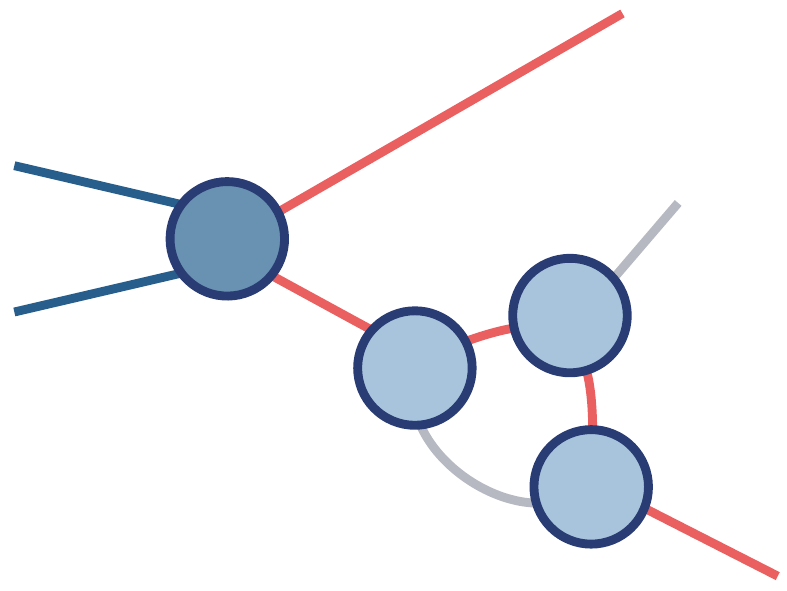}
\caption*{NMC21}
\end{subfigure}
~
\begin{subfigure}[b]{0.22\textwidth}
\includegraphics[scale=0.4]{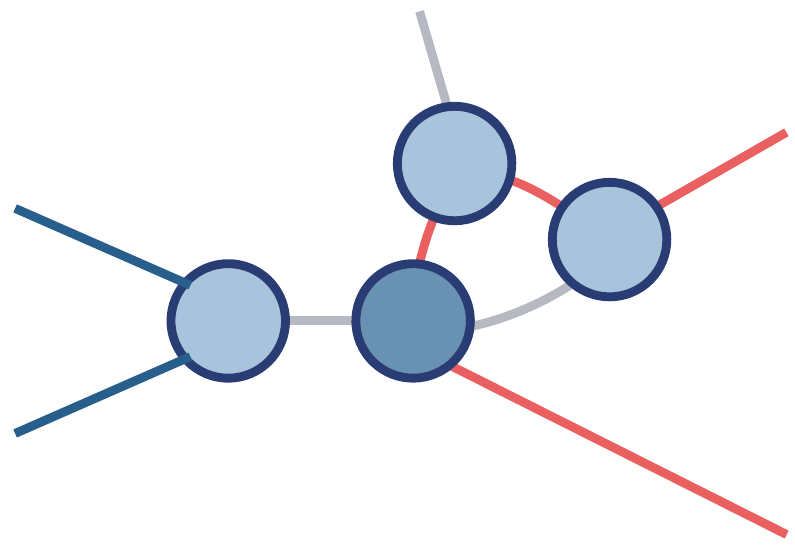}
\caption*{NMC22}
\end{subfigure}
~
\begin{subfigure}[b]{0.22\textwidth}
\includegraphics[scale=0.4]{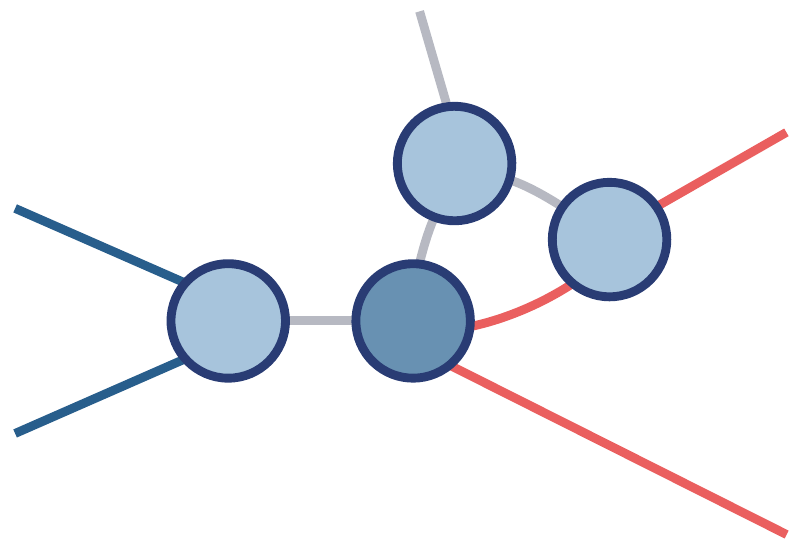}
\caption*{NMC23}
\end{subfigure}

\begin{subfigure}[b]{0.22\textwidth}
\includegraphics[scale=0.4]{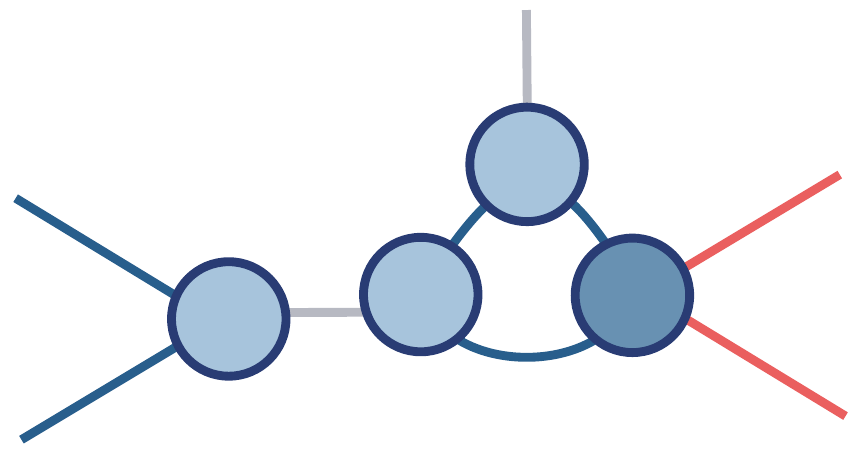}
\caption*{NMC24}
\end{subfigure}
~
\begin{subfigure}[b]{0.22\textwidth}
\includegraphics[scale=0.4]{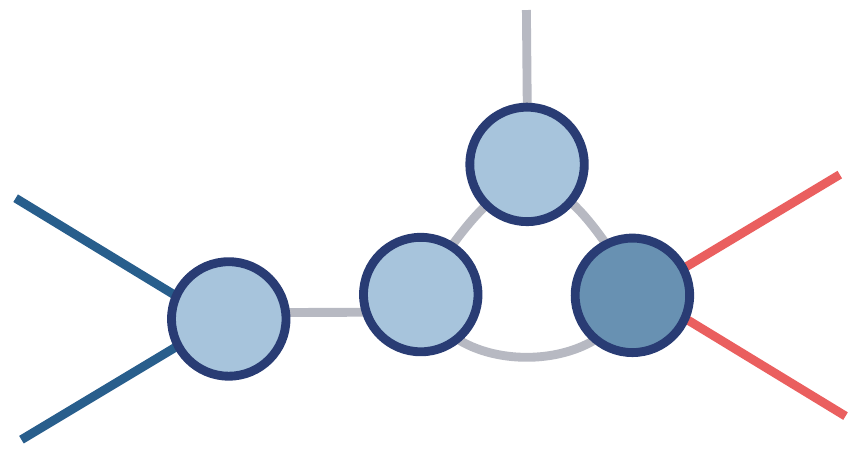}
\caption*{NMC25}
\end{subfigure}
~
\begin{subfigure}[b]{0.22\textwidth}
\includegraphics[scale=0.4]{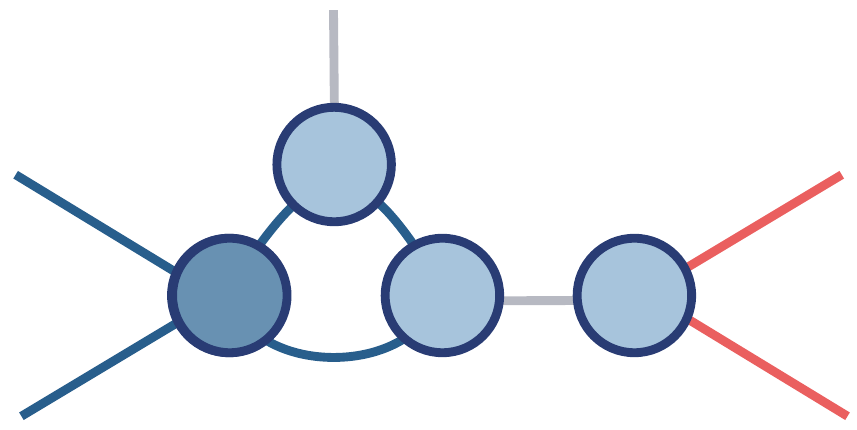}
\caption*{NMC26}
\end{subfigure}
~
\begin{subfigure}[b]{0.22\textwidth}
\includegraphics[scale=0.4]{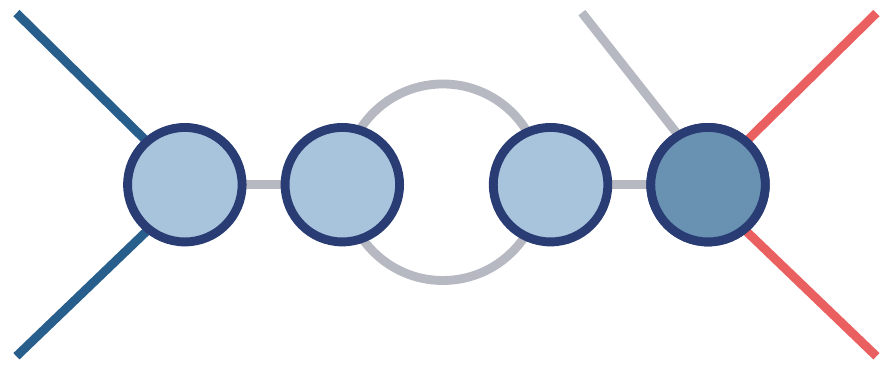}
\caption*{NMC27}
\end{subfigure}

\begin{subfigure}[b]{0.22\textwidth}
\includegraphics[scale=0.4]{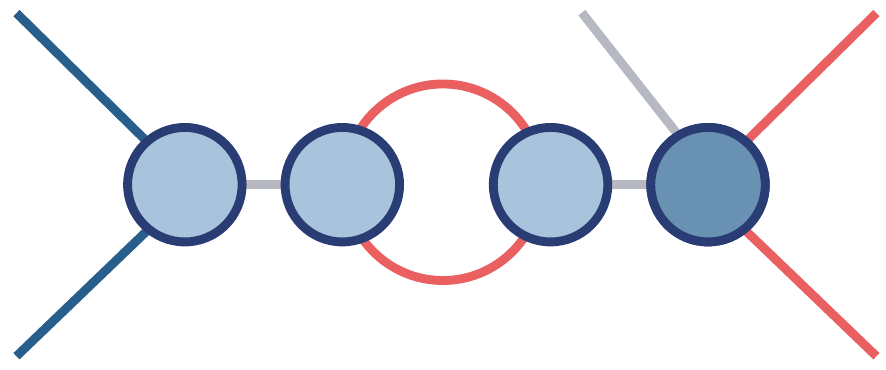}
\caption*{NMC28}
\end{subfigure}
~
\begin{subfigure}[b]{0.22\textwidth}
\includegraphics[scale=0.4]{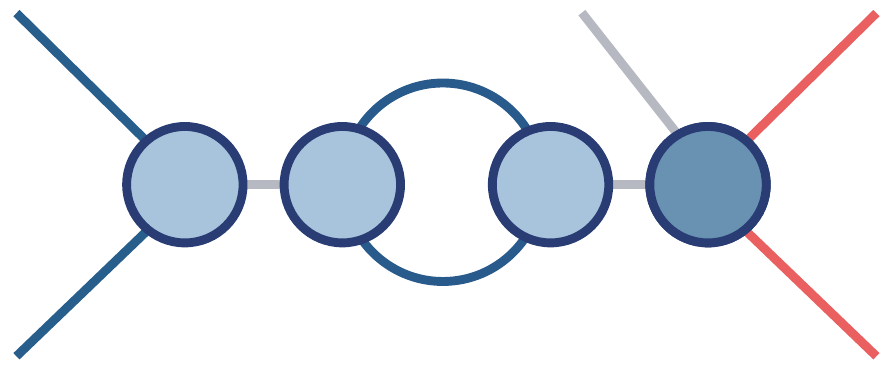}
\caption*{NMC29}
\end{subfigure}
~
\begin{subfigure}[b]{0.22\textwidth}
\includegraphics[scale=0.4]{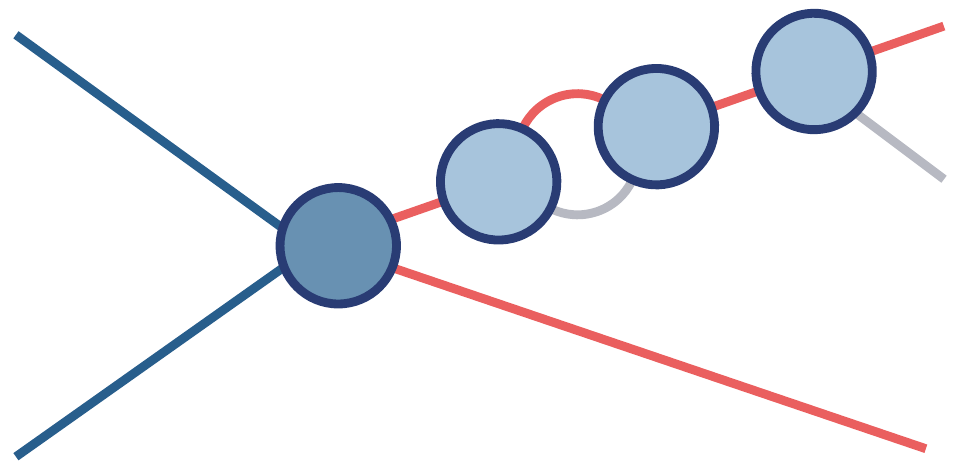}
\caption*{NMC30}
\end{subfigure}

\caption{Next-to-maximal cuts of the one-loop five-point amplitude. Light blue blobs represent three-point amplitudes, and darker blobs represent four and five-point amplitudes. Exposed legs represent on-shell propagators.}
\label{Fig: 5p NMax non-box}
\end{figure}

\begin{figure}
\begin{subfigure}[b]{0.22\textwidth}
\includegraphics[scale=0.4]{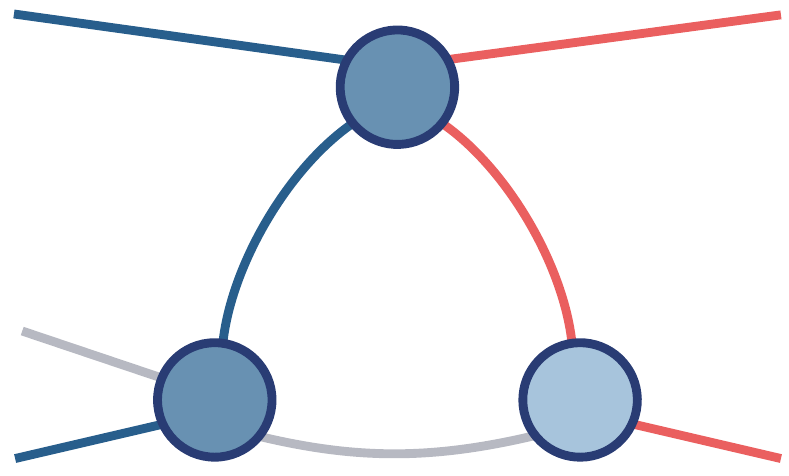}
\caption*{N$^2$MC1}
\end{subfigure}
~
\begin{subfigure}[b]{0.22\textwidth}
\includegraphics[scale=0.4]{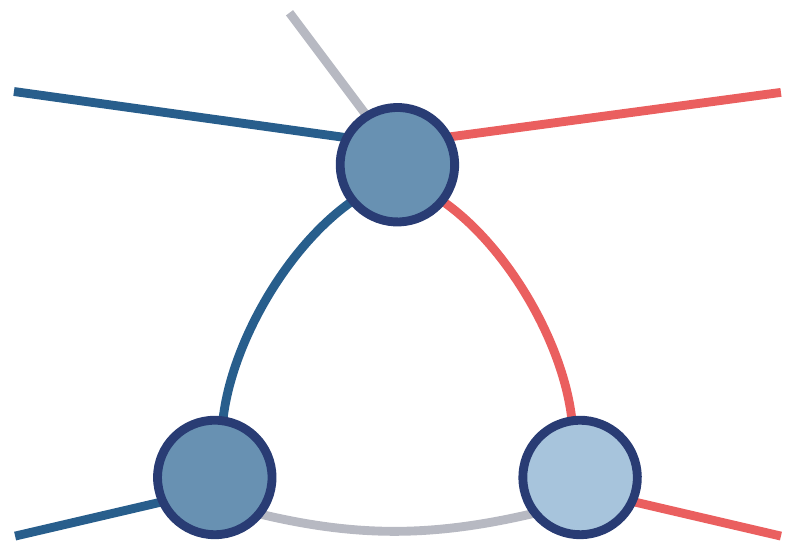}
\caption*{N$^2$MC2}
\end{subfigure}
~
\begin{subfigure}[b]{0.22\textwidth}
\includegraphics[scale=0.4]{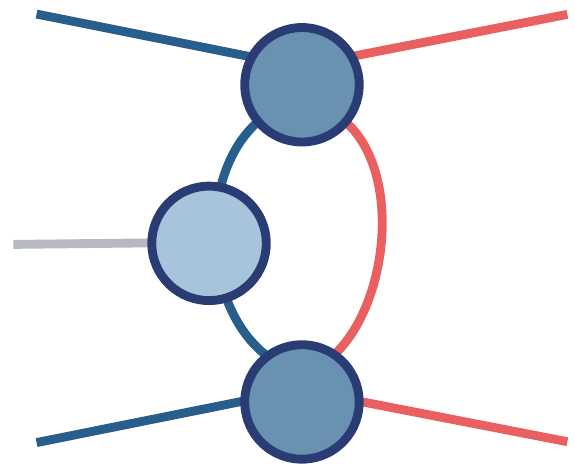}
\caption*{N$^2$MC3}
\end{subfigure}

\begin{subfigure}[b]{0.22\textwidth}
\includegraphics[scale=0.4]{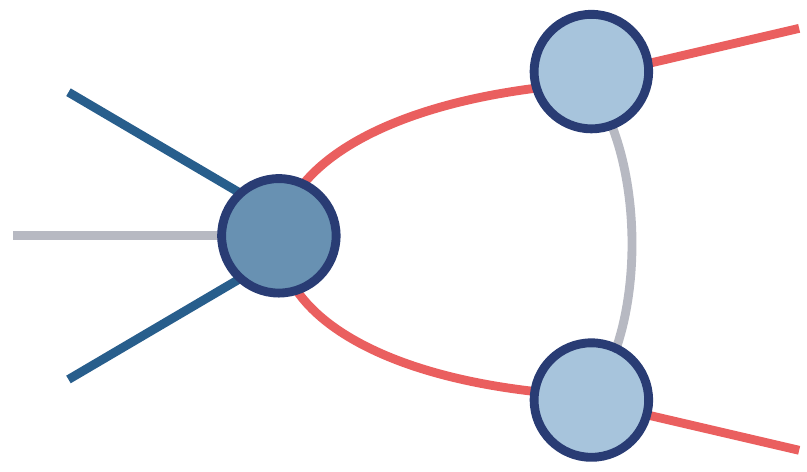}
\caption*{N$^2$MC4}
\end{subfigure}
~
\begin{subfigure}[b]{0.22\textwidth}
\includegraphics[scale=0.4]{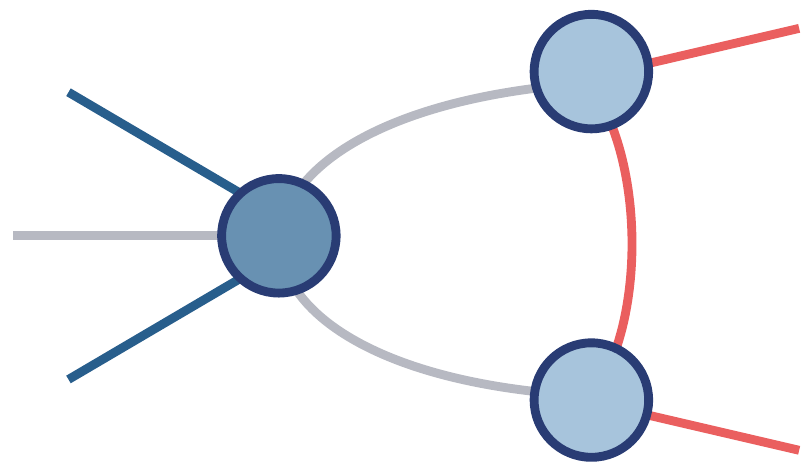}
\caption*{N$^2$MC5}
\end{subfigure}
~
\begin{subfigure}[b]{0.22\textwidth}
\includegraphics[scale=0.4]{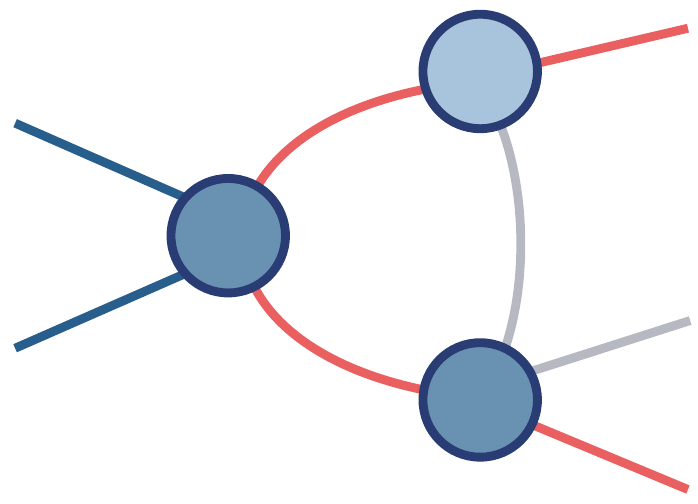}
\caption*{N$^2$MC6}
\end{subfigure}
~
\begin{subfigure}[b]{0.22\textwidth}
\includegraphics[scale=0.4]{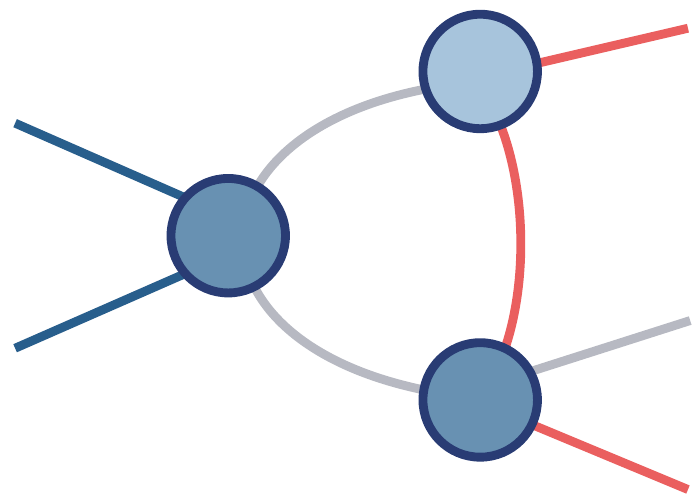}
\caption*{N$^2$MC7}
\end{subfigure}

\begin{subfigure}[b]{0.22\textwidth}
\includegraphics[scale=0.4]{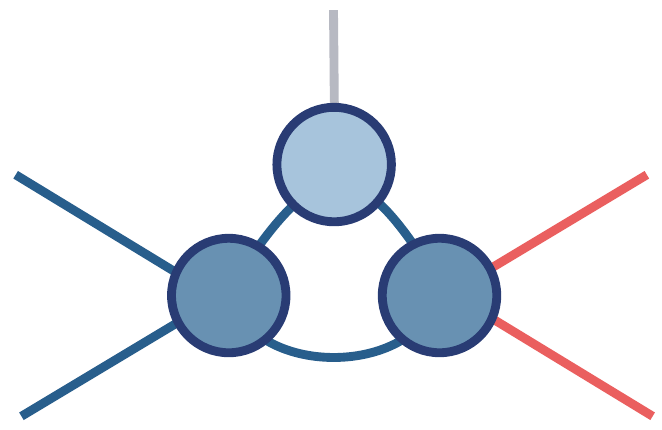}
\caption*{N$^2$MC8}
\end{subfigure}
~
\begin{subfigure}[b]{0.22\textwidth}
\includegraphics[scale=0.4]{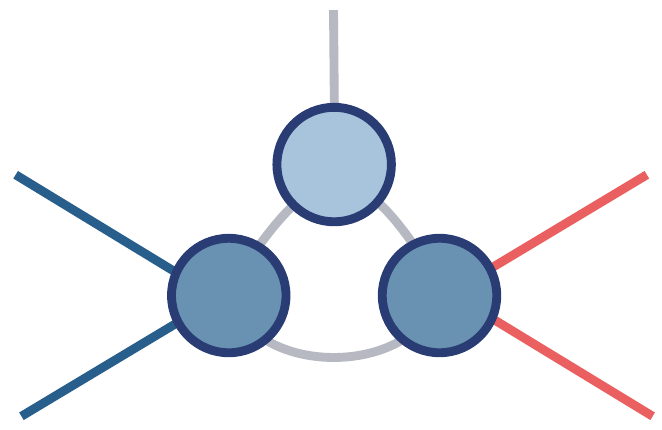}
\caption*{N$^2$MC9}
\end{subfigure}
~
\begin{subfigure}[b]{0.22\textwidth}
\includegraphics[scale=0.4]{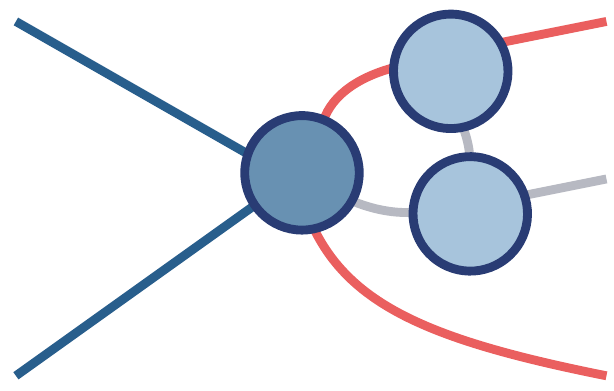}
\caption*{N$^2$MC10}
\end{subfigure}
~
\begin{subfigure}[b]{0.22\textwidth}
\includegraphics[scale=0.4]{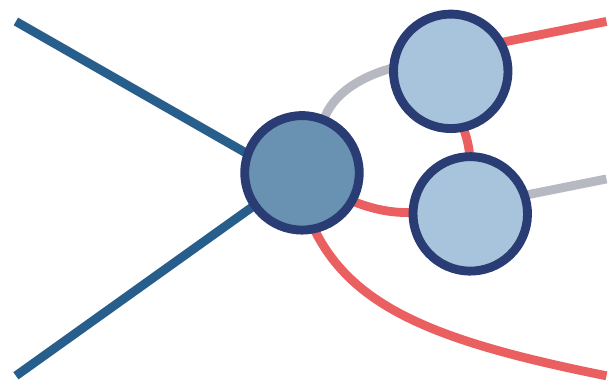}
\caption*{N$^2$MC11}
\end{subfigure}

\caption{N$^2$-maximal triangle cuts of the one-loop five-point amplitude. Light blue blobs represent three-point amplitudes, and darker blobs represent four and five-point amplitudes. Exposed legs represent on-shell propagators.}
\label{Fig: 5p n2max triangle}
\end{figure}

\begin{figure}
\begin{subfigure}[b]{0.22\textwidth}
\includegraphics[scale=0.4]{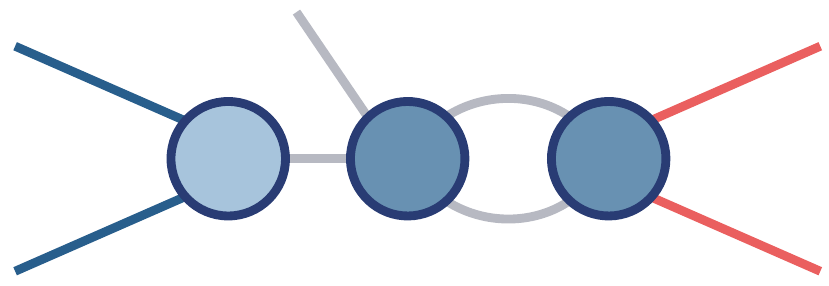}
\caption*{N$^2$MC12}
\end{subfigure}
~
\begin{subfigure}[b]{0.22\textwidth}
\includegraphics[scale=0.4]{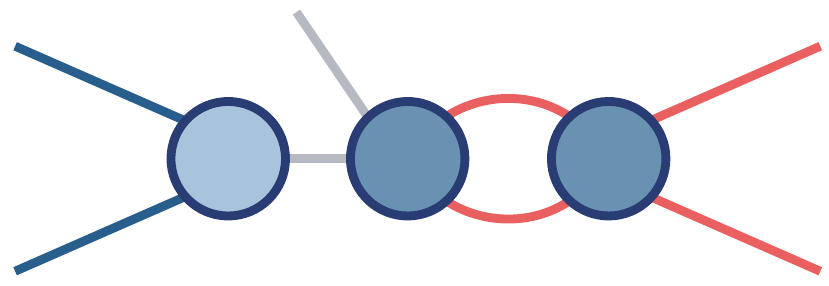}
\caption*{N$^2$MC13}
\end{subfigure}
~
\begin{subfigure}[b]{0.22\textwidth}
\includegraphics[scale=0.4]{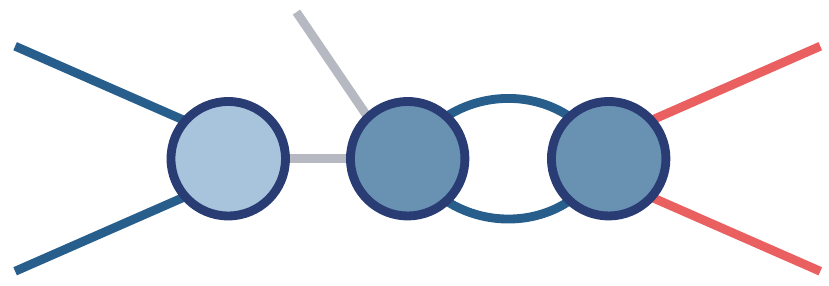}
\caption*{N$^2$MC14}
\end{subfigure}
~
\begin{subfigure}[b]{0.22\textwidth}
\includegraphics[scale=0.4]{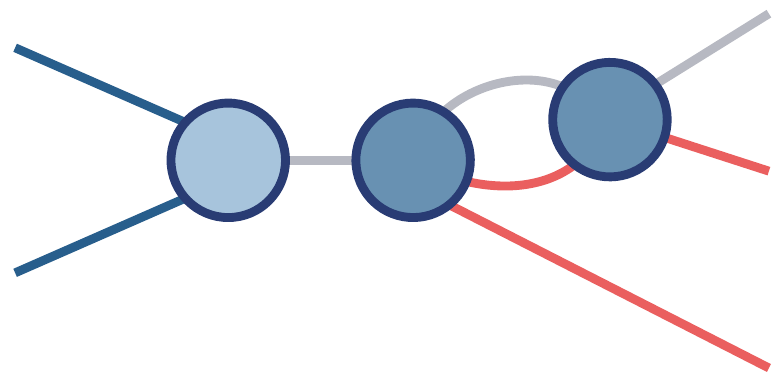}
\caption*{N$^2$MC15}
\end{subfigure}

\begin{subfigure}[b]{0.22\textwidth}
\includegraphics[scale=0.4]{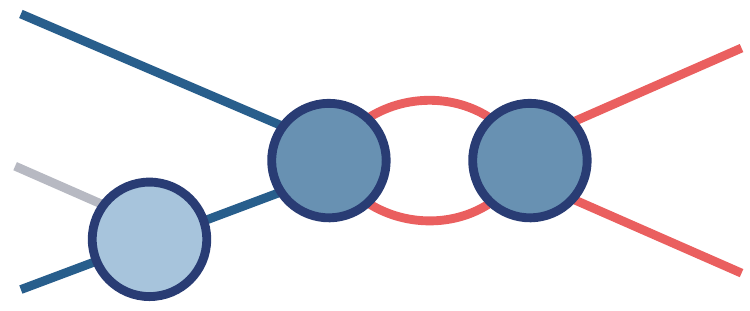}
\caption*{N$^2$MC16}
\end{subfigure}
~
\begin{subfigure}[b]{0.22\textwidth}
\includegraphics[scale=0.4]{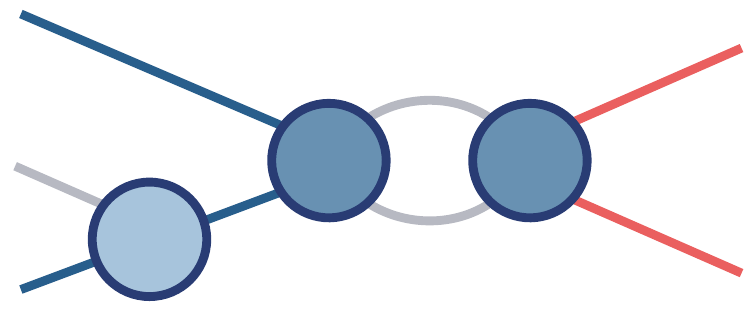}
\caption*{N$^2$MC17}
\end{subfigure}
~
\begin{subfigure}[b]{0.22\textwidth}
\includegraphics[scale=0.4]{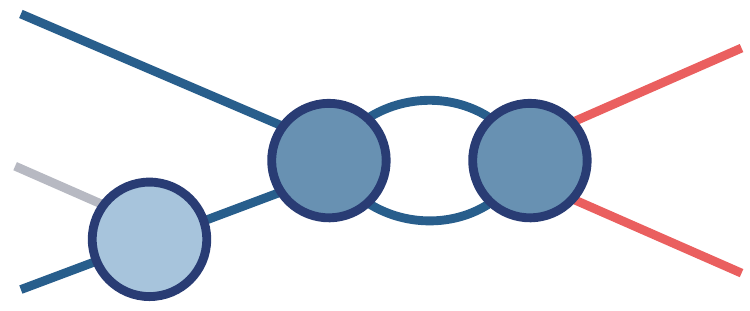}
\caption*{N$^2$MC18}
\end{subfigure}
~
\begin{subfigure}[b]{0.22\textwidth}
\includegraphics[scale=0.4]{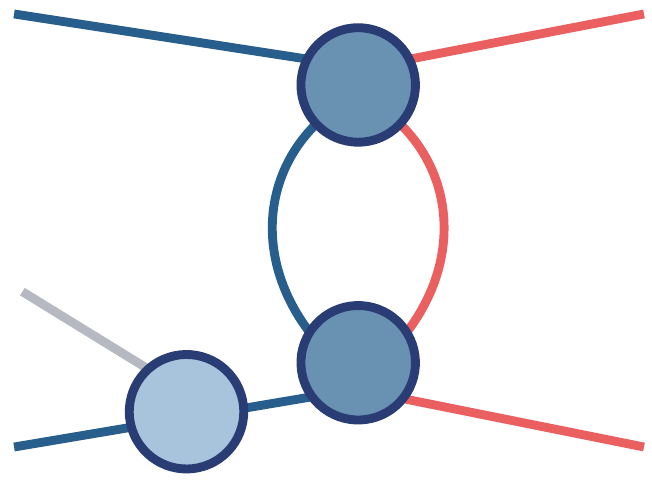}
\caption*{N$^2$MC19}
\end{subfigure}

\caption{N$^2$-maximal cuts of the one-loop five-point amplitude. Light blue blobs represent three-point amplitudes, and darker blobs represent four and five-point amplitudes. Exposed legs represent on-shell propagators.}
\label{Fig: 5p n2max non-triangle}
\end{figure}

\begin{figure}
\begin{subfigure}[b]{0.18\textwidth}
\includegraphics[scale=0.4]{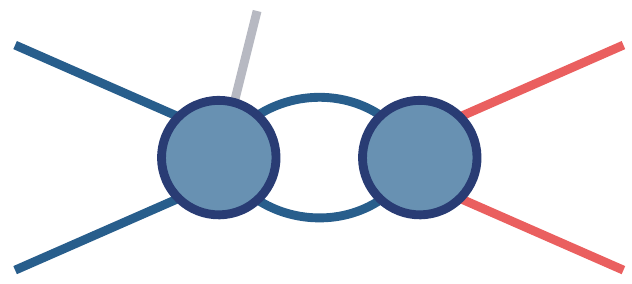}
\caption*{N$^3$MC1}
\end{subfigure}
~
\begin{subfigure}[b]{0.18\textwidth}
\includegraphics[scale=0.4]{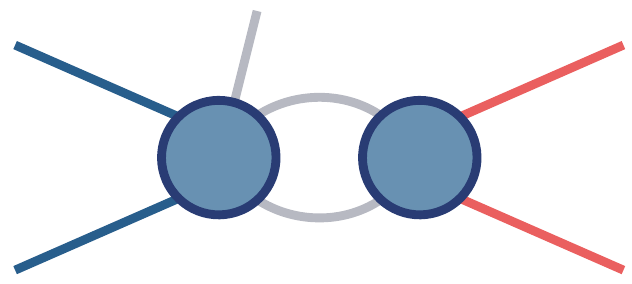}
\caption*{N$^3$MC2}
\end{subfigure}
~
\begin{subfigure}[b]{0.18\textwidth}
\includegraphics[scale=0.4]{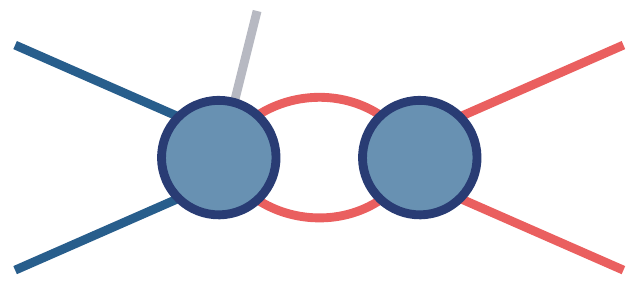}
\caption*{N$^3$MC3}
\end{subfigure}
~
\begin{subfigure}[b]{0.18\textwidth}
\includegraphics[scale=0.4]{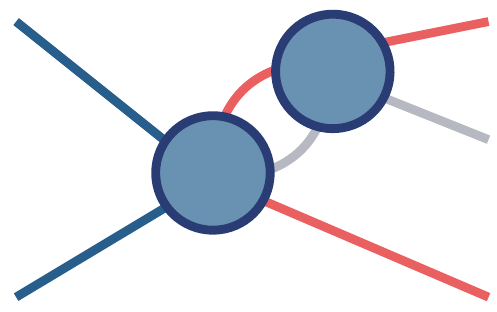}
\caption*{N$^3$MC4}
\end{subfigure}
~
\begin{subfigure}[b]{0.18\textwidth}
\includegraphics[scale=0.4]{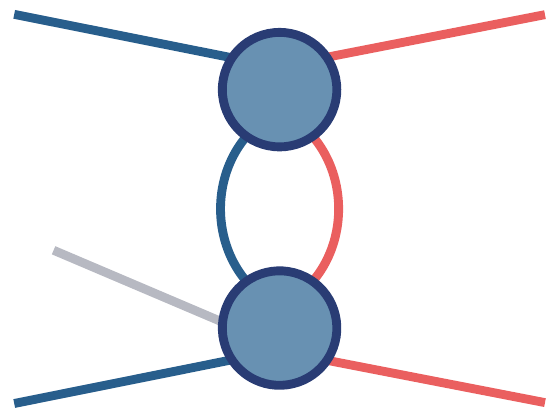}
\caption*{N$^3$MC5}
\end{subfigure}

\caption{N$^3$-maximal bubble cuts of the one-loop five-point amplitude. Dark blue blobs represent four and five-point amplitudes. Exposed legs represent on-shell propagators.}
\label{Fig: 5p n3max cuts}
\end{figure}


\section{Conclusion}
\label{conclusion}

In this paper we demonstrate that the method of maximal cuts in combination with double-copy can be used to bootstrap graph dressings in  Einstein-Hilbert gravity from the double copy of scalar QCD, in a combined procedure we refer to as projective double-copy.  Specifically, we systematize the contribution of unitarity cuts of the amplitudes where only gravitational states are allowed to cross the cuts, while allowing the double-copy to provide all relevant gravitational contact terms.  Starting with the maximal cuts where all propagators are put on-shell, we work our way up to the necessary $N^k$-maximal cuts by systematically releasing cut conditions on propagators. This allows us to project out all non-gravitational states from the graph dressings. For tree-level amplitudes with a single massive scalar we find that no correction is necessary up to five-point, while tree- and one-loop amplitudes with more than one massive scalar do contain excess states that are removed.   We expect these integrands to be useful to understanding and developing optimized approaches to previously uncalculated classical predictions relevant to gravitational wave astrophysics.

The success and relative simplicity of this procedure promises possible verification for future double copy schemes. It would be intriguing to investigate whether including massless ghostly matter in the original theory in a strategic manner will remove the excess states at loop-level, and the method of maximal cuts can here provide a a straight-forward comparison for the resulting amplitudes.

\section{Acknowledgements}
We would like to thank Donal O'Connell for encouragement and detailed discussions regarding classical limits. IVH is supported by the European Union’s Horizon 2020 research and innovation programme under the Marie Skłodowska-Curie grant agreement No. 764850 (SAGEX).  This work was supported by the DOE under contract DE-SC0021485 and by the Alfred P. Sloan Foundation.

\bibliography{GravityProjectiveDoubleCopy}

\end{document}